\documentclass[%
aip,
 sd,%
 amsmath,amssymb,
preprint,
author-year,
]{revtex4-1}
\usepackage{graphicx}
\usepackage{natbib}
\usepackage{tikz}
\usepackage{pgflibrarysnakes}
\usepackage{pgflibraryplothandlers}
\usepackage{dcolumn}%
\usepackage{amsmath}
\usepackage{amsbsy}
\usepackage{amssymb}
\usepackage{mathrsfs}
\usepackage{enumerate}
\usepackage{xcolor}
\usepackage{color}
\usepackage{fancyhdr}
\usepackage{subfigure}
\usepackage{bm}

\newcommand{\BibitemShut}[1]{}

\begin{document}


\title[Oscillating wing-section in transonic flow]{Viscosity, shock-induced separation, and shock reversal---\\Oscillating wing section in transonic flow}

\author{Pradeepa T. Karnick}
\email{pradeepatk@umlaut.com}
\affiliation{Umlaut, Bengaluru}

\author{Kartik Venkatraman}%
\email{kartik@iisc.ac.in}
\affiliation{Department of Aerospace Engineering, Indian Institute of Science, Bengaluru, India 560012}

\date{\today}

\begin{abstract}
We numerically examine the mechanisms that describe the shock-boundary layer interactions in transonic flow past an oscillating wing section. At moderate and high angles of incidence but low amplitudes of oscillation, shock induced flow separation or shock-stall is observed accompanied by shock reversal. Even though the power input to the airfoil by the viscous forces is three orders of magnitude lower than that due to the pressure forces on the airfoil, the boundary layer manipulates the shock location and shock motion and redistributes the power input to the airfoil by the pressure forces. The shock motion is reversed relative to that in an inviscid flow as the boundary layer cannot sustain an adverse pressure gradient posed by the shock, causing the shock to move upstream leading to an early separation. The shock motion shows a phase difference with reference to the airfoil motion and is a function of the frequency of the oscillation. At low angles of incidence, and low amplitudes of oscillation, the boundary layer changes the profile presented to the external flow, leads to a slower expansion of the flow resulting in an early shock, and a diffused shock-foot caused by the boundary layer.
\end{abstract}

\maketitle

\section{\label{sec:intro}Introduction}

Transonic flows are characterized by the presence of mixed regions of subsonic and supersonic flows. The supersonic flow is usually terminated by a shock. These flow features alter the forces exerted on a lifting surface. The appearance of the shock increases the drag force considerably. But perhaps more significant is the interaction of the shock wave with the boundary layer. The viscous boundary layer can change the strength and position of the shock; and if the shock induces a very high pressure gradient across it then this leads to drastic changes in flow features. Shock-stall is one such phenomena wherein shock induced separation leads to aerodynamic stall at  low angles of attack. A prescribed motion to an airfoil that is shock-stalled could significantly alter the direction of shock motion and thereby the aerodynamic loading on the airfoil, when compared to those observed in simulations for an inviscid flow.

A shock on the airfoil surface in a transonic flow drastically changes the aerodynamic forces generated by the flow. A prescribed motion to the airfoil, or the motion of the flap attached to the airfoil, leads to movement of the shock on the airfoil surface. In a seminal set of experiments, \citet[Chap.9]{tijdeman1977}, \citep{tijdeman1980} conducted wind tunnel experiments to study these shock dynamics. The shock motion was simulated by oscillating the flap of an airfoil. Three kinds of shock motion were observed. Of the three, the first was the sinusoidal shock wave motion. This occurred at relatively high super-critical free stream Mach number, and the variation of strength and motion of the shock was almost sinusoidal. The shock was present throughout the cycle of oscillation. These shocks are classified as type A shocks. The second type of shock motion observed, known as type B shock motion, was an interrupted shock wave motion. These were observed at a lower super-critical free stream Mach number flow than the type A shock motion. During a part of the sinusoidal motion of the flap, the shock disappeared from the airfoil surface. The third category of shock wave motion, type C, are the upstream propagated shocks. The flow is at a lower free stream Mach number than type A and B shock motion, but is still super-critical. However the upstream movement of the shock leads to a decrease in the size of the supersonic region over the upper surface that ultimately leads to a decrease in lift. The flap motion causes the shock to form, followed by the upstream motion of the shock with increasing strength. The shock strength starts to decrease with further increase in flap angle, but the upstream motion of the shock continues till it leaves the airfoil from its leading edge. The shock motion over the surface of an oscillating airfoil are similar to those observed by \citet{tijdeman1977}, and therefore we will classify shock oscillations using the three prototype A, B, and C shock oscillations described there. The authors restricted their experiments to flows without stall. But in this study, we have simulated shock dynamics in the shock stall case too. It is important to simulate shock dynamics accurately as its location and strength plays an important role in deciding the aerodynamic loading of the lifting surface.

Viscosity has influence on the transonic flow field especially when the shock is strong enough to cause separation. \citet{martinelli1986} showed that viscosity causes a shift in the shock location by as much as $20\%$ of the airfoil chord length by solving the Reynolds averaged Navier-Stokes (RANS) equations together with the \citet{baldwin1978} turbulence model. Since the shock moves upstream, the shock strength is lower compared to inviscid results. A mention of shock-reversal observed in experiments during shock induced separation is made by \citet{bendiksen2009}, though their own work involved simulating inviscid flow over wings and airfoils. 

Transonic buffeting over a stationary airfoil or wing is similar to transonic flow over an oscillating wing. Buffeting or self-sustained shock motion on a stationary airfoil surface due to unsteady transonic flow has been the subject of intense study for more than fifty years, and wind-tunnel experiments as well as numerical simulations are well documented \citep{lee2001}. The shock motion in buffeting is very similar to those observed on oscillating airfoils in a transonic flow, and are caused by shock boundary layer interactions and shock induced separation of the turbulent boundary layer. 

The interaction between shock motion due to transonic buffeting over a and that due to an oscillating wing in a transonic flow could be similar to the entrainment phenomenon that occurs in oscillating solid bodies wherein the vortex shedding frequency gets entrained by the oscillating frequency.  With that motivation, \citet{raveh2008} numerically studied the unsteady flow over a stationary and an oscillating airfoil at large angles of attack using a RANS solver. Large shock oscillations were found at a combination of free stream Mach number and mean angle of attack even though the airfoil was held stationary. Low amplitude airfoil oscillations about this buffeting condition revealed two frequencies in the fluid dynamic response---one was the buffeting frequency and the other due to airfoil oscillation. But beyond a certain amplitude of oscillation the buffeting frequency component vanishes and only the airfoil oscillating frequency remains. The critical oscillation amplitude when this transition occurs is shown to be proportional to the ratio between the airfoil oscillation frequency and the buffeting frequency. 

Turbulent boundary layer interaction with shock motion on oscillating airfoil is likely to give rise to steep and rapid changes in pressure on the airfoil surface, and together with its phase relation to the airfoil motion will determine the aerodynamic power input to the airfoil. In this study, firstly we focus on relating the shock motion with boundary layer separation, and subsequently the influence of shock-boundary layer interaction on the skin-friction and pressure on the airfoil surface, and finally the influence of the pressure and skin-friction forces on the instantaneous power input to the airfoil. We have used an unsteady Reynolds averaged Navier-Stokes solver based on the Spalart-Almaras turbulence model to numerically simulate the flow over pitching airfoils. This solver is validated with experimental test cases wherever possible. We also have investigated the effect of viscosity in changing aerodynamic forces on airfoils by comparing the flow features of both viscous and inviscid results.

The paper is organized as follows. Section~\ref{sec:floweq} outlines the fluid flow equations, the turbulence modeling, and numerical procedures used. The influence on shock and boundary layer is first investigated for a stationary airfoil in Section~\ref{sec:staf} for two test cases. Shock displacement and boundary layer thickening for pitching oscillations of an airfoil, with two different geometries, as well as different Mach numbers and kinematic parameters are the subject of Section~\ref{sec:ch_ust_af}. The shock is not strong enough to cause separation of the boundary layer. However, at a higher mean angle of incidence, shock induced boundary layer separation occurs and is studied in Section~\ref{sec:ch_shck_bl_sep}. Section~\ref{sec:concl} is a summary with some concluding remarks.

\section{\label{sec:floweq}Fluid flow equations and turbulence modeling}

Though the compressible Navier-Stokes equations satisfy instantaneously the complete turbulent flow field, the spatial and temporal scales needed to resolve the turbulent flow is extremely small leading to very large computational simulation time. For engineering applications, computation of moderately gross flow characteristics that leads to an accurate computation of loads on the airfoil are sufficient. Therefore, instead of solving the Navier-Stokes equations for the flow variables, these governing equations are solved for statistically averaged quantities including the effect of fluctuations on mean quantities. 

The conservation equations for the flow in terms of the two-dimensional Favre averaged compressible Navier-Stokes equations are solved together with Sutherland's law that relates dynamic viscosity with temperature, Fourier's law that relates the heat flux with the temperature gradient in the fluid, and turbulence closure approximations for a compressible flow \citep[Chapter 5]{wilcox1994}. The turbulent eddy viscosity term is estimated by solving the \citet{spalart1992} one-equation model. The model equation is further modified in terms of dynamic eddy viscosity $\tilde{\mu}^T = \bar{\rho}\tilde{\nu}^T$ which takes into account variable density \citep[][Chapter 2]{philippe1999}

These equations were non-dimensionalized with free-stream pressure, free-stream density,  and the airfoil chord length as the reference parameters.

A finite volume approach is used for the spatial discretization of the integral form of the governing equations as it satisfies discrete conservation which is necessary for accurate flow computation. Roe's \citep{roe1981} approximate Riemann solver is used for the spatial discretization of the convective part of density weighted time averaged Navier-Stokes equations. Monotone upstream-centered scheme for conservation laws proposed by \citet{vanleer1979} is used to achieve second-order accurate spatial discretization in smooth regions of flow. The van Albada limiter \citep[Chapter 21]{hirsch1994} is used for shock resolution without oscillations. Viscous terms are treated using second-order accurate central-difference scheme.

A second-order accurate implicit time discretization of the flow governing equations is carried out. A relaxation scheme which uses the lower-upper implicit factorization proposed by \citet{jameson1981_c} is implemented for steady flow computations. An implicit dual time-stepping is performed for the unsteady flow computations which solves the algebraic equations obtained from both spatial and temporal discretization as a steady flow problem using the relaxation technique in pseudo-time.

Special care is required while discretizing the turbulence model equation considering stability of the numerical scheme into account. The solution to this equation should always produce a nonnegative turbulent viscosity. \citet{spalart1992} introduced strategies at the discretization stage to achieve this. In our implementation, the Spalart-Allmaras turbulence model equation is decoupled from the Navier-Stokes equation during the solution process. The advection part is discretized using first-order accurate upwind scheme. The dissipative term is discretized using second-order accurate central-difference scheme. To guarantee positivity of $\tilde{\nu}^T$, the production term is integrated explicitly and the destruction term is integrated implicitly. Lastly, the trip term is neglected by assuming the free stream flow to be turbulent.

The fluid velocity on the solid boundary is set equal to the moving solid boundary velocity. The solid wall is assumed adiabatic. The turbulent viscosity is set equal to zero on the solid wall because the Reynolds stresses are zero on the wall. Pressure is extrapolated on the solid surface from the computational domain. Assuming that the flow is subsonic in the free stream, the far-field boundary conditions are imposed using Riemann invariants. If $V_n$ represents the velocity normal to the far-field boundary and $C$ as the speed of sound; with subscript $\infty$ represent free stream quantities and subscript $e$ represent extrapolated quantities from the computational domain, then the Riemann invariants

\begin{equation}
\begin{aligned}
& R_\infty = V_{n_\infty} + \dfrac{2 C_\infty}{\gamma-1}, \\
& R_e = V_{n_e} - \dfrac{2 C_e}{\gamma-1},
\end{aligned}
\end{equation}

correspond to incoming and outgoing waves. The above equations are added and subtracted to give

\begin{equation}
\begin{aligned}
& V_{n_b} =  \dfrac{1}{2} \left( R_\infty + R_e \right), \\
& C_{b} =  \dfrac{\gamma-1}{4} \left( R_\infty - R_e \right).
\end{aligned}
\end{equation}

$V_{n_b}$ and $C_{b}$ are the normal velocity and the speed of sound to be specified at the far-field boundary. At an outflow boundary the tangential velocity and entropy are specified by extrapolation from the computational domain whereas for an inflow boundary they are the free stream values. These quantities give a complete definition of the flow in the far-field. If the flow is supersonic then all the flow quantities are specified as free stream values at the inflow boundary and are extrapolated at the outflow boundary.

Wall bounded turbulence computations require the resolution of flow near the solid surface. The normal grid spacing required for a given wall bounded turbulence computation uses the law of the wall which states that the velocity near a smooth wall is a universal function of the non-dimensional distance $y^+$ from the wall \citep[Chapter 13]{kundu2007}. The approximate grid spacing required for the resolution of flow in terms of the required non-dimensional wall distance is given by

\begin{equation}
\frac{y}{c} = 5.27 Re_\infty^{-\frac{9}{5}} y^+.
\end{equation}

Here $c$ is the length scale for the problem. The viscous sublayer next to the wall which is dominated by viscous effects will range over a distance of five non-dimensional wall distance. Choosing the normal grid spacing depends on the turbulence model used. The Spalart-Allmaras turbulence model requires a grid spacing of less than $2.5y^+$.

The flow solver is validated first for flow over stationary airfoils and then flow over pitching airfoils. Structured grids of C-type are generated around the airfoil in all viscous flow simulations. C-type grids are represented by three numbers $n1 \times n2 \times An3$ with first number $n1$ representing the number of grids around the airfoil and cut-lines, the second number $n2$ representing the number of grids normal to the surface of the airfoil, and the third number $An3$  represents the number of grids on the airfoil surface. Far-field boundary is set at $20$ chord-lengths away from the airfoil for all cases.

\section{\label{sec:staf}Influence of boundary layer on shock for a stationary airfoil}

\begin{figure}
\subfigure[Coefficient of pressure variation on the chord]{\label{steady2_cp}\includegraphics[scale=0.40]{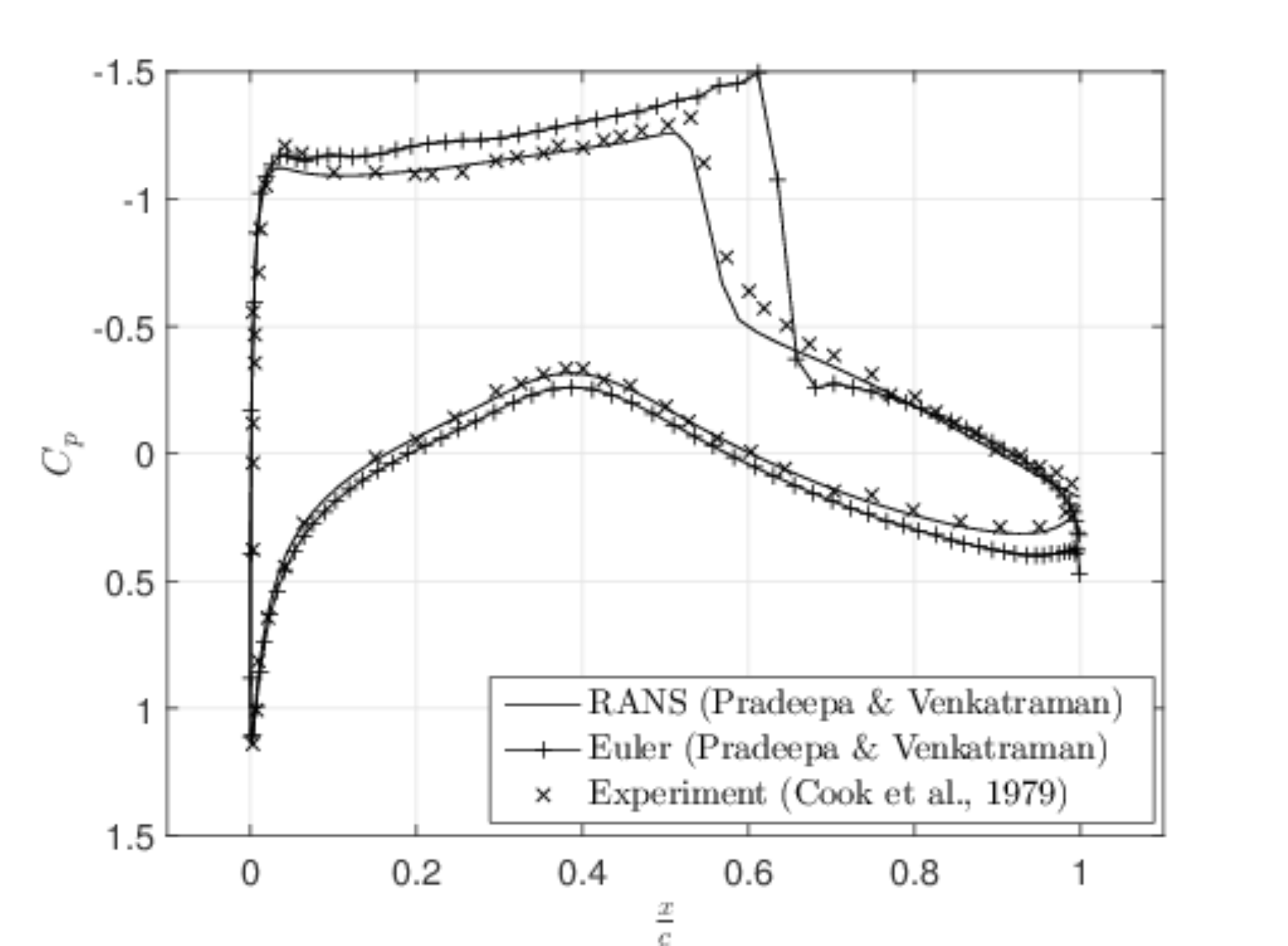}}
\subfigure[Coefficient of friction variation on the chord]{\label{steady2_cf}\includegraphics[scale=0.40]{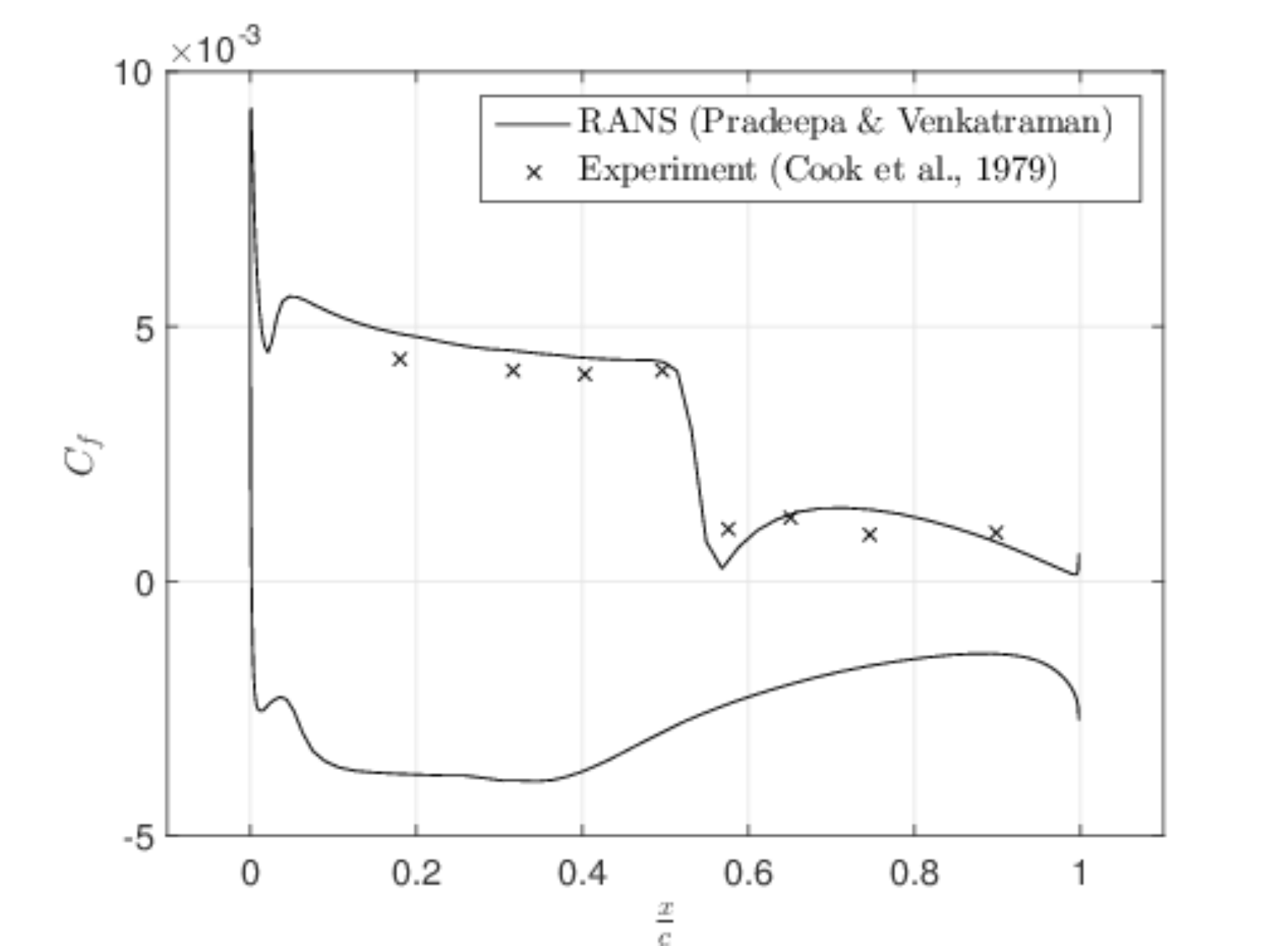}}
\subfigure[Mach contours over the airfoil]{\label{steady2_mach}\includegraphics[scale=0.40]{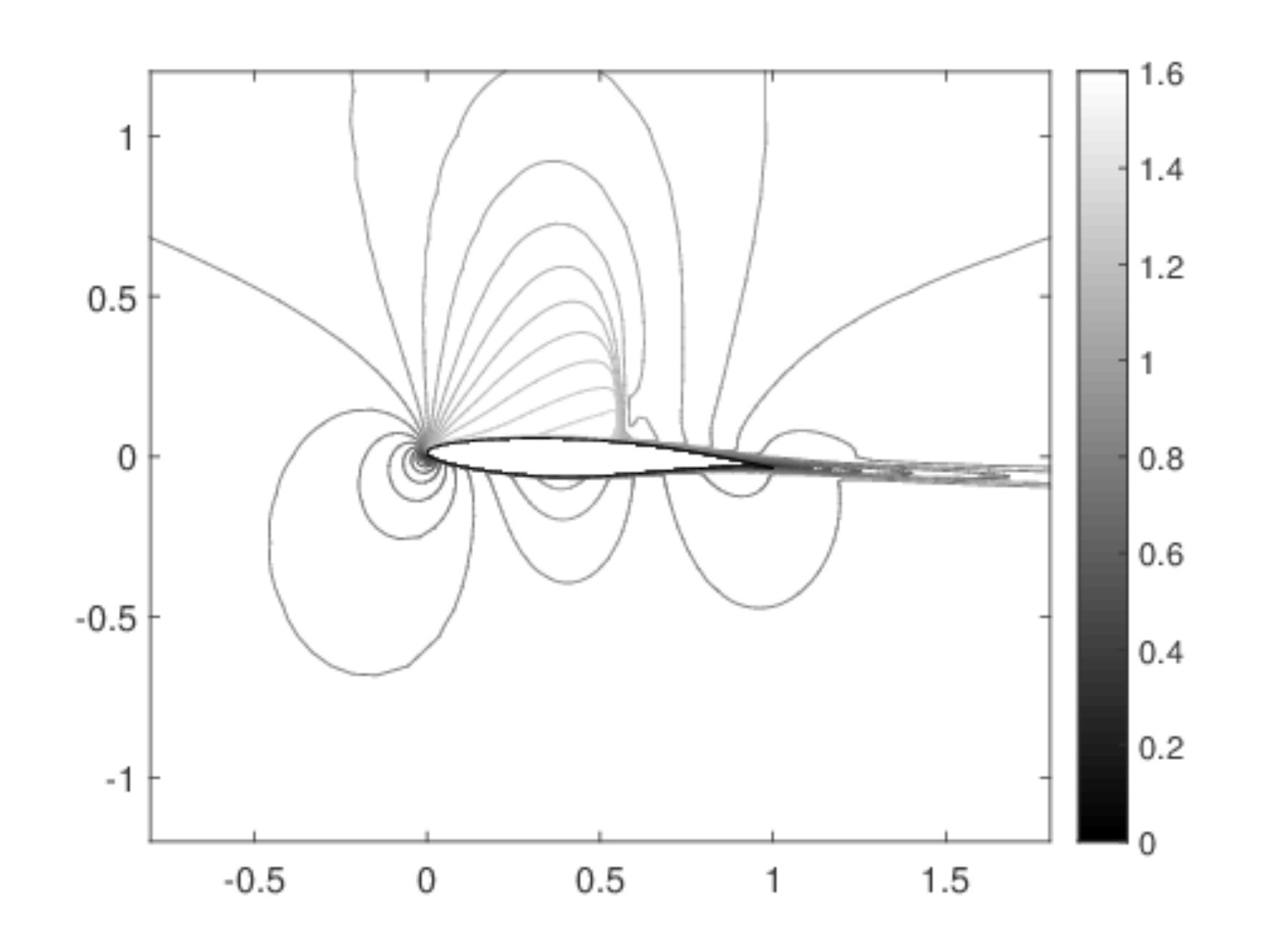}}
\subfigure[Residue decay with iterations]{\label{steady2_residue}\includegraphics[scale=0.40]{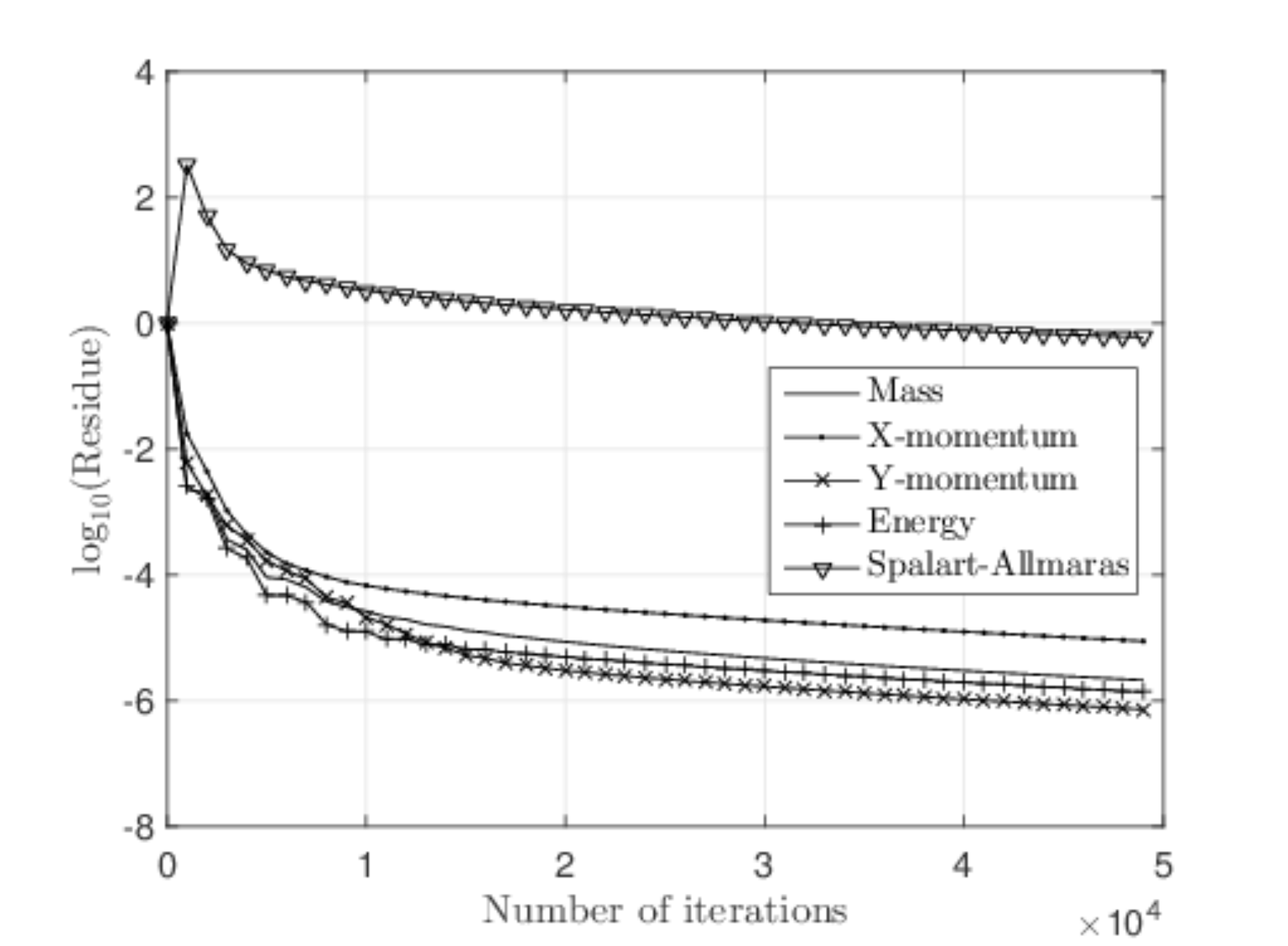}}
\caption{Transonic flow over an RAE2822 airfoil; $M_\infty = 0.73$, $Re_\infty = 6.5\times10^6$ at $\alpha=2.79^\circ$.}
\label{steady2}
\end{figure}

\begin{figure}
\subfigure[Mach contours viscous case]{\label{steady2_mach_close}\includegraphics[scale=0.40]{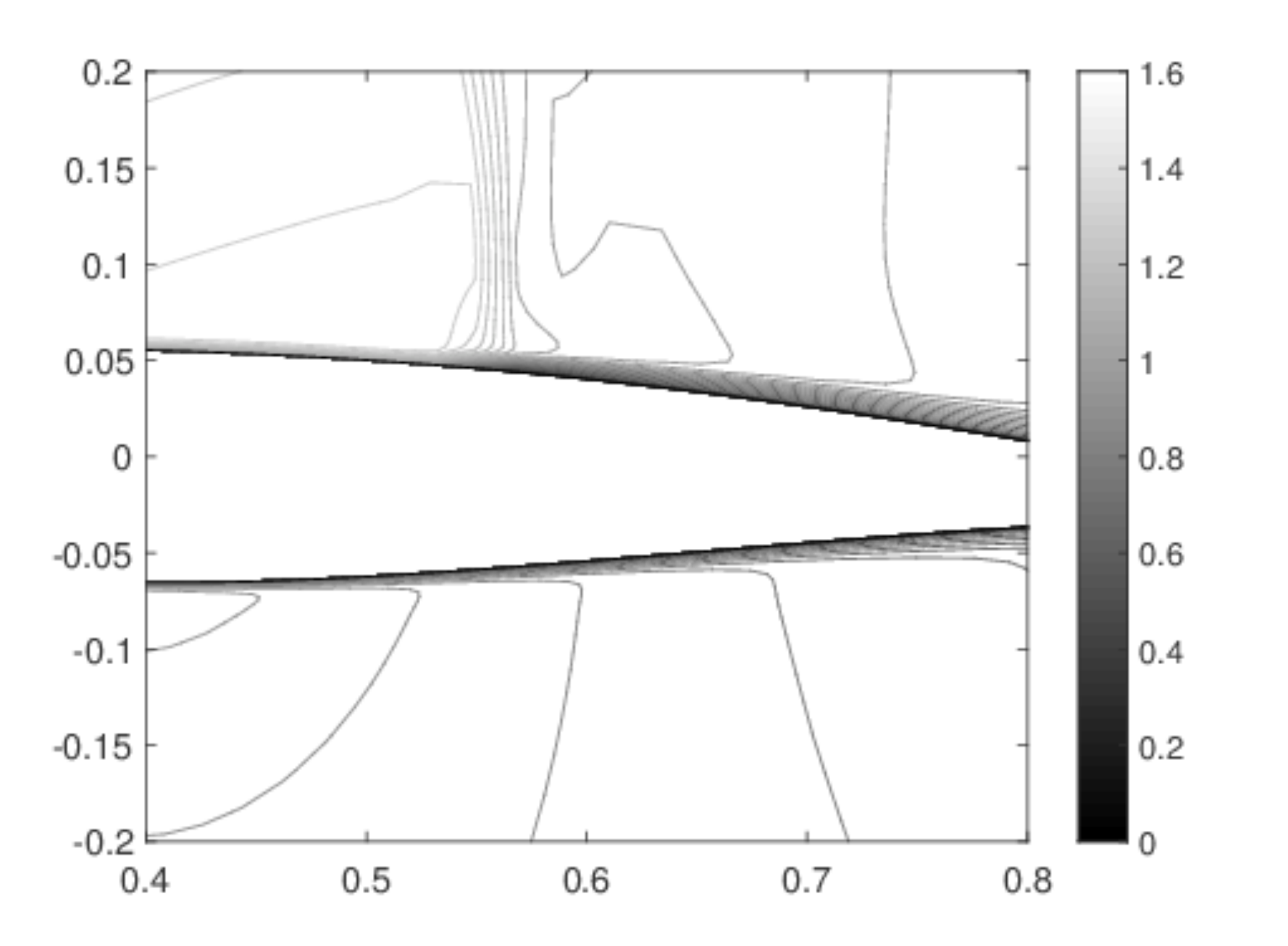}}
\subfigure[Mach contours inviscid case]{\label{steady2_mach_close_euler}\includegraphics[scale=0.40]{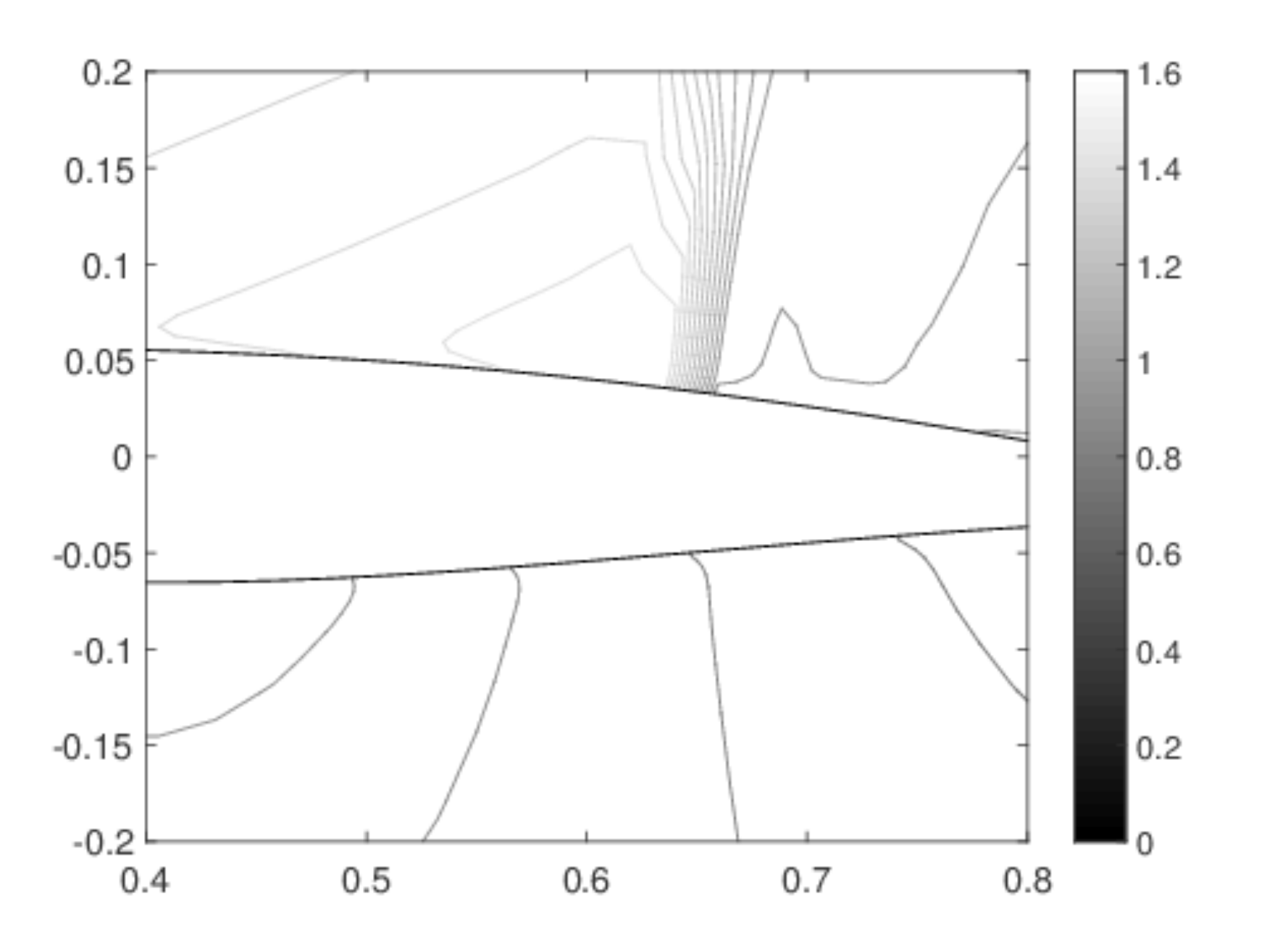}}
\subfigure[Pressure contours viscous case]{\label{steady2_pressure_close}\includegraphics[scale=0.40]{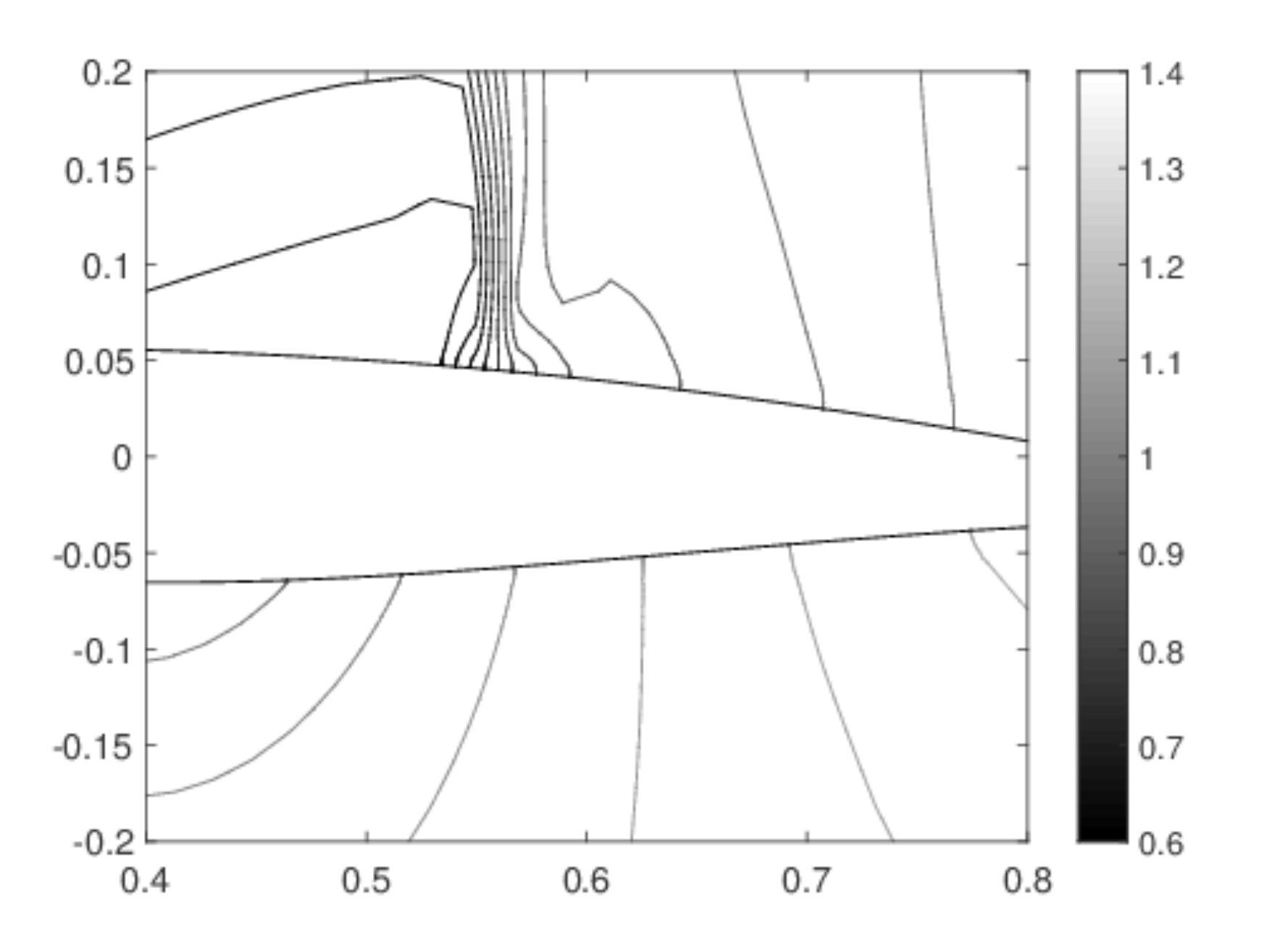}}
\subfigure[Pressure contours inviscid case]{\label{steady2_pressure_close_euler}\includegraphics[scale=0.40]{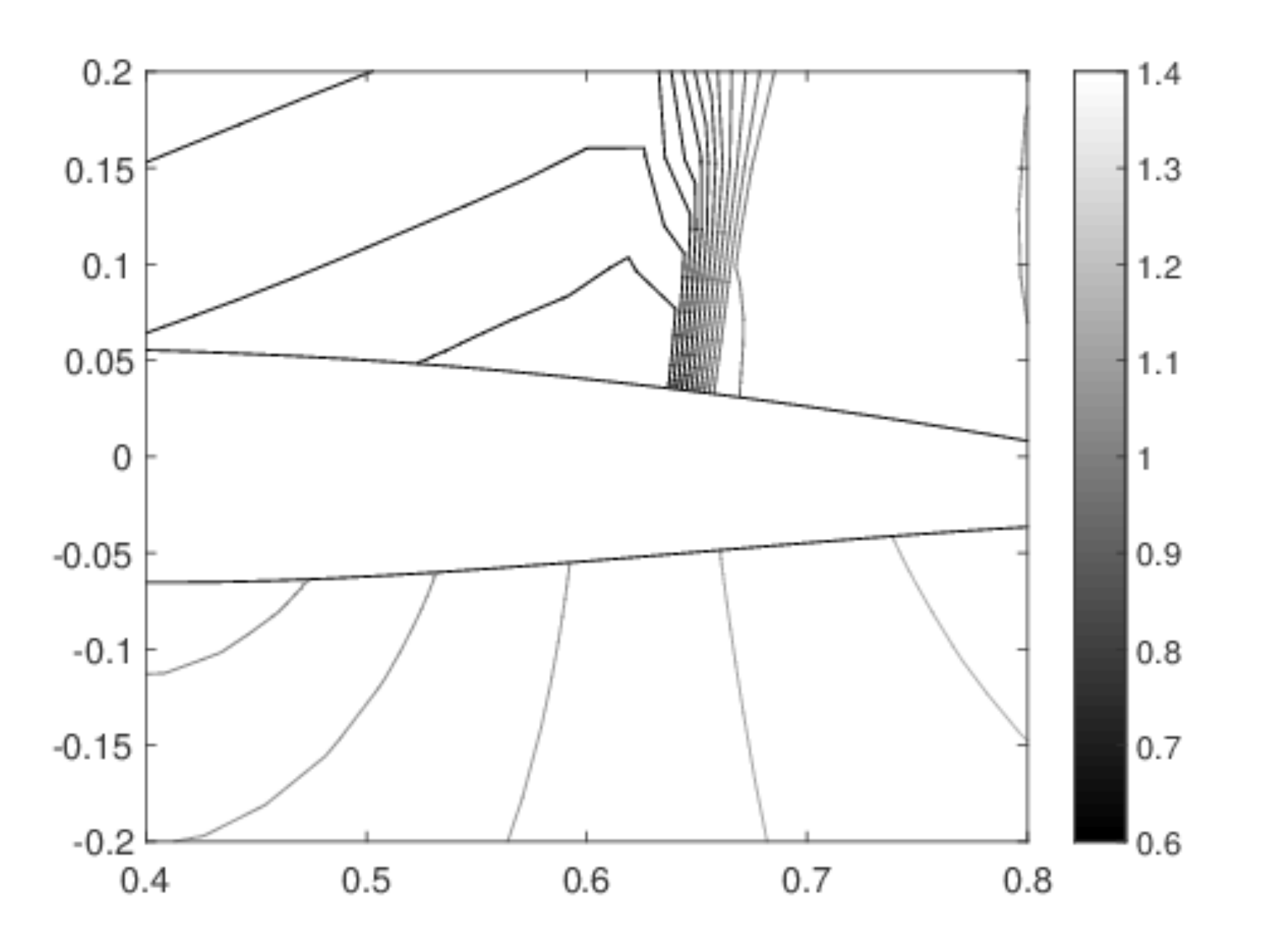}}
\caption{Close-up of pressure and Mach contours for both RANS and Euler solutions for RAE2822 airfoil; $M_\infty = 0.73$, $Re_\infty = 6.5\times10^6$ at $\alpha=2.79^\circ$}
\label{steady2_close}
\end{figure}

We first investigate transonic flow past a stationary airfoil with a shock present on its upper surface. We have results from experiment for this example \citep{cook1979}, and therefore it serves as a means of validating our numerical procedures. An RAE 2822 airfoil is held in a flow of Mach number $0.73$, Reynolds number $6.5\times10^6$, and at an angle of $2.79^{\circ}$ to the free stream. Figure \ref{steady2} shows the variations of coefficient of pressure and skin friction on the airfoil surface, Mach contour over the airfoil, and residue decay with iterations.  The convergence of the numerical computations is ensured with residues of each governing equation, except the turbulence model equation, dropping by almost six decades---shown in Figure \ref{steady2_residue}.

A sudden change in pressure distribution can be seen in Figure \ref{steady2_cp} indicating the presence of a shock, almost close to mid-chord, on the airfoil surface. The Mach contours in Figure \ref{steady2_mach} indicate that the local fluid velocity on the airfoil surface increases because of the airfoil geometry. This supersonic pocket ends with a shock shown by a sudden decrease in Mach number. The pressure and skin friction distribution are compared with the available experimental results of \citet{cook1979} and are shown together in Figures \ref{steady2_cp} and \ref{steady2_cf}. The computed numerical viscous results are in close agreement with the experimental results. The shock strength and its location are captured accurately.  Also note that the fluid has not separated anywhere on the airfoil, as can be seen from the variation of the skin friction coefficient over the airfoil in Figure \ref{steady2_cf}.

In Figure \ref{steady2_cp} we have also superimposed the inviscid flow numerical simulation results over the viscous flow results. The shock strength and its location as predicted by the inviscid flow simulations are quantitatively far and away from both experiment and RANS solutions. This could be attributed to the presence of the boundary layer that changes the airfoil profile as seen by the external flow field. 

Figure~\ref{steady2_close} shows a close-up view of the effect of shock on the boundary layer in terms of the pressure and Mach contours for the RANS and Euler simulations. An early shock with a diffused shock-foot can be seen in the viscous case---Figures~\ref{steady2_mach_close} and \ref{steady2_pressure_close}---when compared with the inviscid case---Figures~\ref{steady2_mach_close_euler} and \ref{steady2_pressure_close_euler}. In the outer region, the flow reaches supersonic speeds and the only possible way for it to reach subsonic flow at the downstream is through a shock. But the flow in the inner region, that is the boundary layer region, has not reached supersonic speeds because of viscous effects. Hence the information that there is a high pressure region in the post-shock region passes up the stream through the boundary layer, making the boundary layer thicker upstream of the shock---as seen in Figure~\ref{steady2_mach_close} close to $x/c=0.55$. This results in a gradual pressure rise in the shock region of the boundary layer leading to a diffused shock-foot as seen around $x/c=0.55$. Increase in the boundary layer thickness downstream of the shock, presents a lower pressure gradient to the external flow compared to that in the inviscid flow. So, not only does the presence of the boundary layer change the shock location significantly, but it also changes the pressure rise across the shock.

\begin{figure}
\subfigure[Coefficient of lift versus angle of attack]{\label{unsteady2_cl}\includegraphics[scale=0.40]{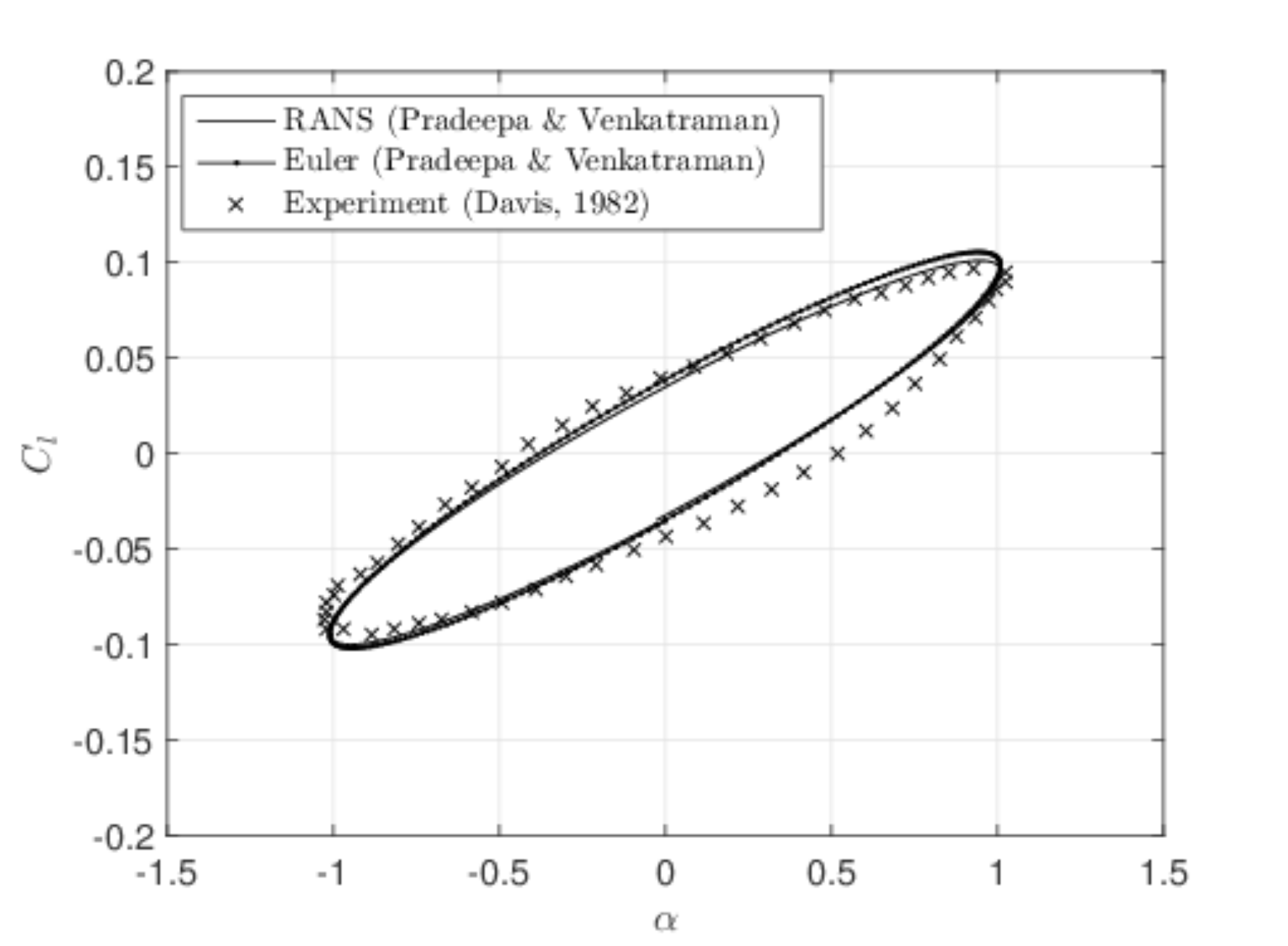}}
\subfigure[Coefficient of moment versus angle of attack]{\label{unsteady2_cm}\includegraphics[scale=0.40]{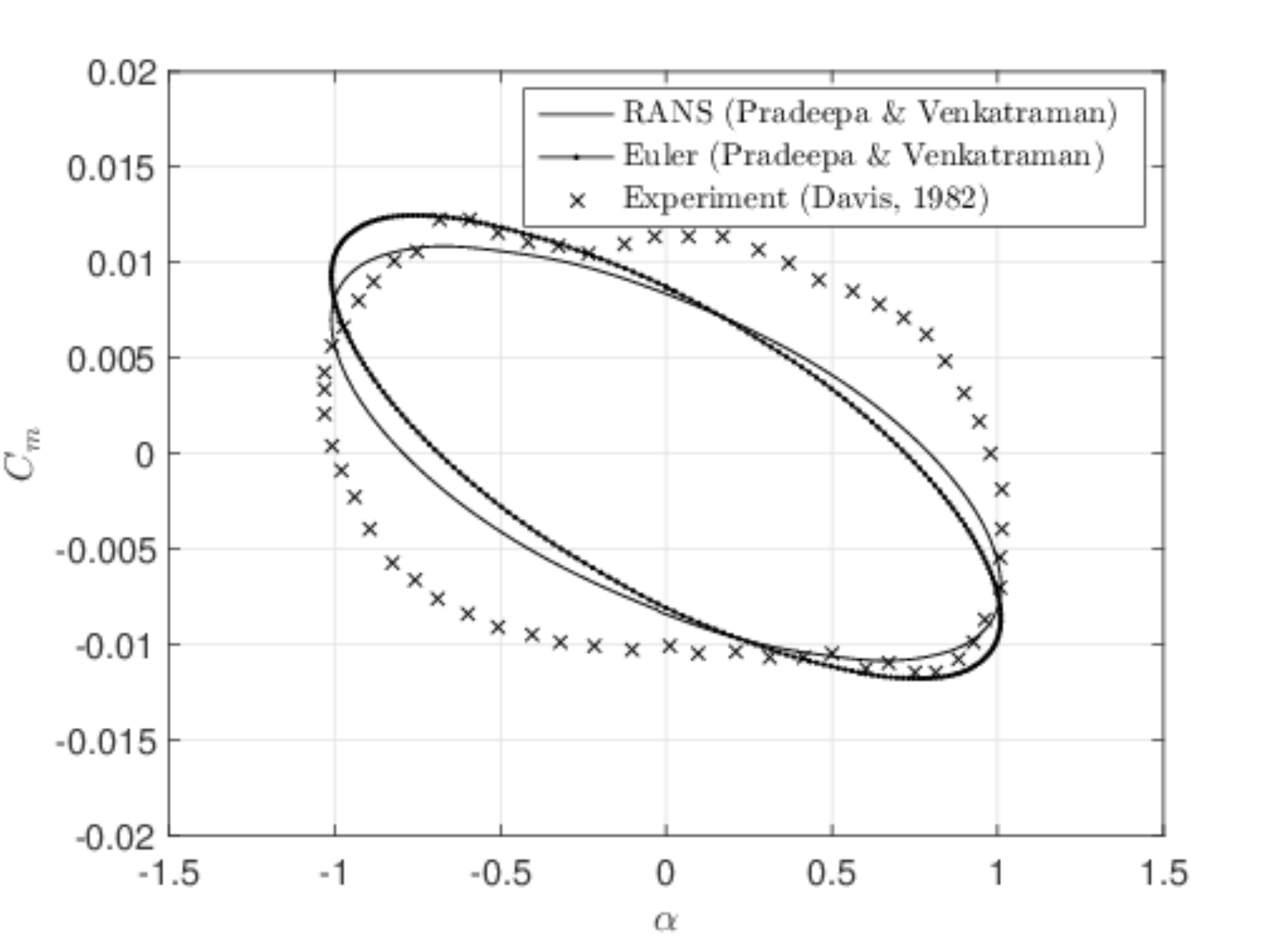}}
\caption{Lift and moment coefficient history---AGARD CT6 test case}
\label{unsteady2}
\end{figure}

\section{Shock displacement and thickening in an oscillating airfoil}
\label{sec:ch_ust_af}

\begin{figure}
\subfigure[$\alpha = 0.99^\circ, \varphi = 101.25^\circ$]{\label{u2_11}\includegraphics[scale=0.40]{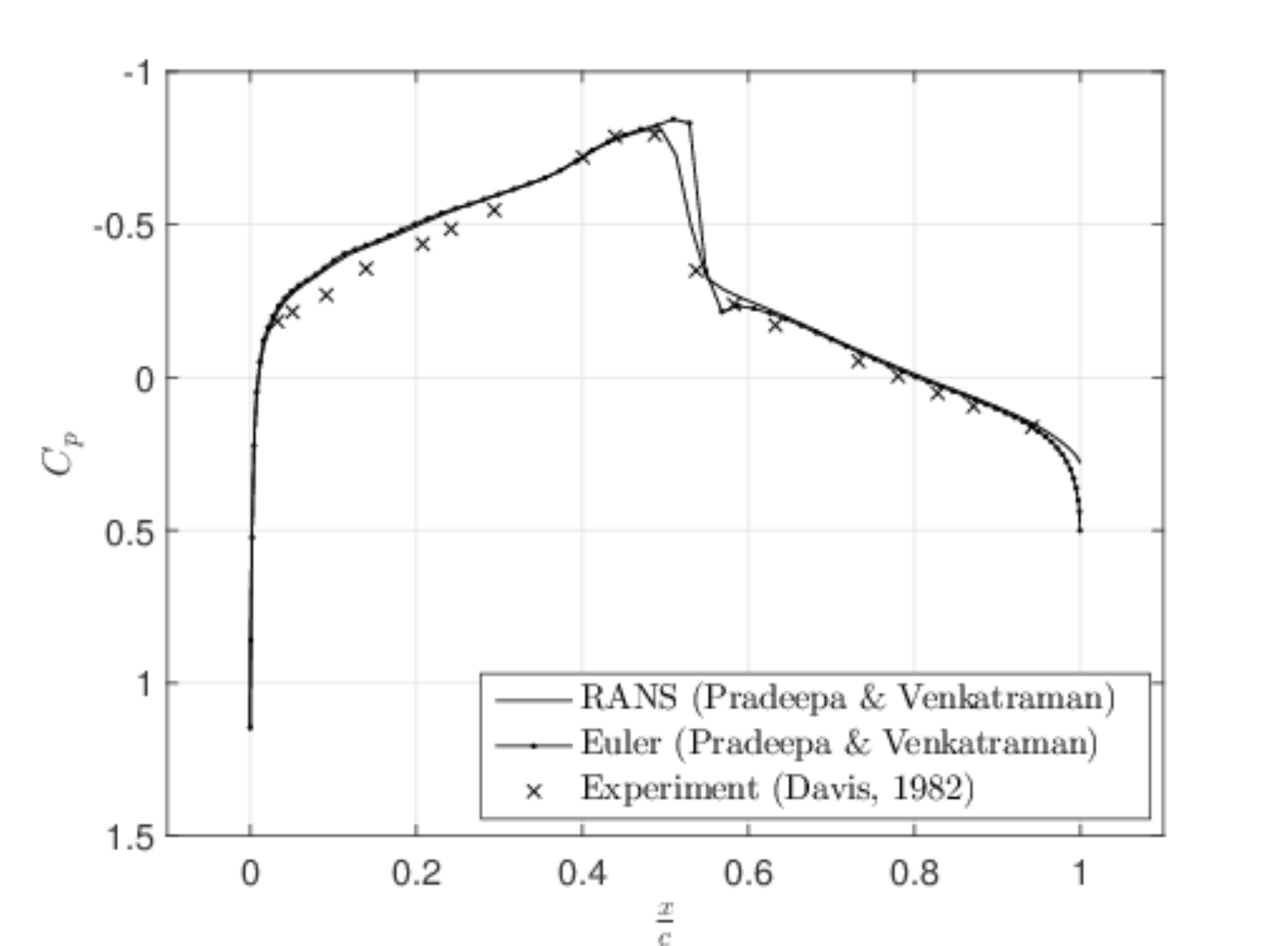}}
\subfigure[$\alpha = 0.57^\circ, \varphi = 146.25^\circ$]{\label{u2_12}\includegraphics[scale=0.40]{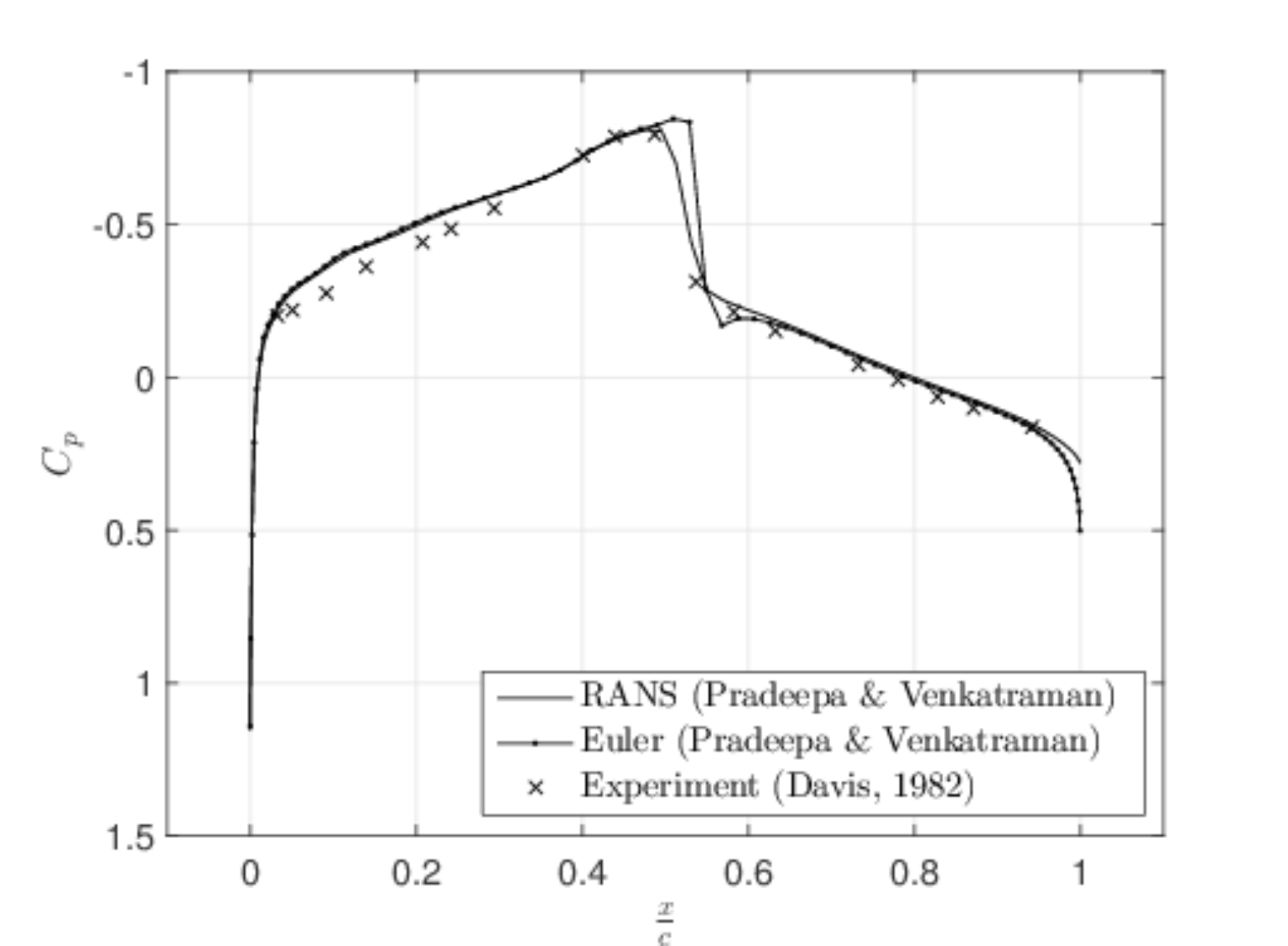}}
\subfigure[$\alpha = 0.08^\circ, \varphi = 174.37^\circ$]{\label{u2_13}\includegraphics[scale=0.40]{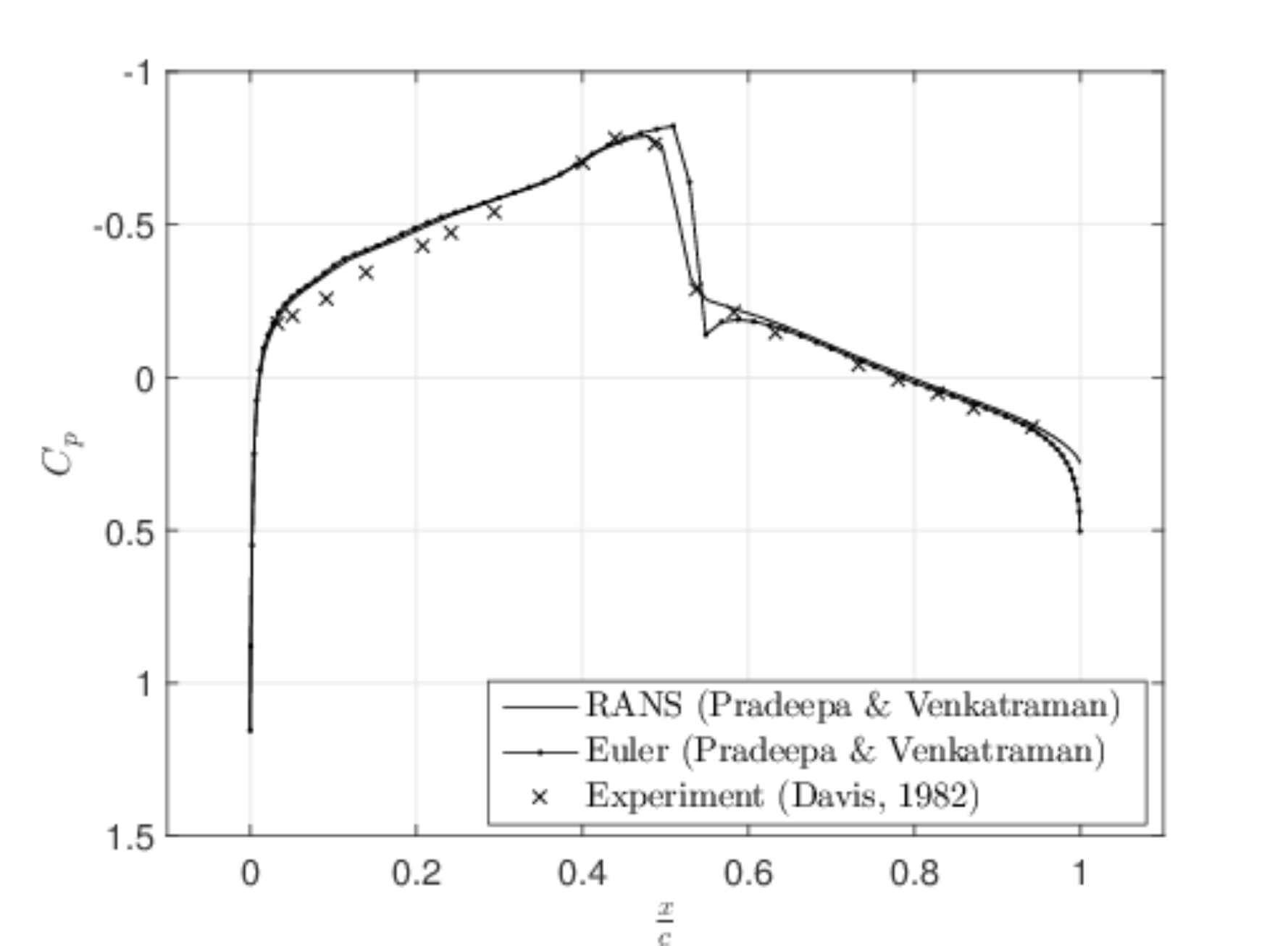}}
\subfigure[$\alpha = -0.49^\circ, \varphi = 209.53^\circ$]{\label{u2_14}\includegraphics[scale=0.40]{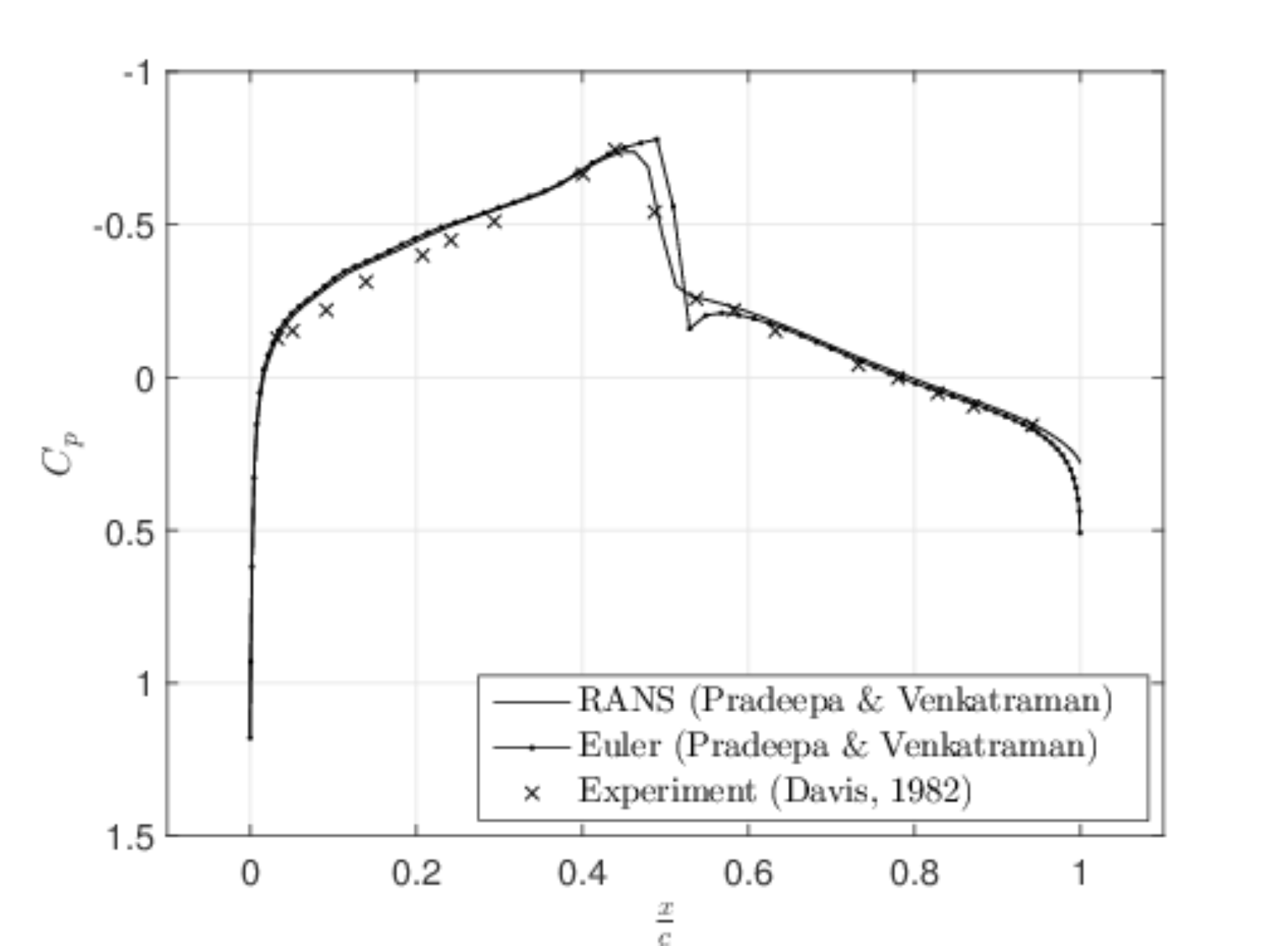}}
\subfigure[$\alpha = -1.01^\circ, \varphi = 269.99^\circ$]{\label{u2_15}\includegraphics[scale=0.40]{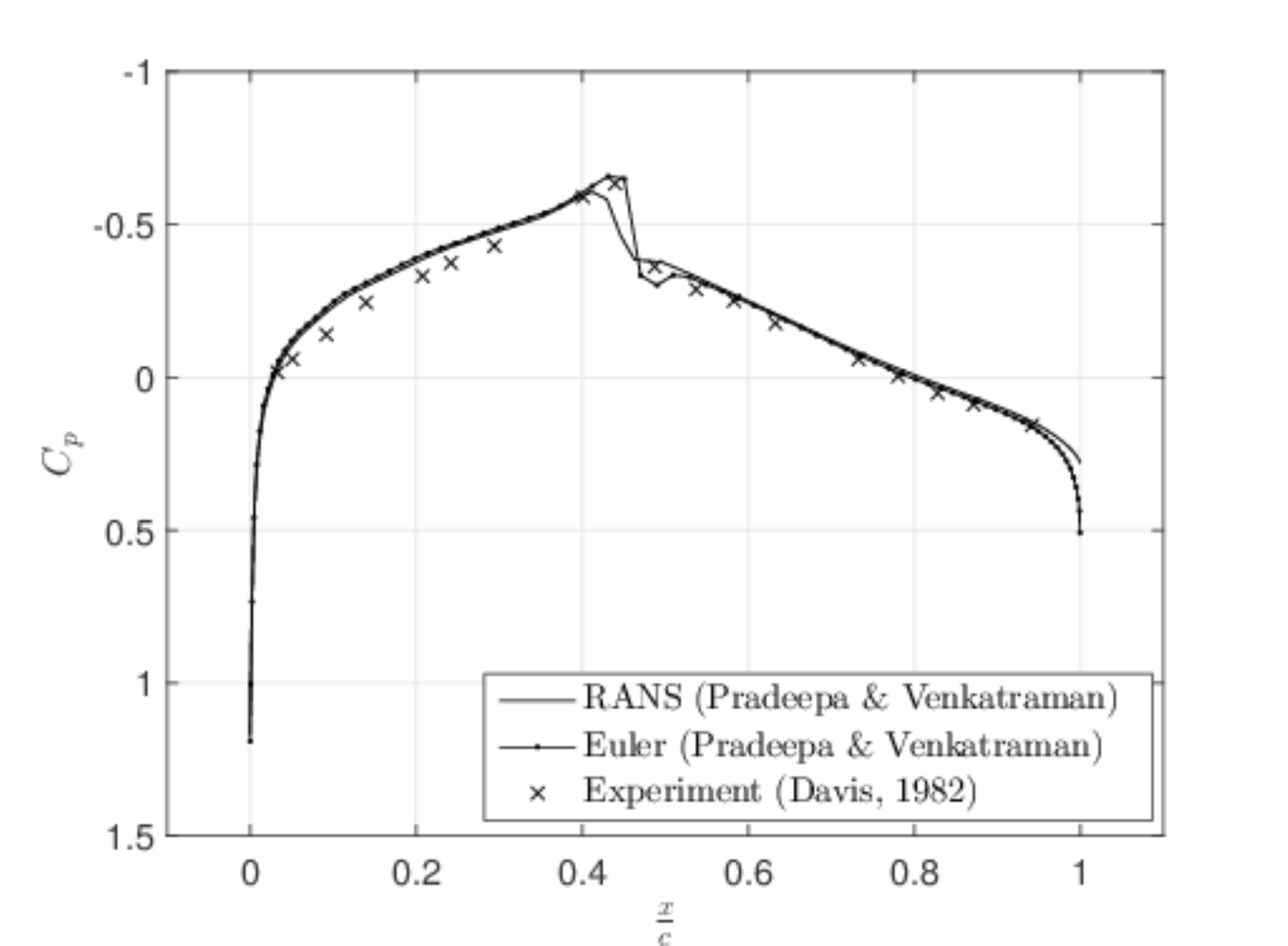}}
\subfigure[$\alpha = -0.40^\circ, \varphi = 336.09^\circ$]{\label{u2_16}\includegraphics[scale=0.40]{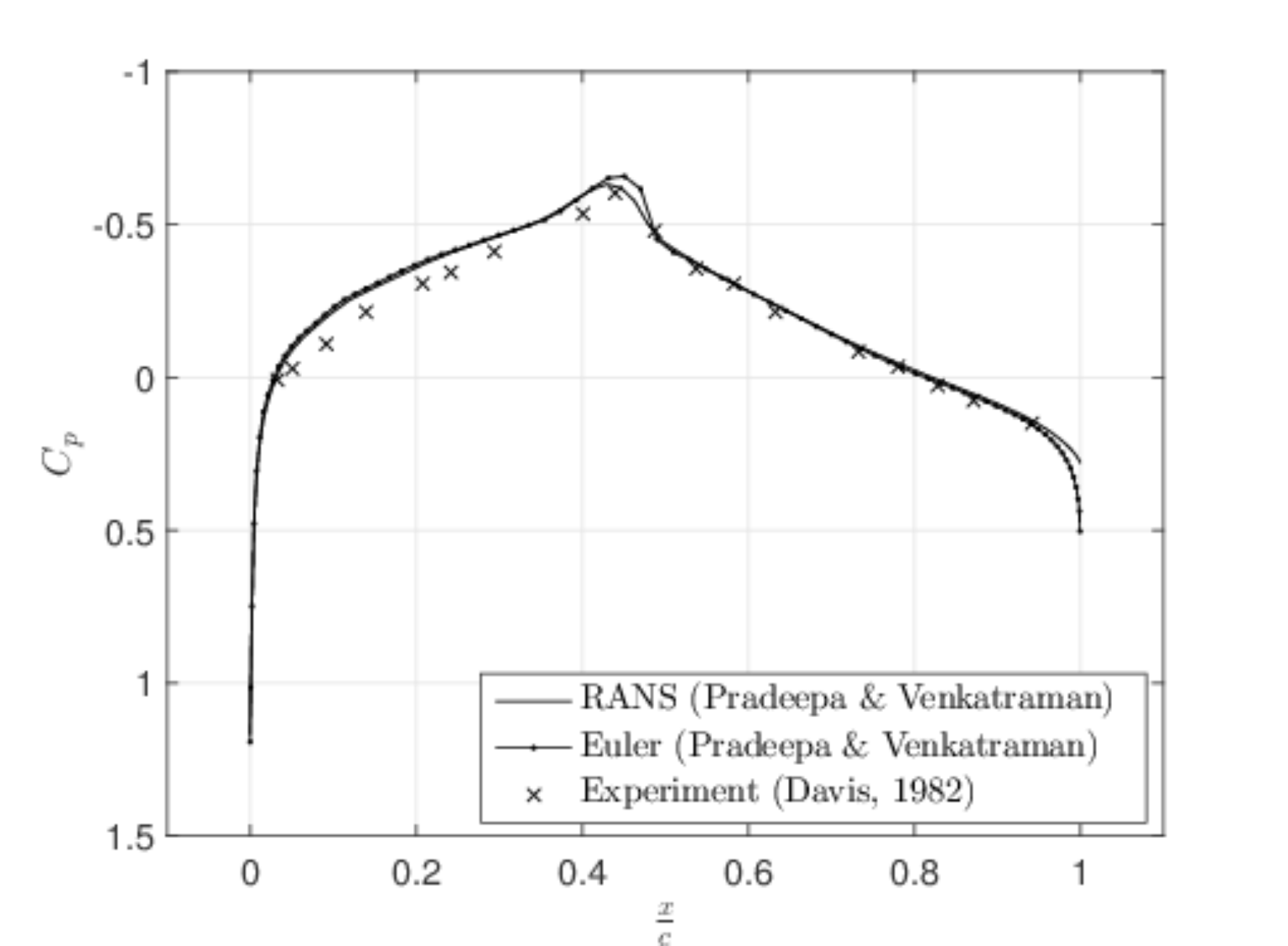}}
\caption{Instantaneous pressure distribution on the airfoil upper surface---AGARD CT6 test case}
\label{unsteady2_inst}
\end{figure}

We now investigate an example involving a NACA64A010  airfoil pitching about its quarter chord in a flow of Mach number $0.796$ and Reynolds number $12.56 \times 10^6$; the mean angle oscillation is $0^\circ$ with an amplitude of oscillation of $1.01^\circ$ at a reduced frequency of $0.202$. This example is known as the AGARD CT6 test case \citep{davis1982}. 

The airfoil pitching motion is defined as 

\begin{equation}
\label{unst1}
\alpha(t) = \alpha_m + \alpha_0 \sin \omega t.
\end{equation}

$\alpha$ is the instantaneous angle of attack, $\alpha_m$ is the mean angle of attack, and $\alpha_0$ is the amplitude of oscillation in pitch. $t$ denote time. The frequency of oscillation $\omega$ is related to the reduced frequency $\kappa$ as

\begin{equation}
\label{unst2}
\kappa = \dfrac{\omega b}{U_\infty}.
\end{equation}

Here $b$ is the semi-chord and $U_\infty$ is the free stream velocity of the fluid. First the steady state solution is calculated over the stationary airfoil at its mean angle of attack. Thereafter the airfoil is set in motion as defined by Equation (\ref{unst1}). 

Unsteady calculations are performed and the lift and moment variations are observed after the transients in the solution die down. All unsteady flow computations are performed using grids of C-type and of size $256 \times 72 \times A160$. Grid spacing normal to the solid surface is set to $y^+ \approx 1.0$.

Lift and moment coefficients are shown in Figure \ref{unsteady2}. Inviscid flow numerical results are also superimposed on these figures which shows a slight variation with RANS results. The reasons for both inviscid and viscous solutions to be in close agreement with the experimental results are because of high Reynolds number, low angle of attack, and the airfoil being thin. As the Reynolds number increases, the boundary layer thickness decreases leading to approximately the same profile for the external flow field of viscous and inviscid simulations. The mean angle of attack and the amplitude of oscillation are also both low which in turn leads to reduced viscous effects. Note that the NACA64A010 airfoil used in the AGARD CT6 case being thin has a higher critical Mach number. Therefore the effect of higher Mach number leading to shock induced viscous effects are also reduced.

Pressure distribution on the airfoil surface at different instants for AGARD CT6 test case are shown in Figure \ref{unsteady2_inst}. Inviscid pressure distribution are also superimposed over these results along with the experimental results of \citet{davis1982}. The shock motion is of  class B type where the shock appears intermittently over a cycle of pitching oscillation \citep[Chap.9,pp.62]{tijdeman1977}. It can be seen that the shock location and strength captured by the RANS solver are close to those from experiment when compared to those of the inviscid results. The inviscid flow solver captures a stronger shock aft of the experimentally predicted shock location. The effect of viscosity in changing the strength and location of the shock when the airfoil is moving is clearly seen once again

\begin{figure}
\subfigure[$\alpha = 0.99^\circ, \varphi = 101.25^\circ$]{\label{u2_21}\includegraphics[scale=0.40]{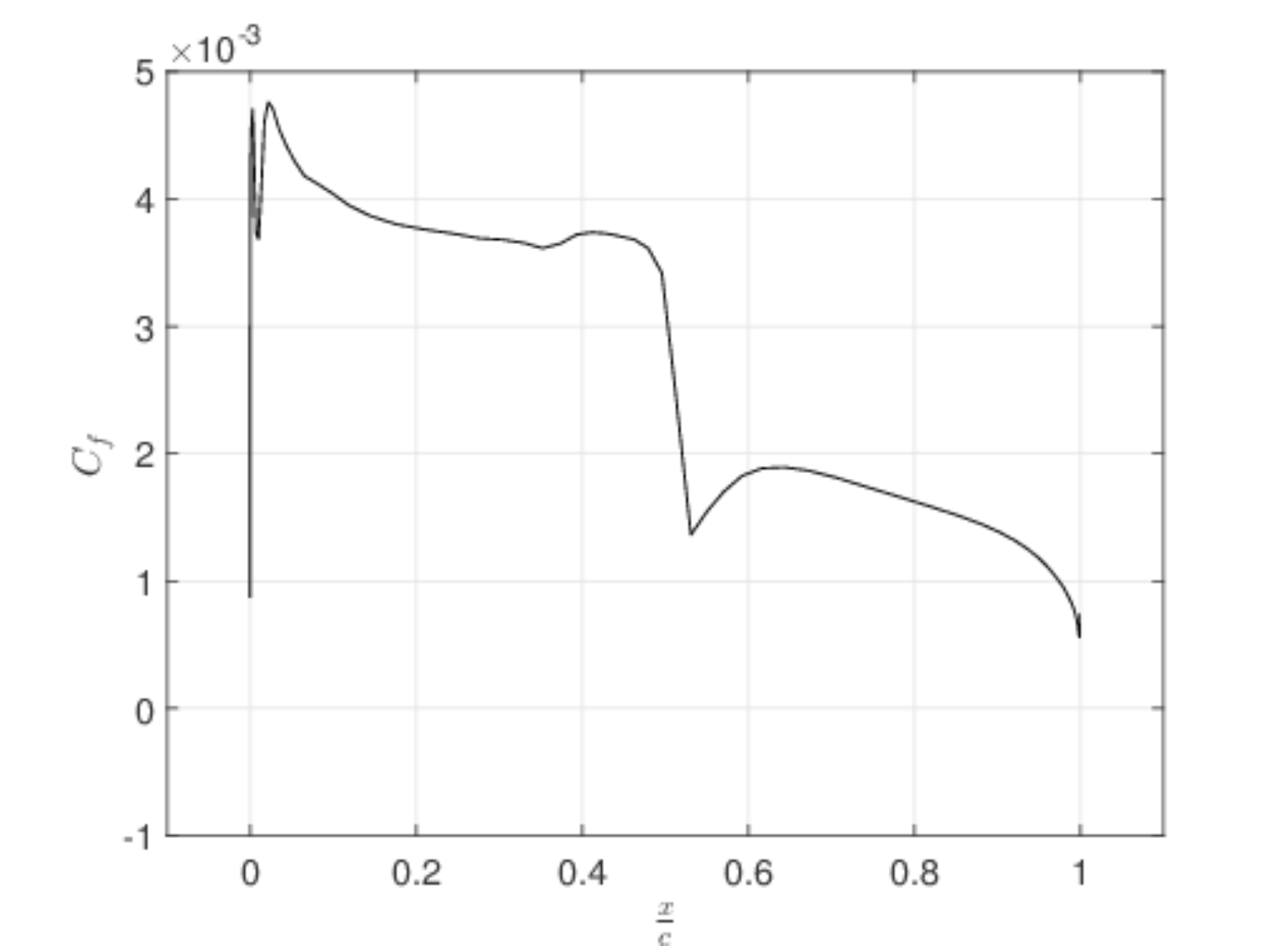}}
\subfigure[$\alpha = 0.57^\circ, \varphi = 146.25^\circ$]{\label{u2_22}\includegraphics[scale=0.40]{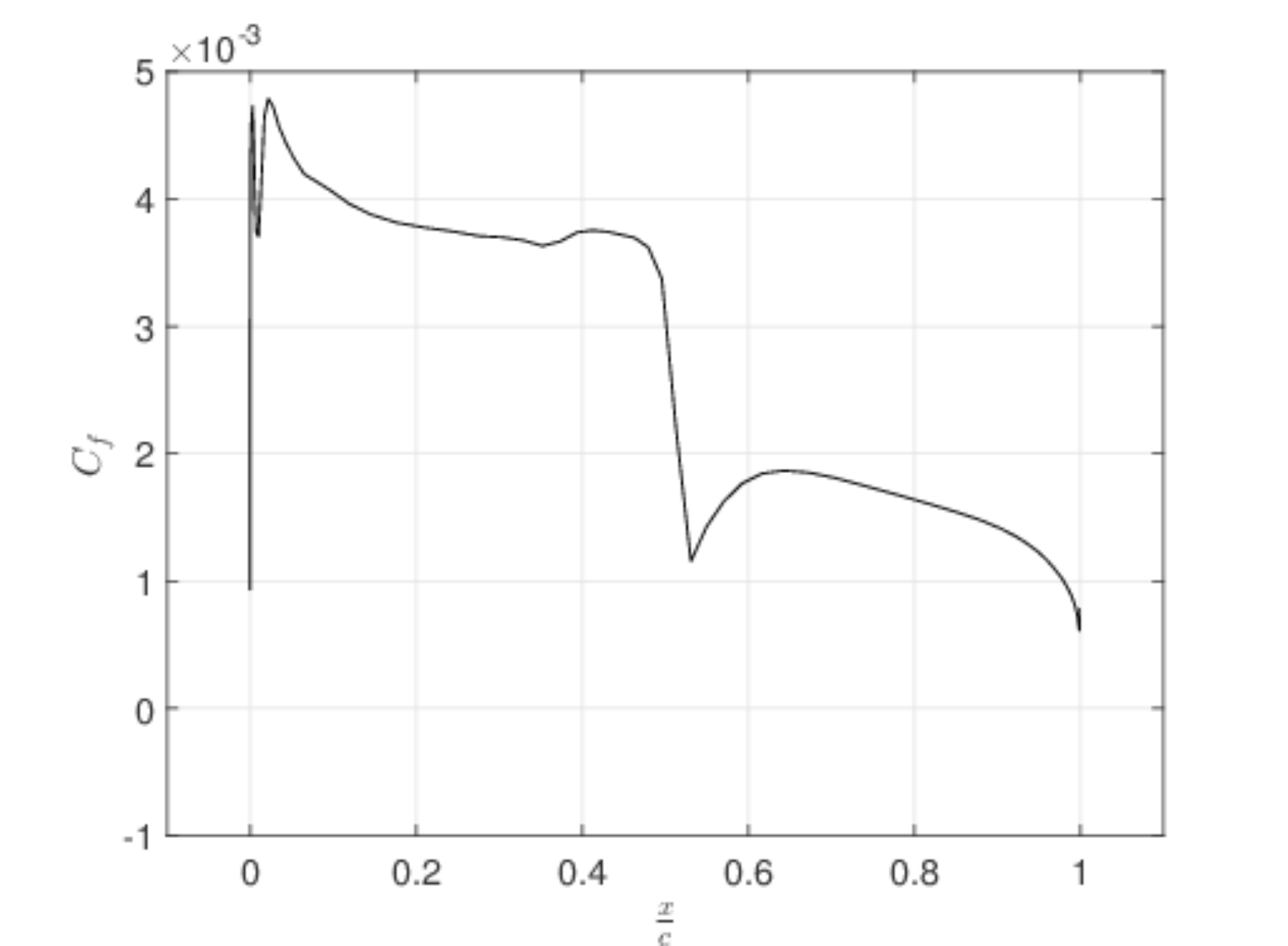}}
\subfigure[$\alpha = 0.08^\circ, \varphi = 174.37^\circ$]{\label{u2_23}\includegraphics[scale=0.40]{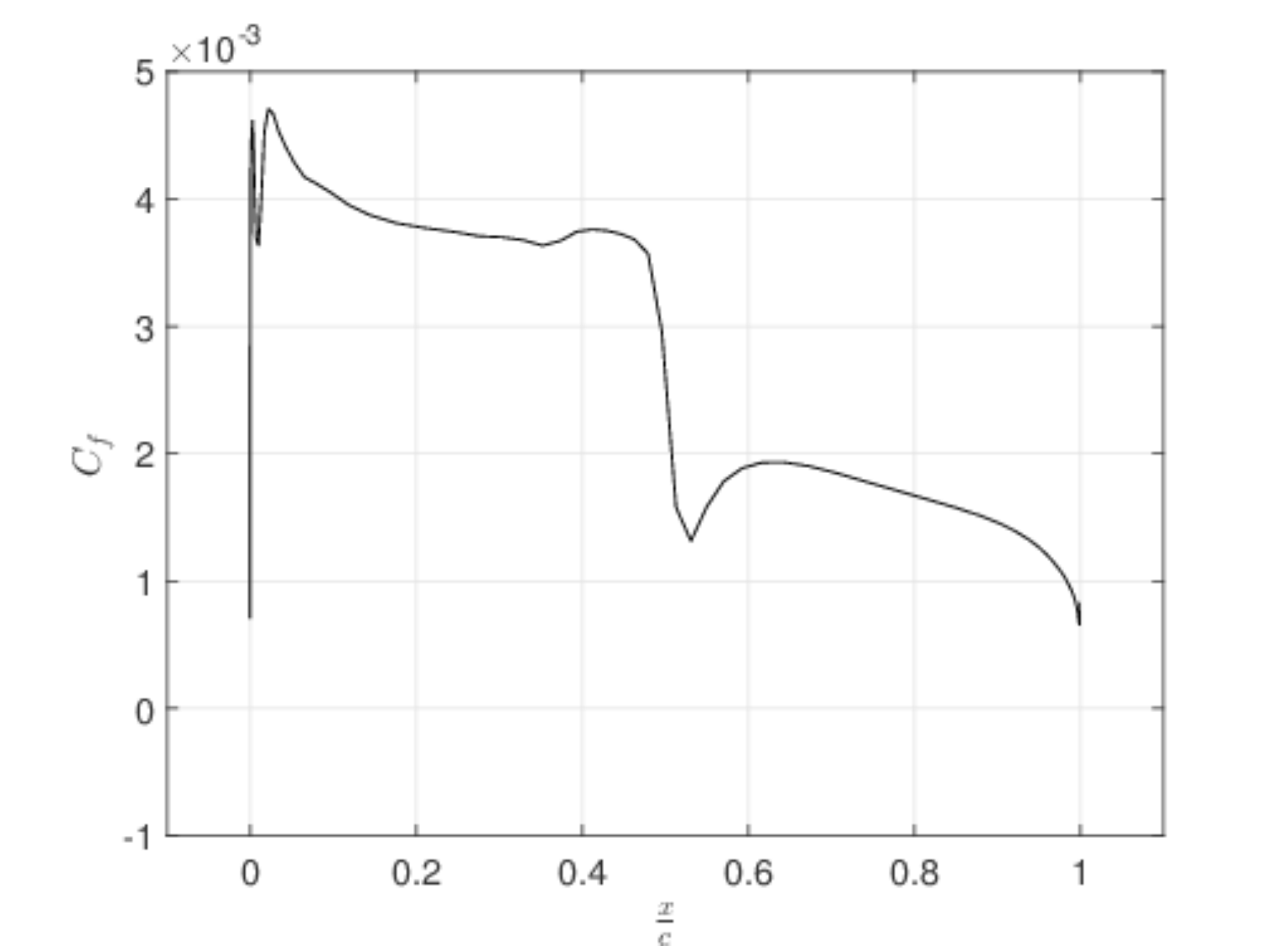}}
\subfigure[$\alpha = -0.49^\circ, \varphi = 209.53^\circ$]{\label{u2_24}\includegraphics[scale=0.40]{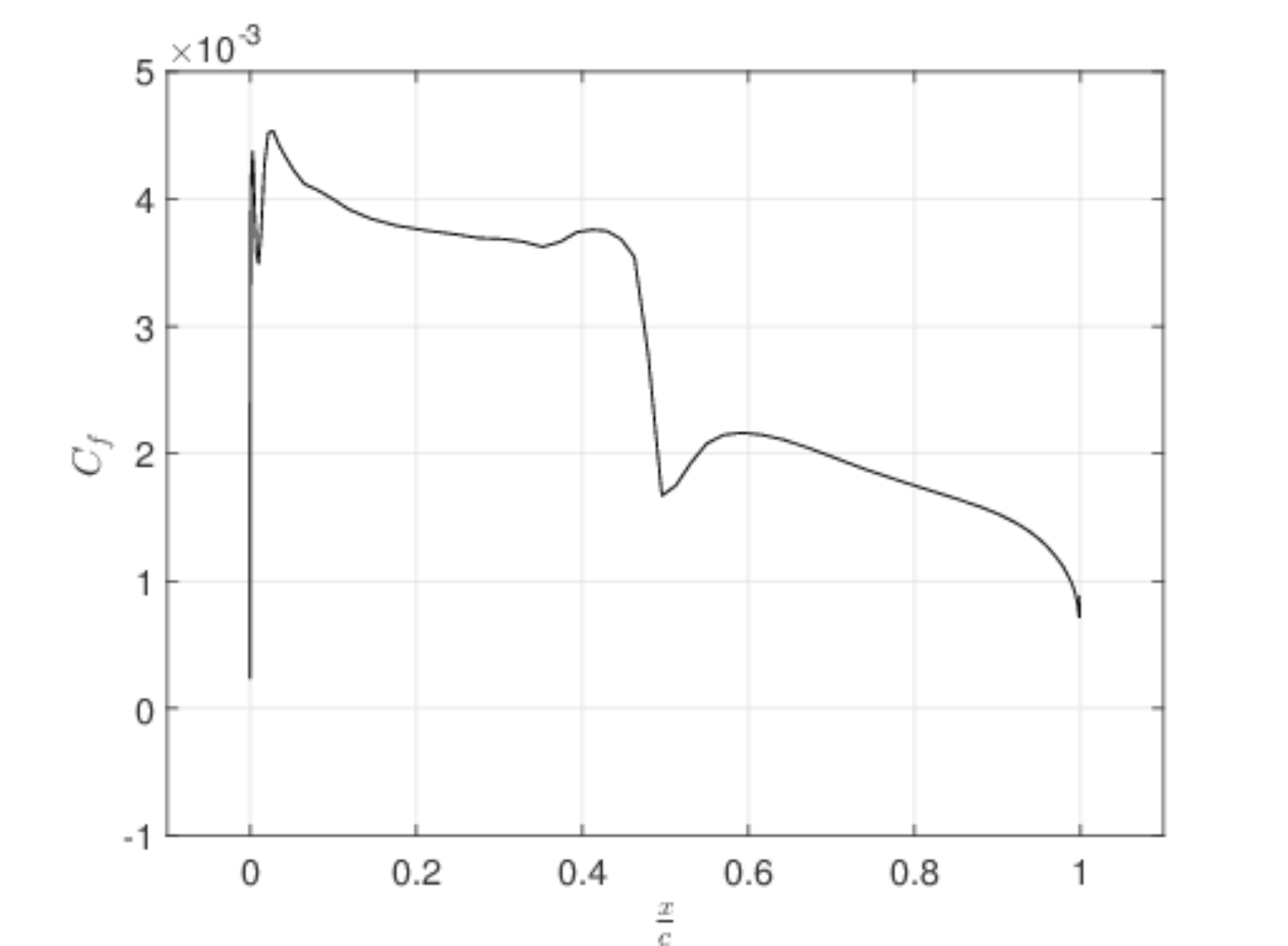}}
\subfigure[$\alpha = -1.01^\circ, \varphi = 269.99^\circ$]{\label{u2_25}\includegraphics[scale=0.40]{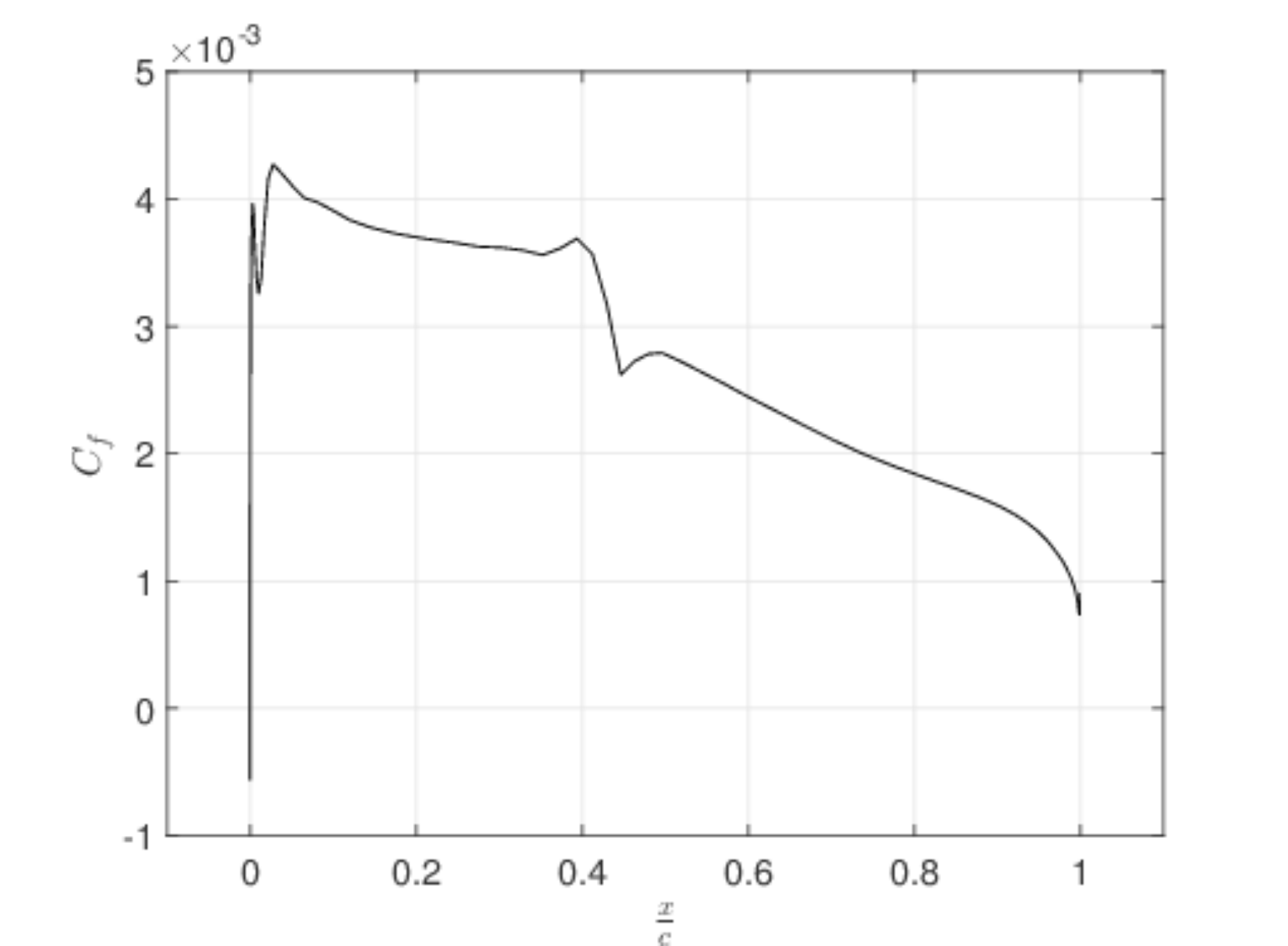}}
\subfigure[$\alpha = -0.40^\circ, \varphi = 336.09^\circ$]{\label{u2_26}\includegraphics[scale=0.40]{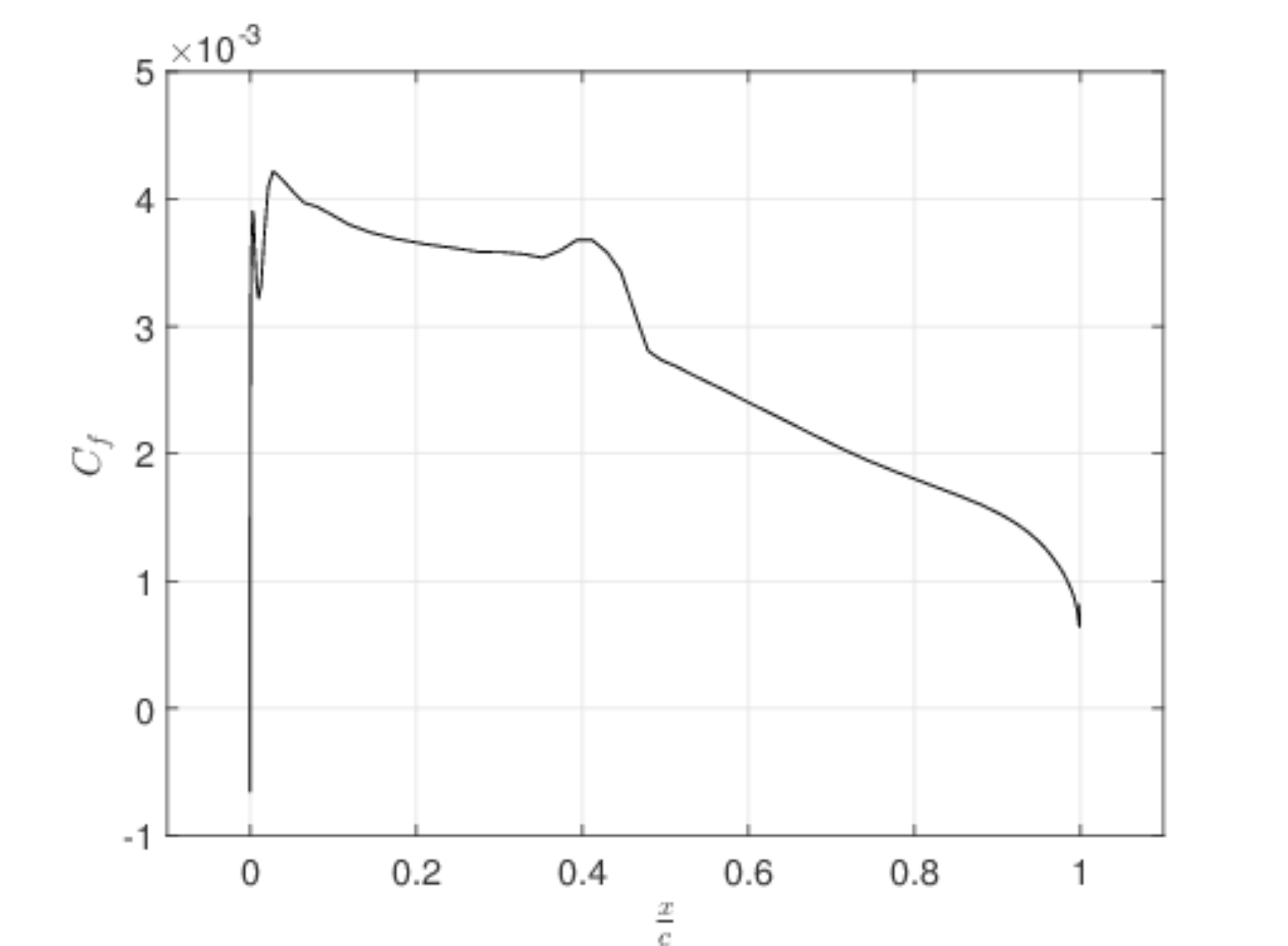}}
\caption{Instantaneous surface friction distribution on airfoil upper surface---AGARD CT6 test case}
\label{unsteady2_inst_f}
\end{figure}

Figure \ref{unsteady2_inst_f} shows the variation of skin friction coefficient calculated on the upper surface of the airfoil at different instants during AGARD CT6 unsteady cycle. Sudden changes in $C_f$ are observed near the leading edge and shocks---regions of high velocity gradients. The shock does not cause enough adverse pressure gradient for the flow to separate as the skin friction distribution curves do not cross the horizontal axis or $C_f = 0$ line.

\begin{figure}
\begin{center}
\includegraphics[scale=0.50]{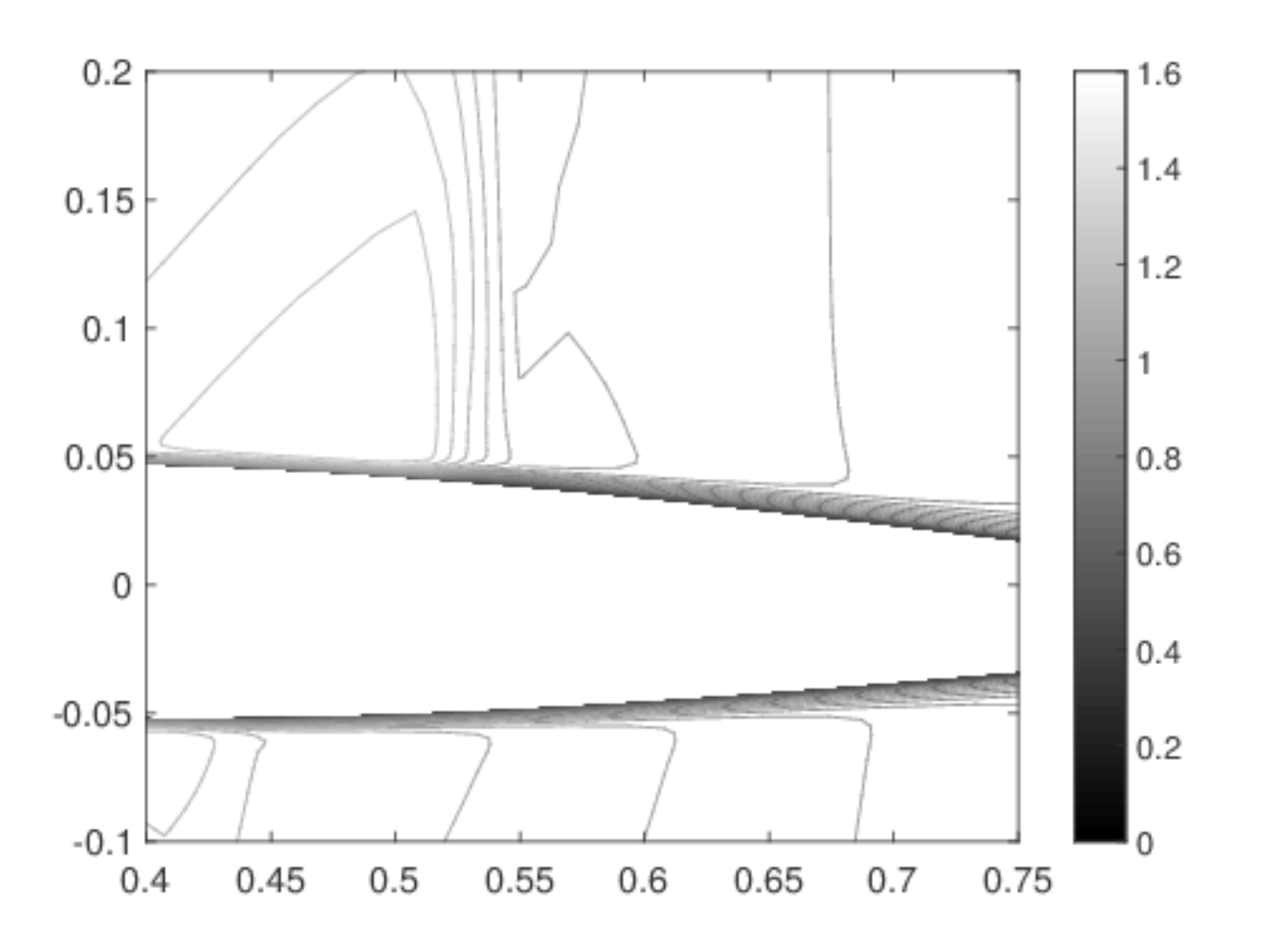}
\caption{Instantaneous Mach contours at $\alpha = 0.99^\circ, \varphi = 101.25^\circ$---AGARD CT6 test case}
\label{u2_31_close}
\end{center}
\end{figure}

Figure~\ref{u2_31_close} shows Mach contours close to the airfoil surface at an instant when it is nearly at its highest nose up position. The presence of shock causes the boundary layer to thicken; the boundary layer on the bottom surface starts to increase in thickness much earlier than the upper surface. The supersonic region around the upper surface of the airfoil, which is also the region of expansion, leads to acceleration of the flow that offers a favorable pressure gradient. And hence the boundary layer has a very low rate of growth in this supersonic region of the upper surface. But the shock at the downstream end of the supersonic pocket increases the pressure abruptly leading to a sudden increase in boundary layer thickness.

In summary, the inclusion of viscosity improves the capability of the flow solver to predict accurately the transonic aerodynamic flow features when compared to an inviscid flow solver. Unsteady shock parameters such as shock strength and its motion, with proper phase difference with the airfoil motion, are accurately captured. The shock, though, was not strong enough to create enough adverse pressure gradient for the flow to separate.

\section{Shock induced boundary layer separation}
\label{sec:ch_shck_bl_sep}

\begin{figure}
\subfigure[$\alpha = 4.99^\circ, \varphi = 98.43^\circ$]{\label{u3_11}\includegraphics[scale=0.40]{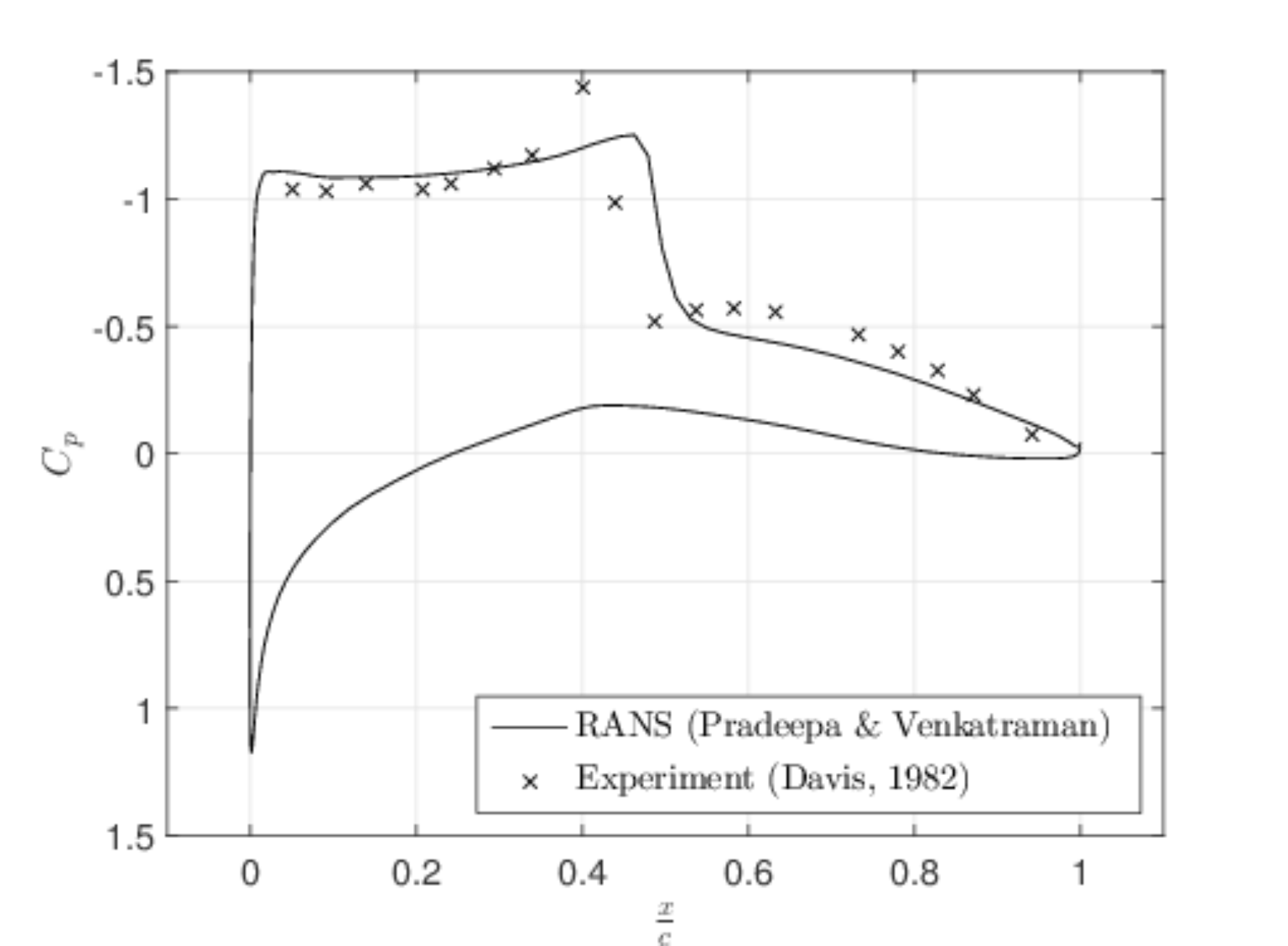}}
\subfigure[$\alpha = 4.58^\circ, \varphi = 144.84^\circ$]{\label{u3_12}\includegraphics[scale=0.40]{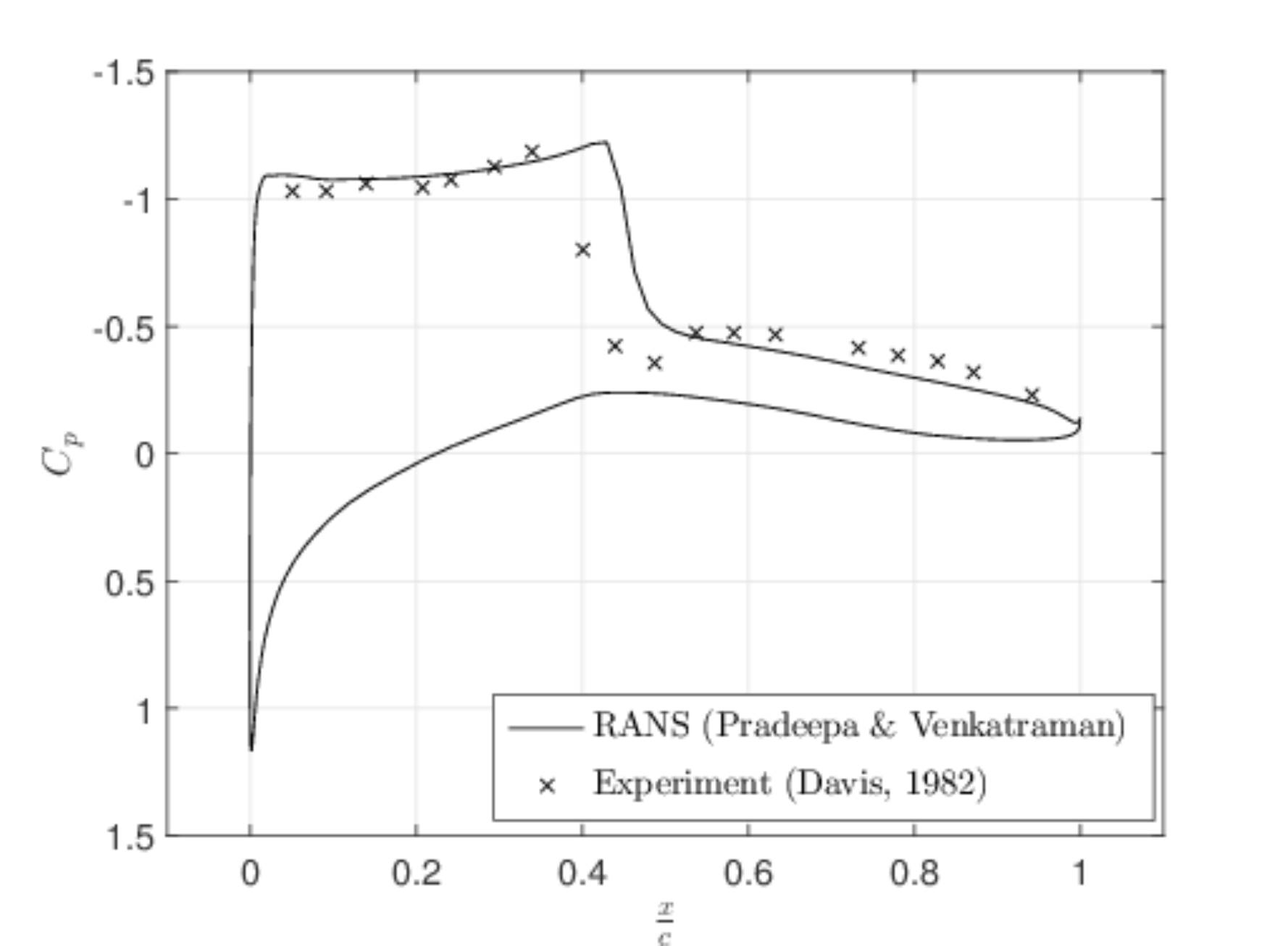}}
\subfigure[$\alpha = 4.08^\circ, \varphi = 175.10^\circ$]{\label{u3_13}\includegraphics[scale=0.40]{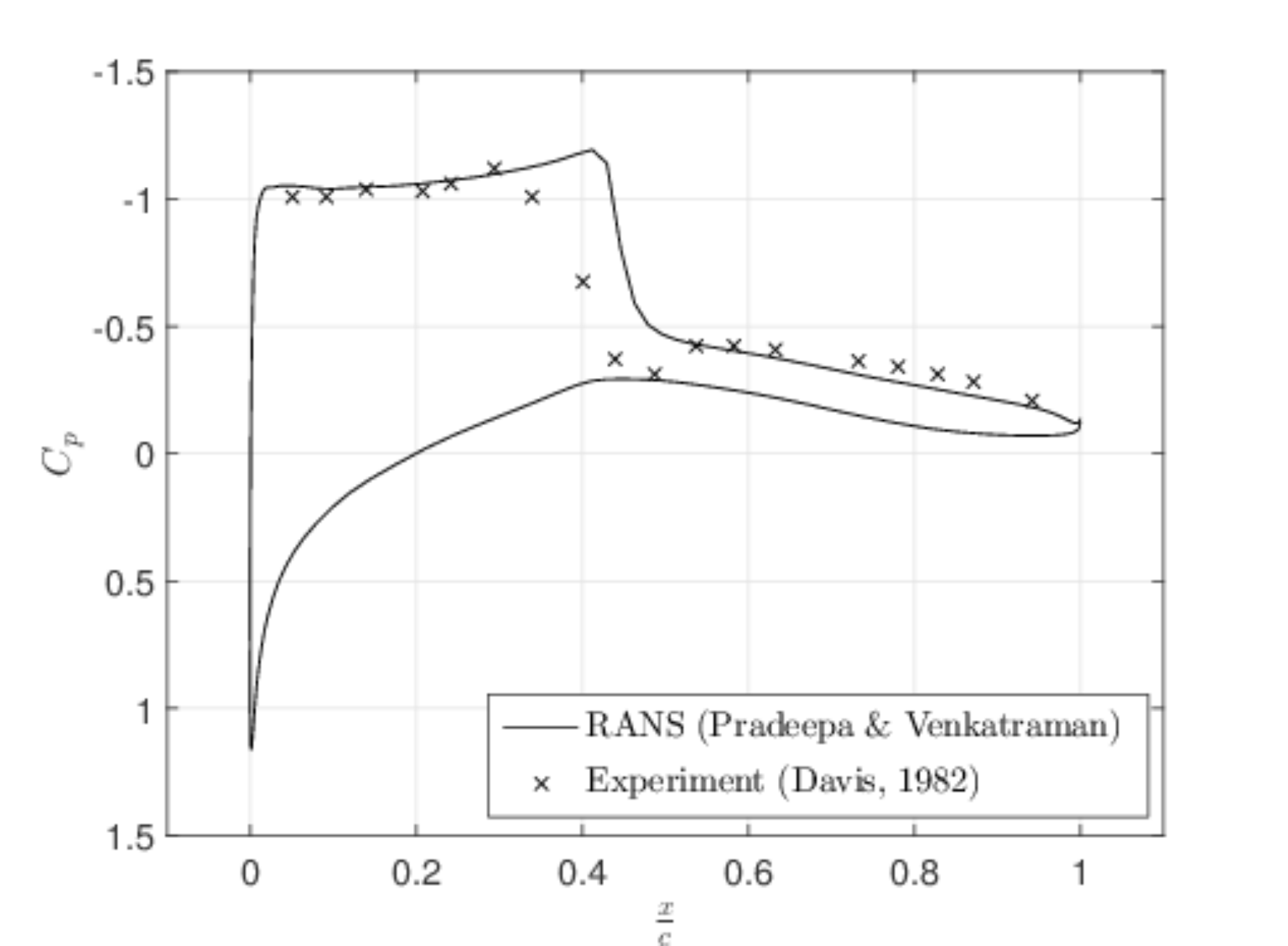}}
\subfigure[$\alpha = 3.58^\circ, \varphi = 204.61^\circ$]{\label{u3_14}\includegraphics[scale=0.40]{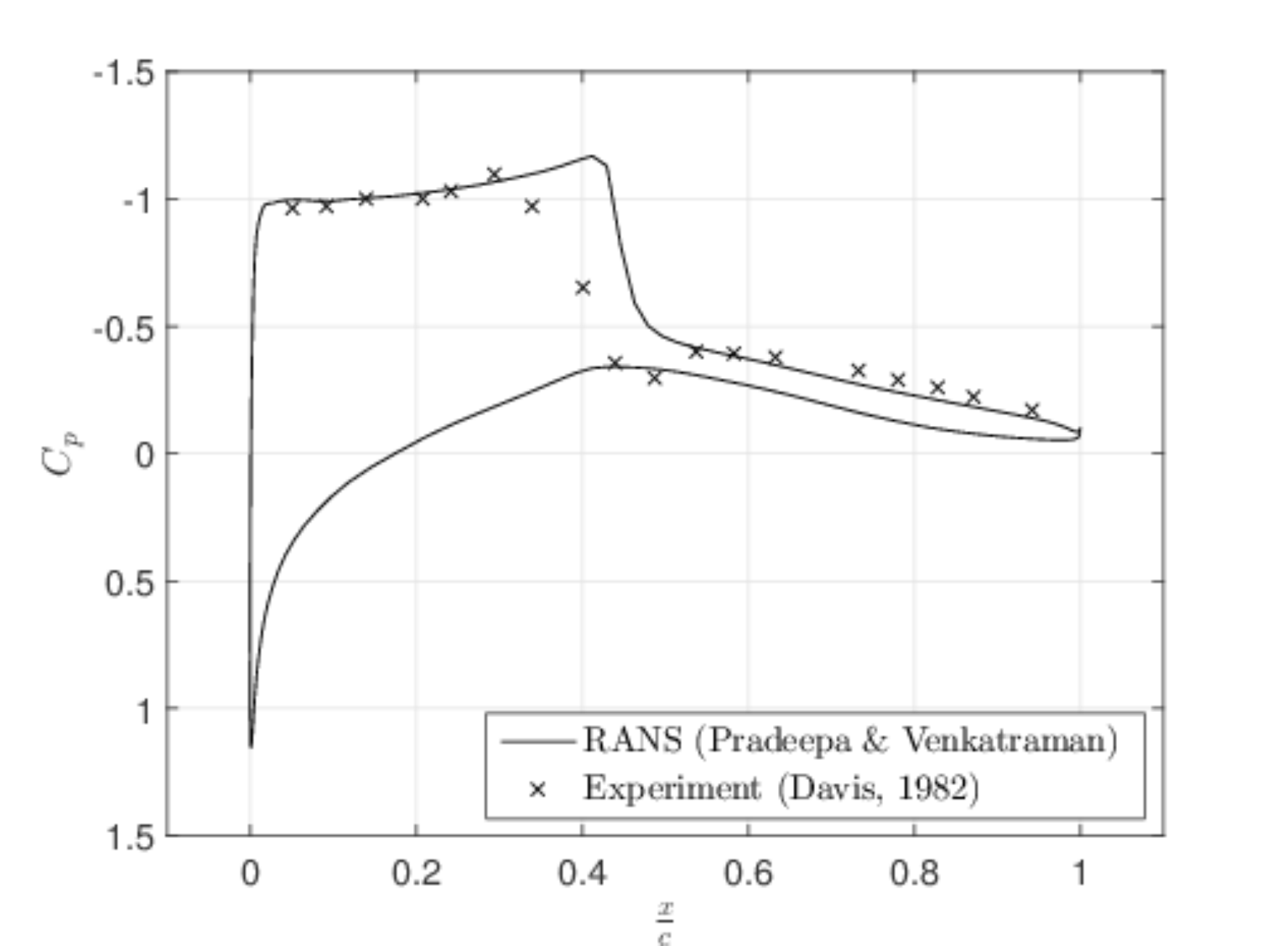}}
\subfigure[$\alpha = 2.99^\circ, \varphi = 269.98^\circ$]{\label{u3_15}\includegraphics[scale=0.40]{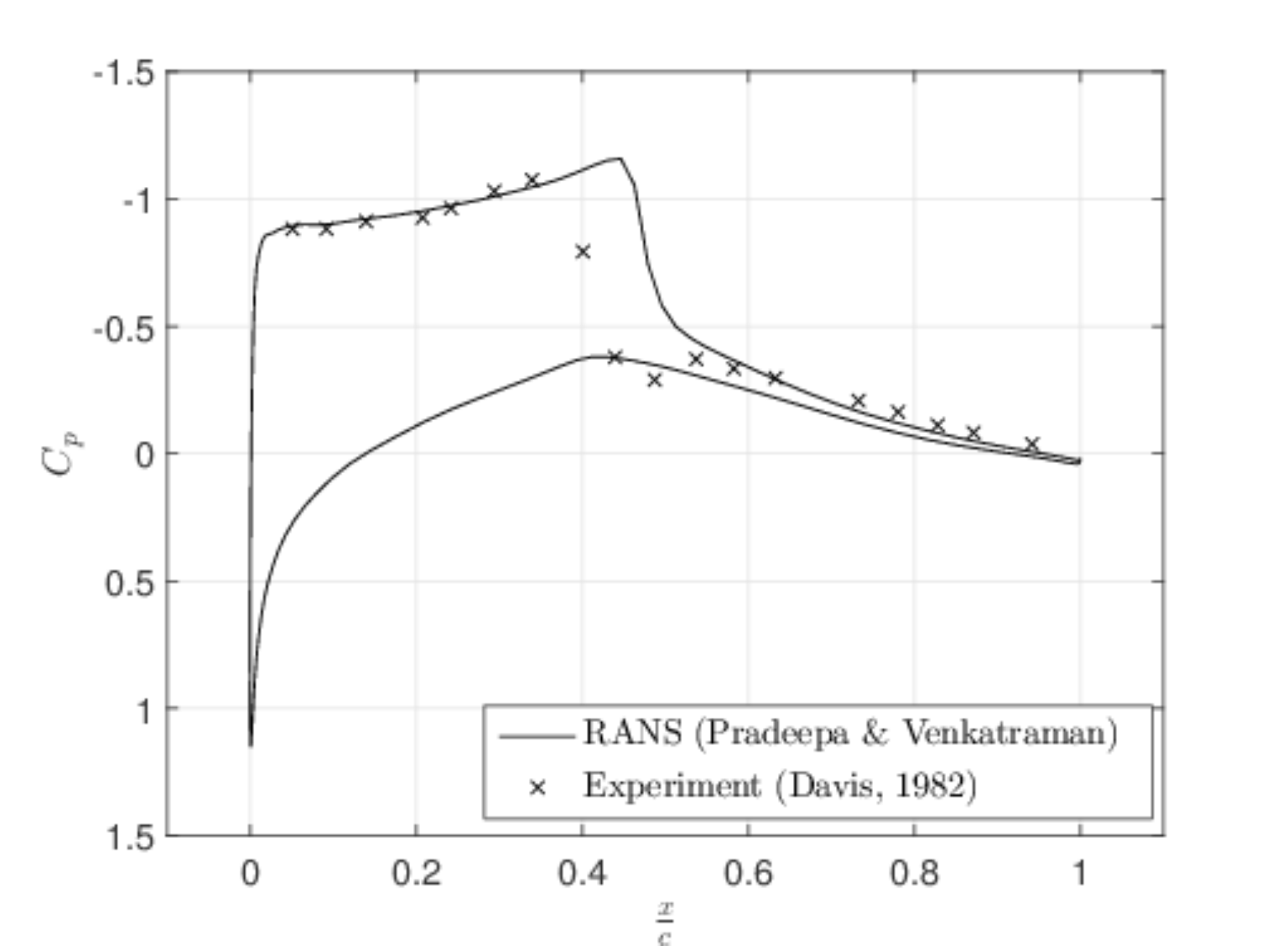}}
\subfigure[$\alpha = 3.51^\circ, \varphi = 331.69^\circ$]{\label{u3_16}\includegraphics[scale=0.40]{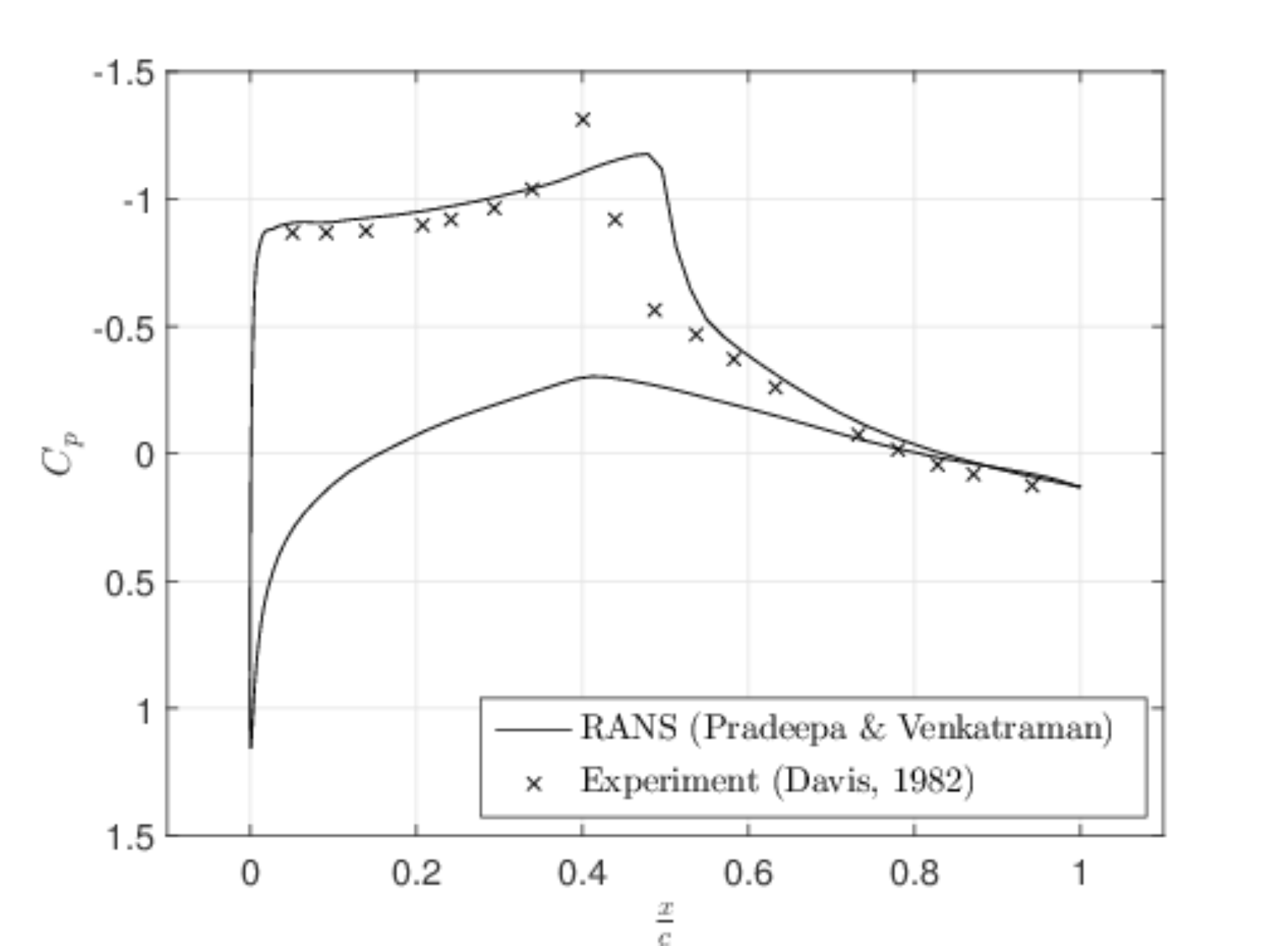}}
\caption{Instantaneous pressure distribution for shock-stall test case on the airfoil}
\label{unsteady3_inst}
\end{figure}

\begin{figure}
\subfigure[$\alpha = 4.99^\circ, \varphi = 98.43^\circ$]{\label{u3_21}\includegraphics[scale=0.40]{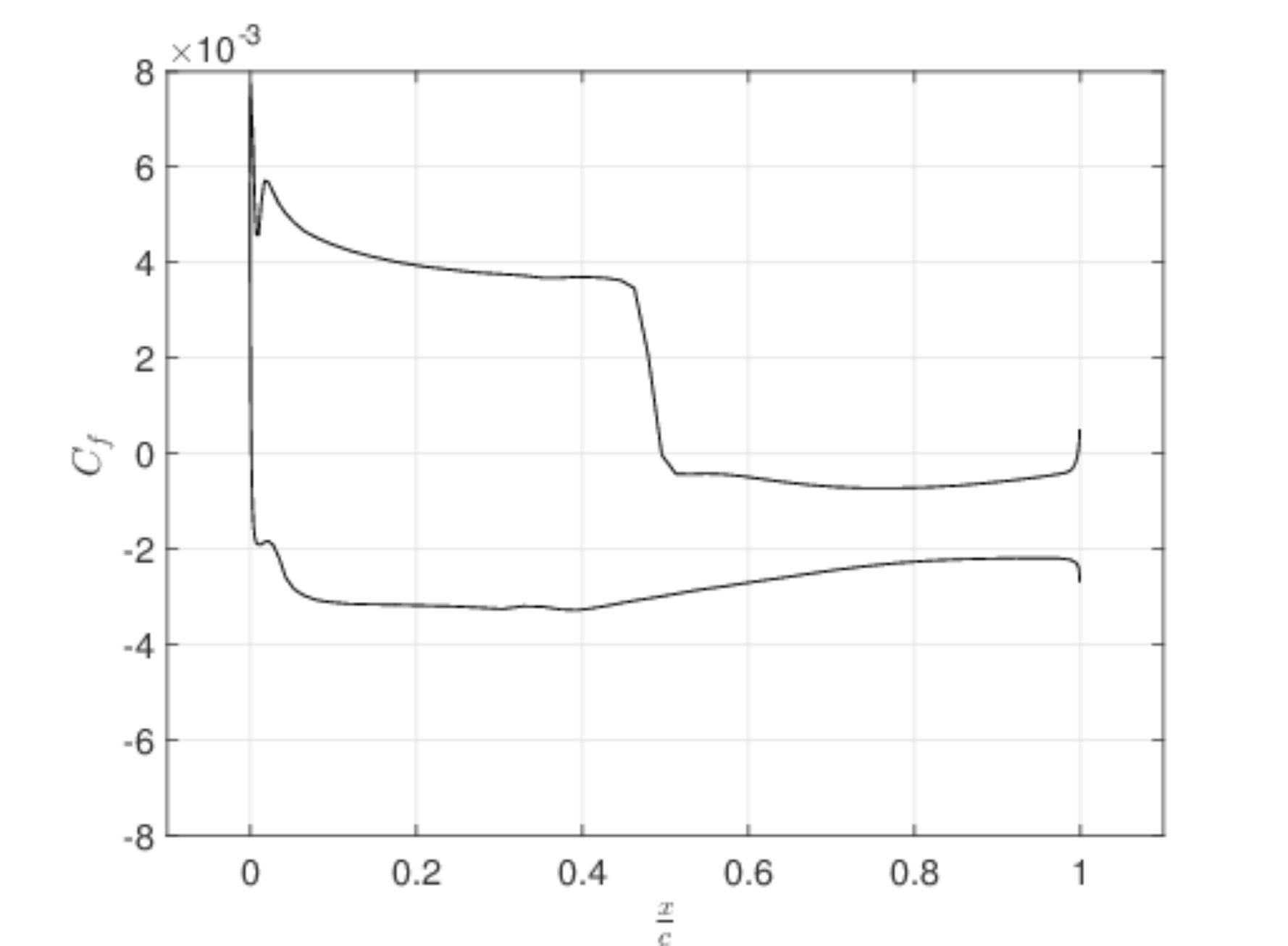}}
\subfigure[$\alpha = 4.58^\circ, \varphi = 144.84^\circ$]{\label{u3_22}\includegraphics[scale=0.40]{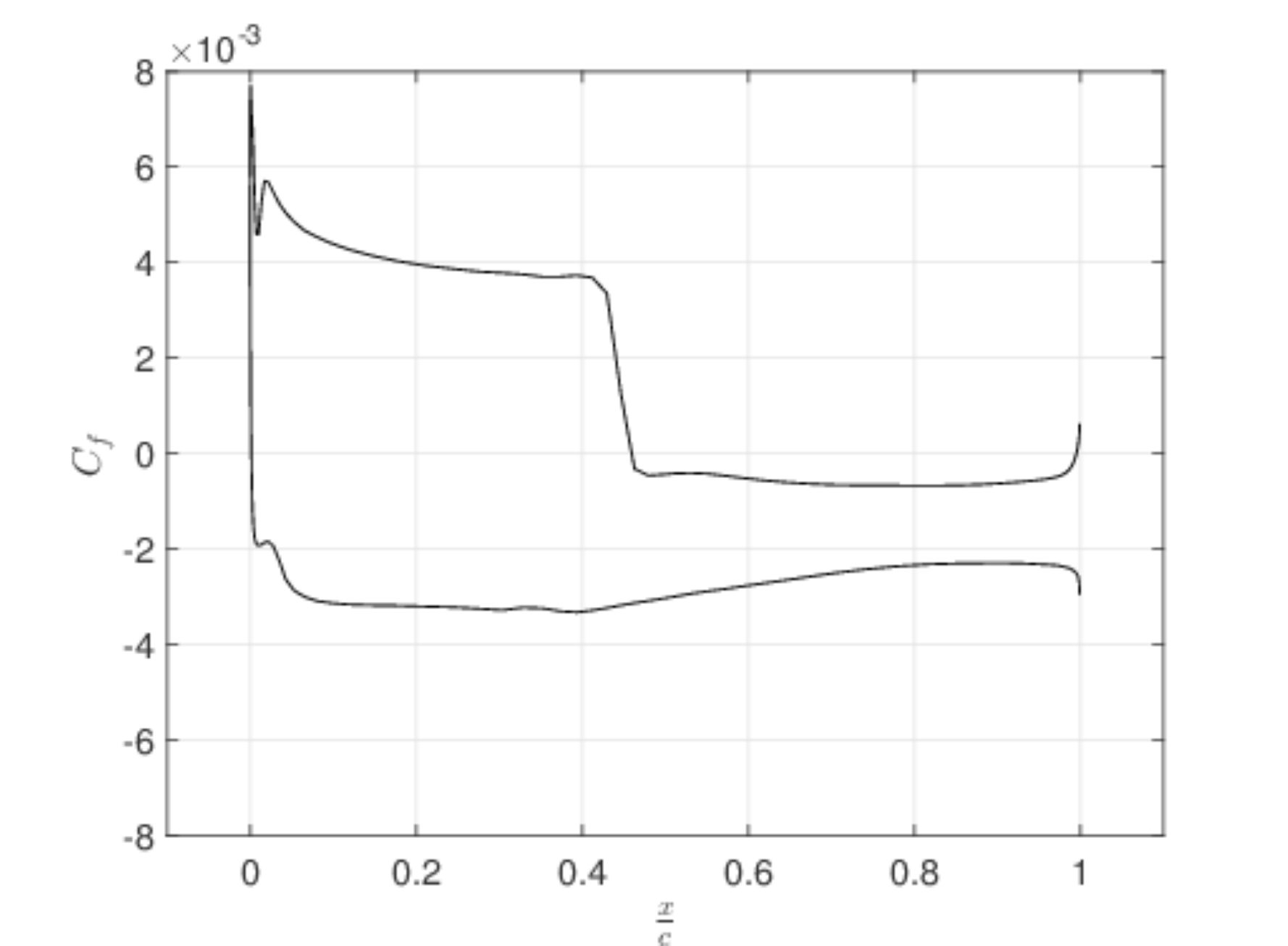}}
\subfigure[$\alpha = 4.08^\circ, \varphi = 175.10^\circ$]{\label{u3_23}\includegraphics[scale=0.40]{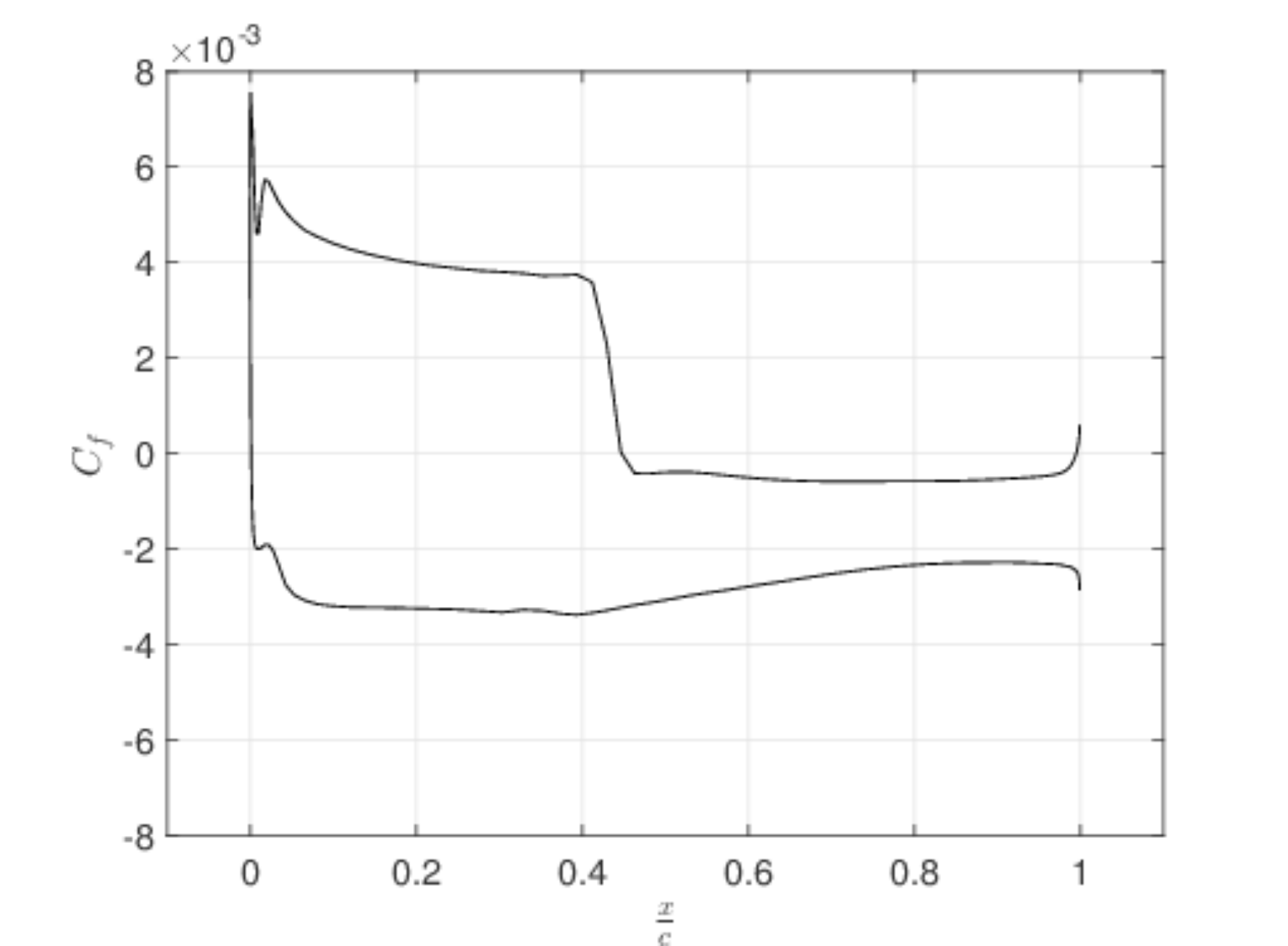}}
\subfigure[$\alpha = 3.58^\circ, \varphi = 204.61^\circ$]{\label{u3_24}\includegraphics[scale=0.40]{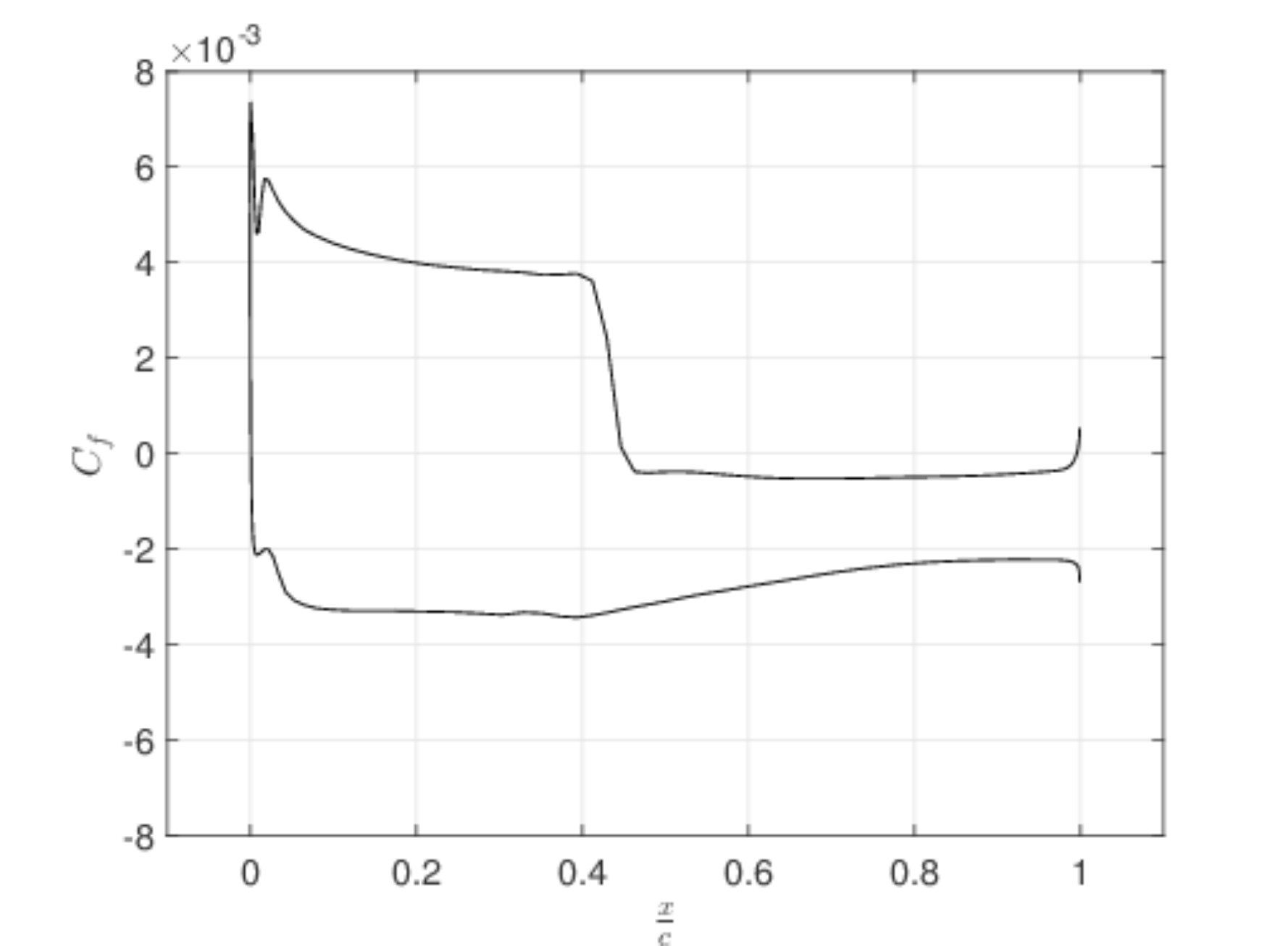}}
\subfigure[$\alpha = 2.99^\circ, \varphi = 269.98^\circ$]{\label{u3_25}\includegraphics[scale=0.40]{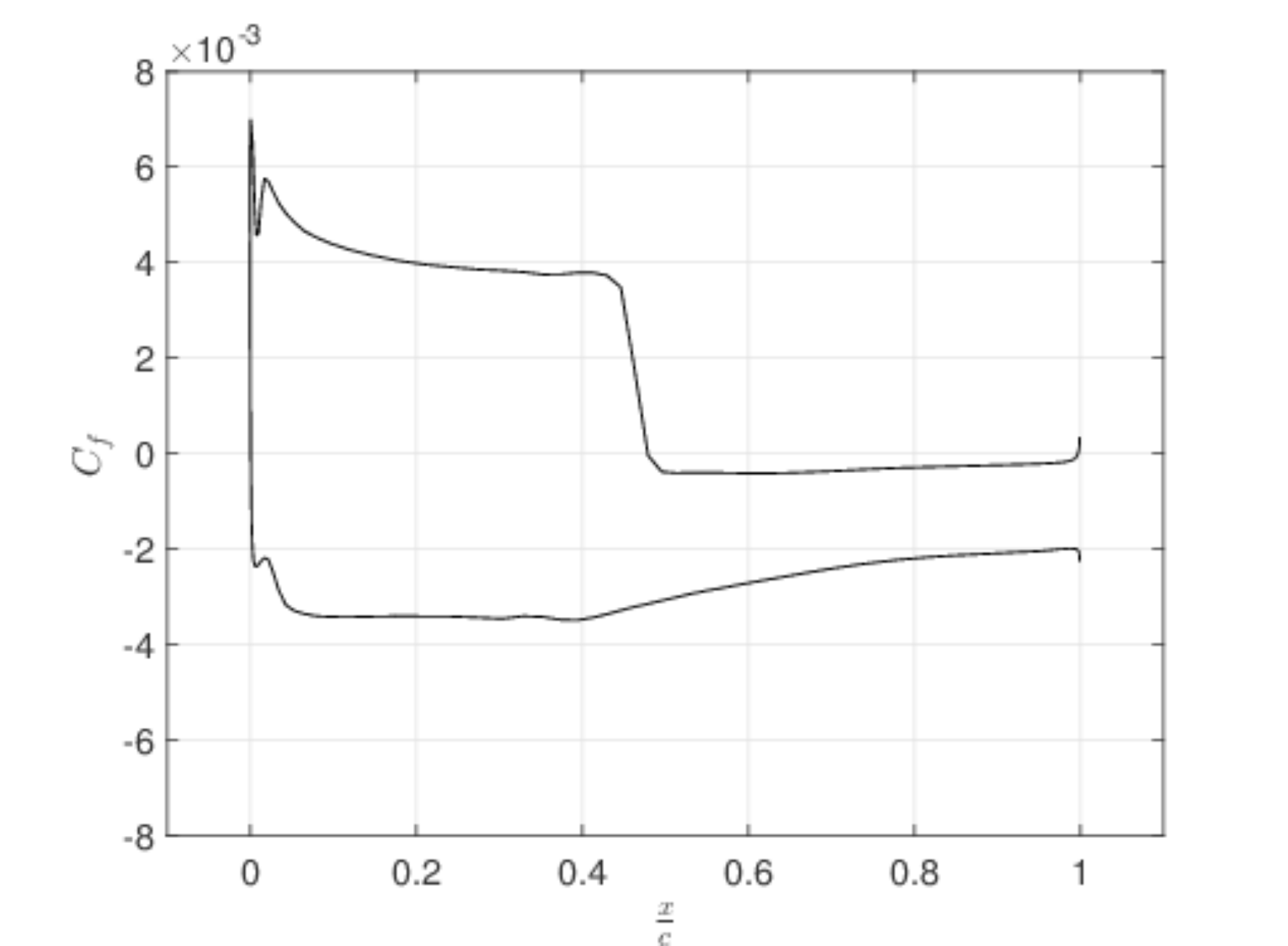}}
\subfigure[$\alpha = 3.51^\circ, \varphi = 331.69^\circ$]{\label{u3_26}\includegraphics[scale=0.40]{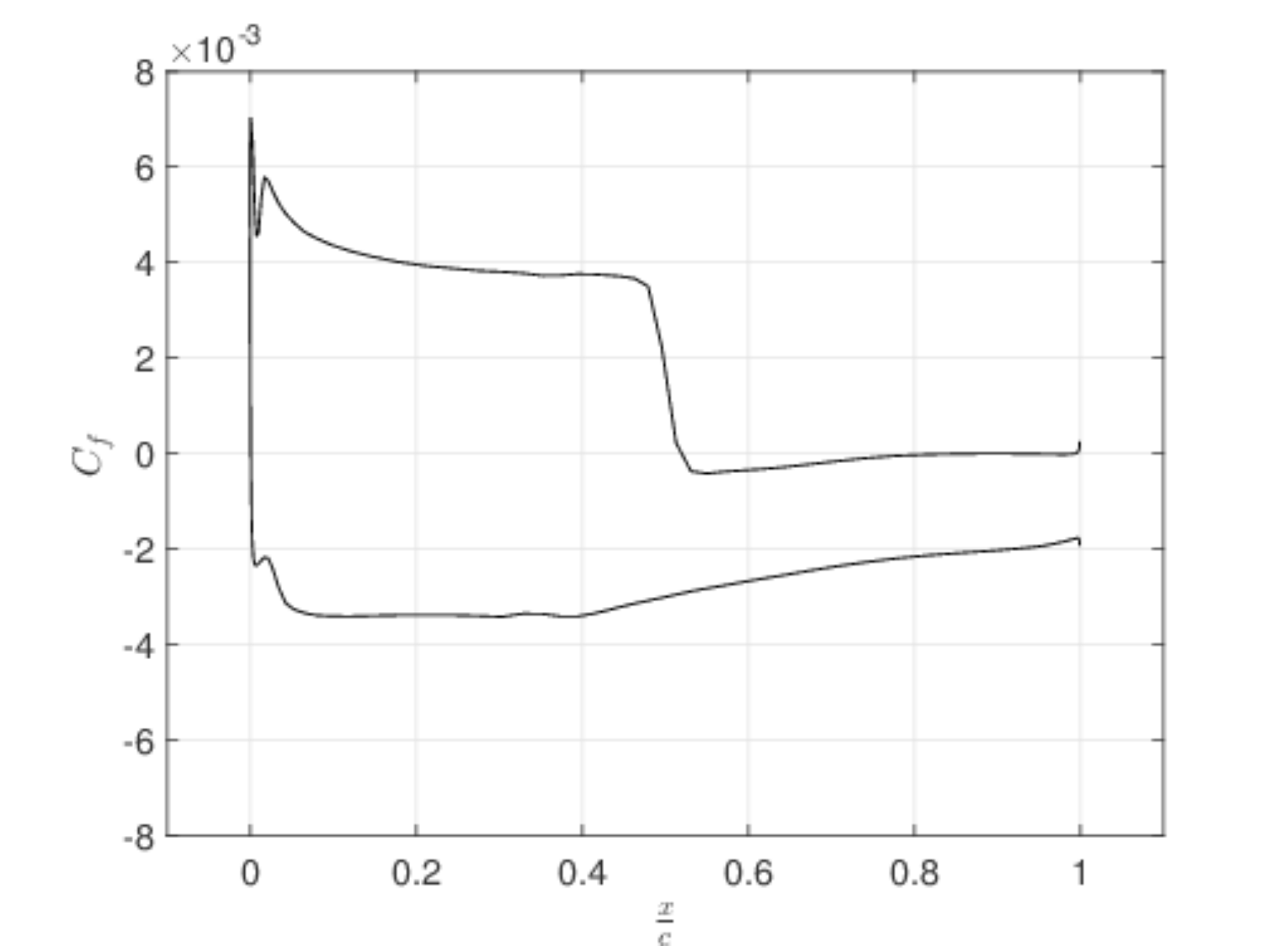}}
\caption{Instantaneous surface friction distribution on the airfoil for shock-stall test case}
\label{unsteady3_inst_f}
\end{figure}

In Section~\ref{sec:ch_ust_af}, the unsteady aerodynamic coefficients computed using RANS equations compared well with corresponding experimental results. Mach contours suggested the thickening of boundary layer because of the sudden increase in pressure at the shock. The shock was not strong enough to cause separation. Skin friction estimates over the surface of the airfoil testify to that. The governing time averaged equations together with the Spalart-Allmaras model equation was able to predict the flow with considerable accuracy. 

Now we consider a case of shock induced separation \citep{davis1982}. The airfoil considered is also a NACA64A010, pitching about its quarter chord with an amplitude of $1.01^\circ$, but now about a mean angle of $4.0^\circ$ while held in a fluid flow of Mach number $0.789$ and Reynolds number of $11.88 \times 10^6$. The reduced frequency of pitching is $\kappa = 0.204$.

\begin{figure}
\subfigure[$\alpha = 4.99^\circ, \varphi = 98.43^\circ$]{\label{u3_31}\includegraphics[scale=0.40]{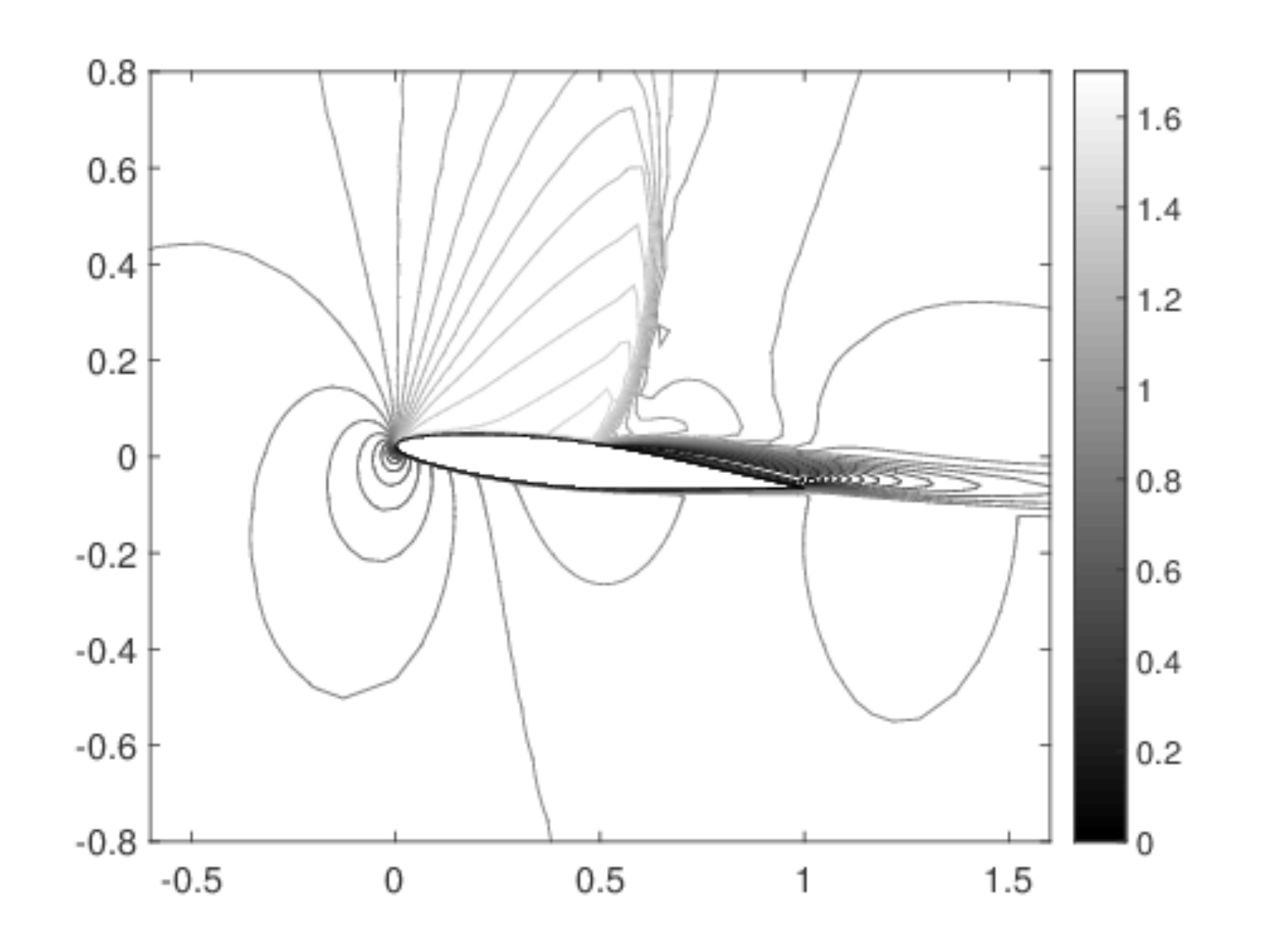}}
\subfigure[$\alpha = 4.58^\circ, \varphi = 144.84^\circ$]{\label{u3_32}\includegraphics[scale=0.40]{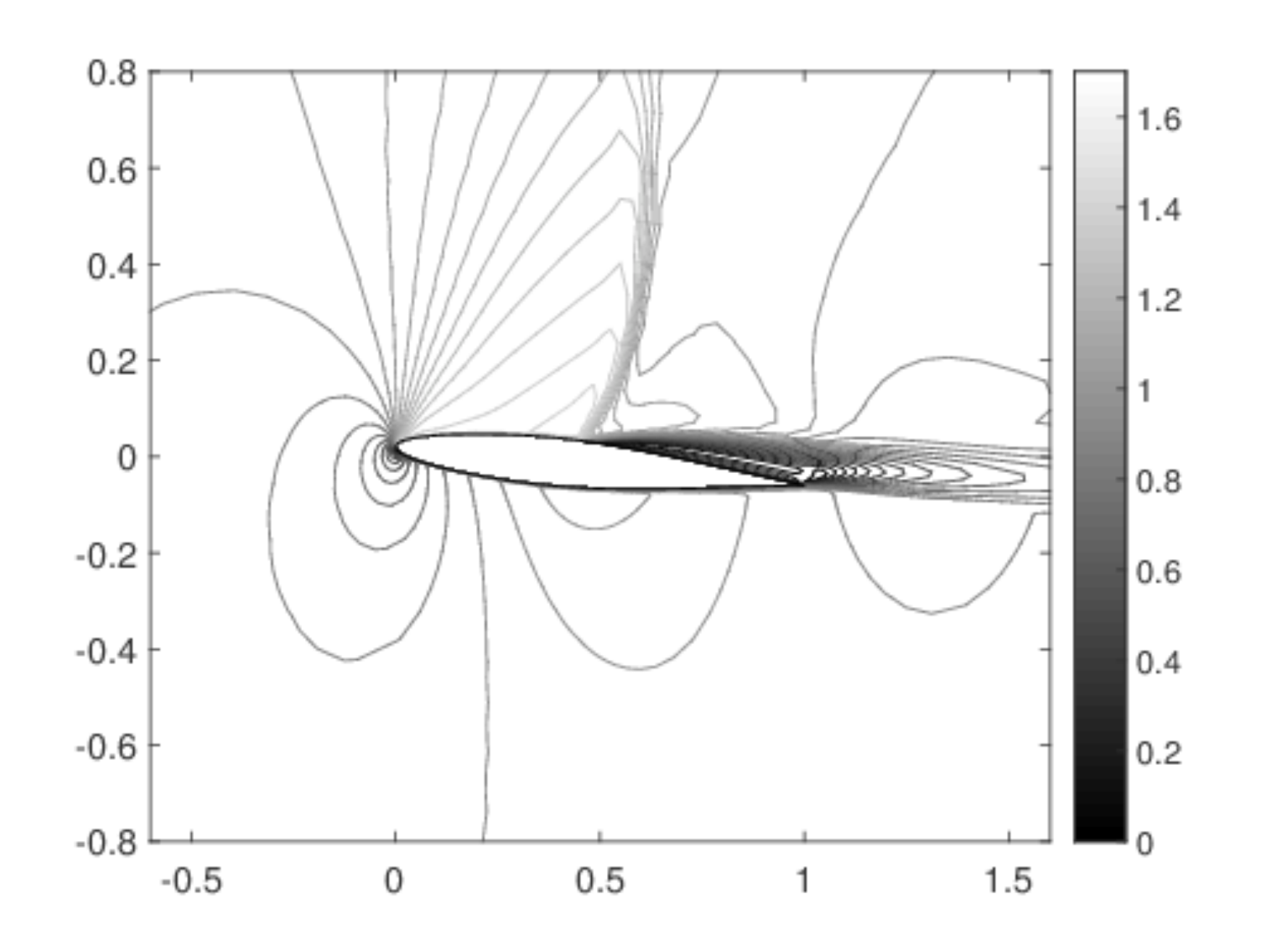}}
\subfigure[$\alpha = 4.08^\circ, \varphi = 175.10^\circ$]{\label{u3_33}\includegraphics[scale=0.40]{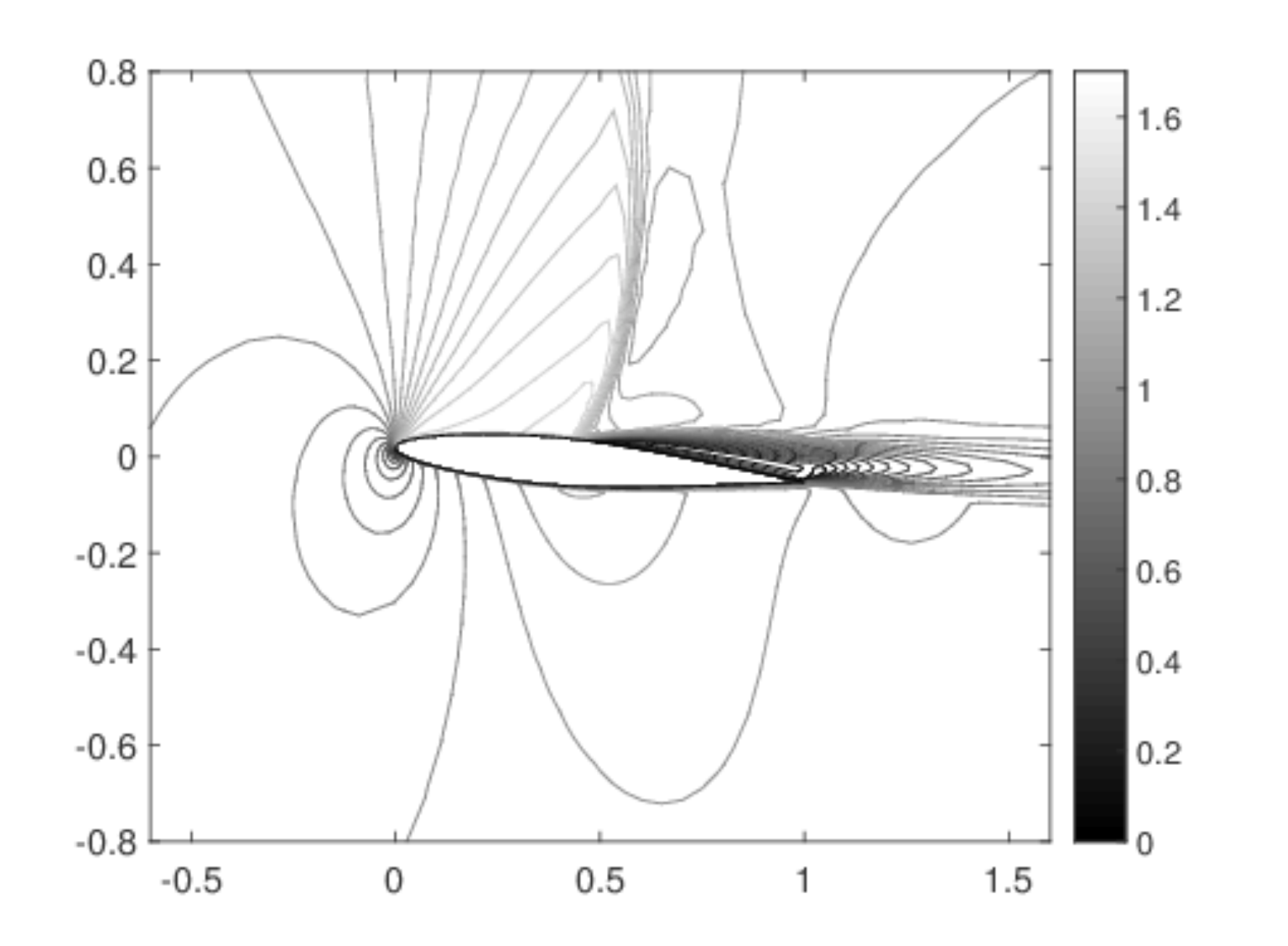}}
\subfigure[$\alpha = 3.58^\circ, \varphi = 204.61^\circ$]{\label{u3_34}\includegraphics[scale=0.40]{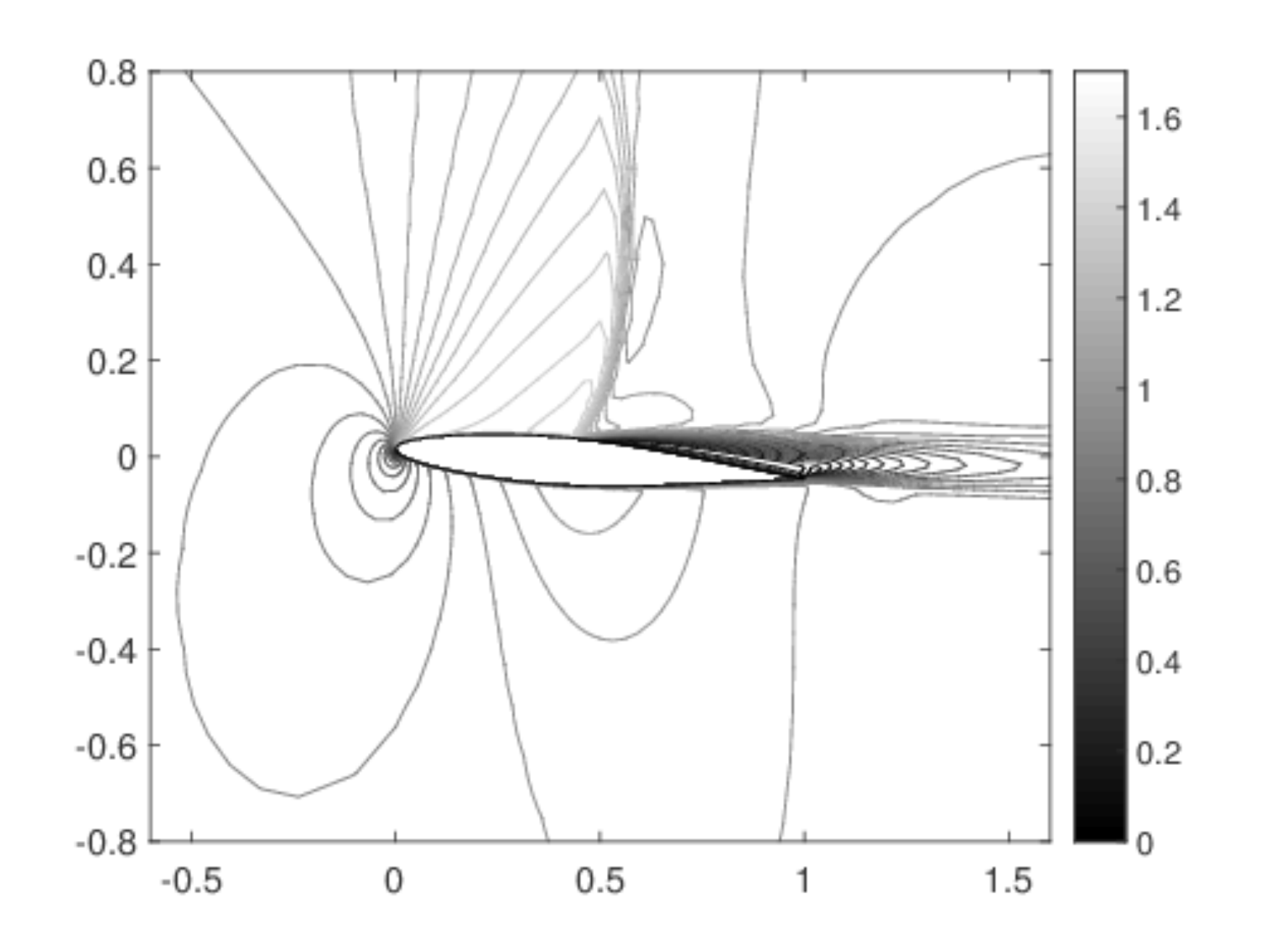}}
\subfigure[$\alpha = 2.99^\circ, \varphi = 269.98^\circ$]{\label{u3_35}\includegraphics[scale=0.40]{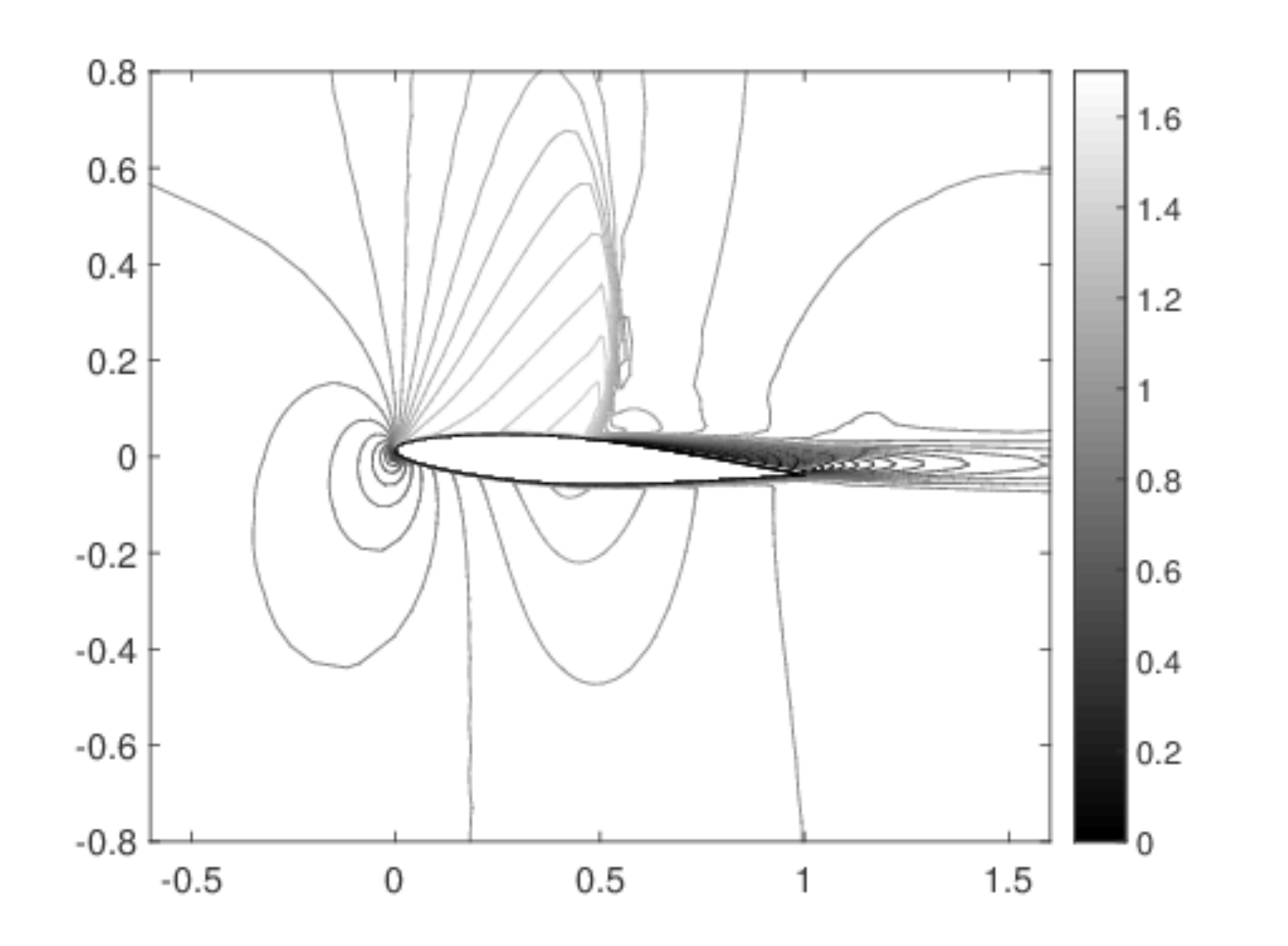}}
\subfigure[$\alpha = 3.51^\circ, \varphi = 331.69^\circ$]{\label{u3_36}\includegraphics[scale=0.40]{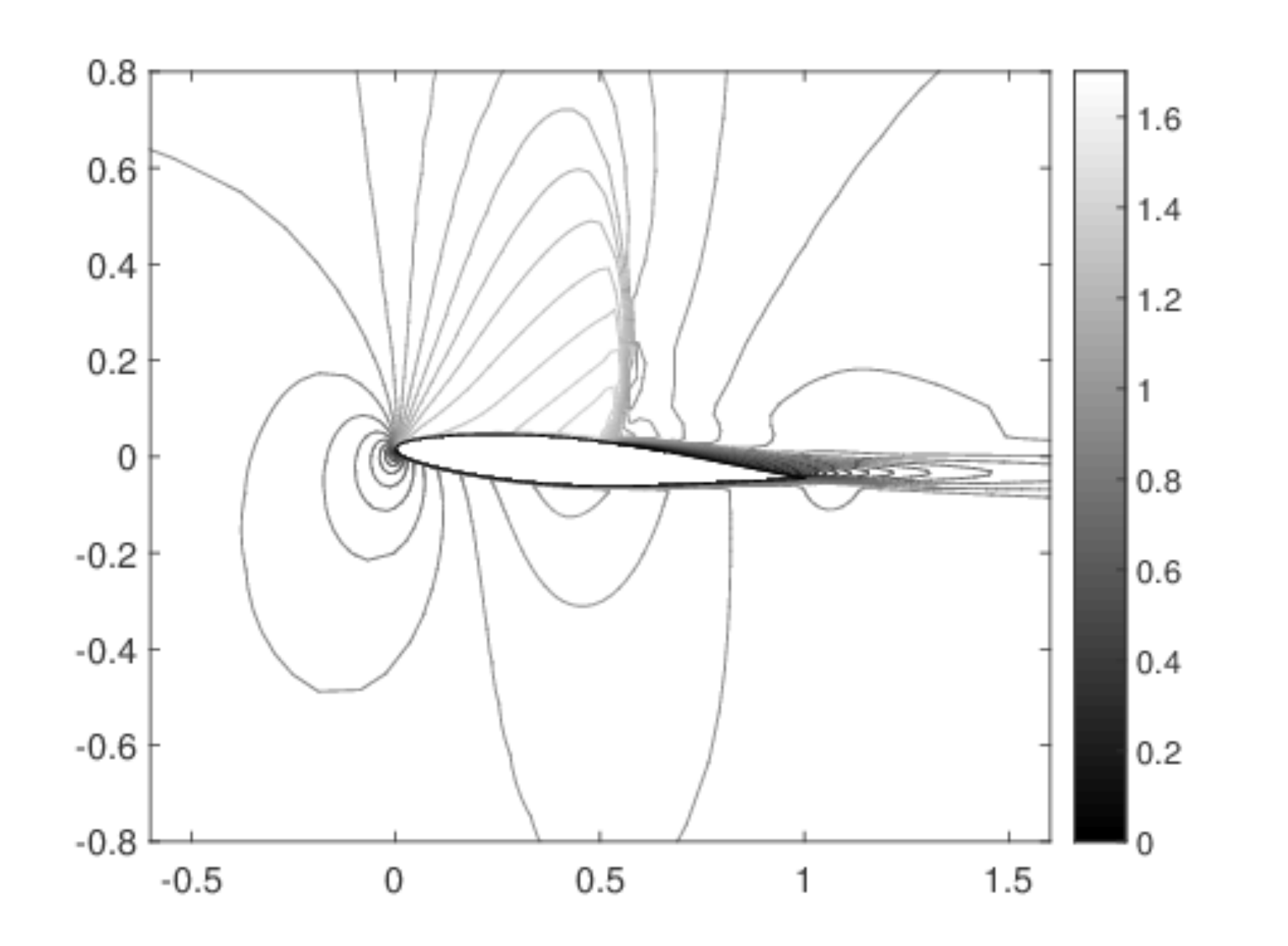}}
\caption{Instantaneous Mach contours on the airfoil for shock-stall test case}
\label{unsteady3_inst_mach}
\end{figure}

The pressure distribution on the airfoil surface at different instants in a pitching cycle are shown in Figure~\ref{unsteady3_inst}. It can be seen that the shock is present throughout the pitching cycle and hence is a class A type of shock motion \citep[Chap.9, pp.62]{tijdeman1977}. Our RANS computation using the Spalart-Allmaras turbulence model captures the flow accurately everywhere except near the shock. The shock location is always predicted slightly aft of the experimentally predicted shock location of \citet{davis1982}. The reason for this discrepancy is attributed to the turbulence model used. The \citet{spalart1992} turbulence model used in the turbulence flow computation cannot predict the shock induced separation point accurately because of the Boussinesq approximation used to model the Reynolds stress. The shock on the airfoil upper surface is strong enough to cause the flow to separate. This is seen when the $C_f$ curve changes from a positive to a negative value as shown in the skin friction distribution at different instants in Figure \ref{unsteady3_inst_f}. Mach contours at these instants, Figure~\ref{unsteady3_inst_mach}, shows the supersonic pocket being abruptly terminated by a shock, as well as the boundary layer separation due to the adverse pressure gradient created by the shock.

\begin{figure}
\subfigure[$\alpha = 4.99^\circ, \varphi = 98.43^\circ$]{\label{u3_51}\includegraphics[scale=0.40]{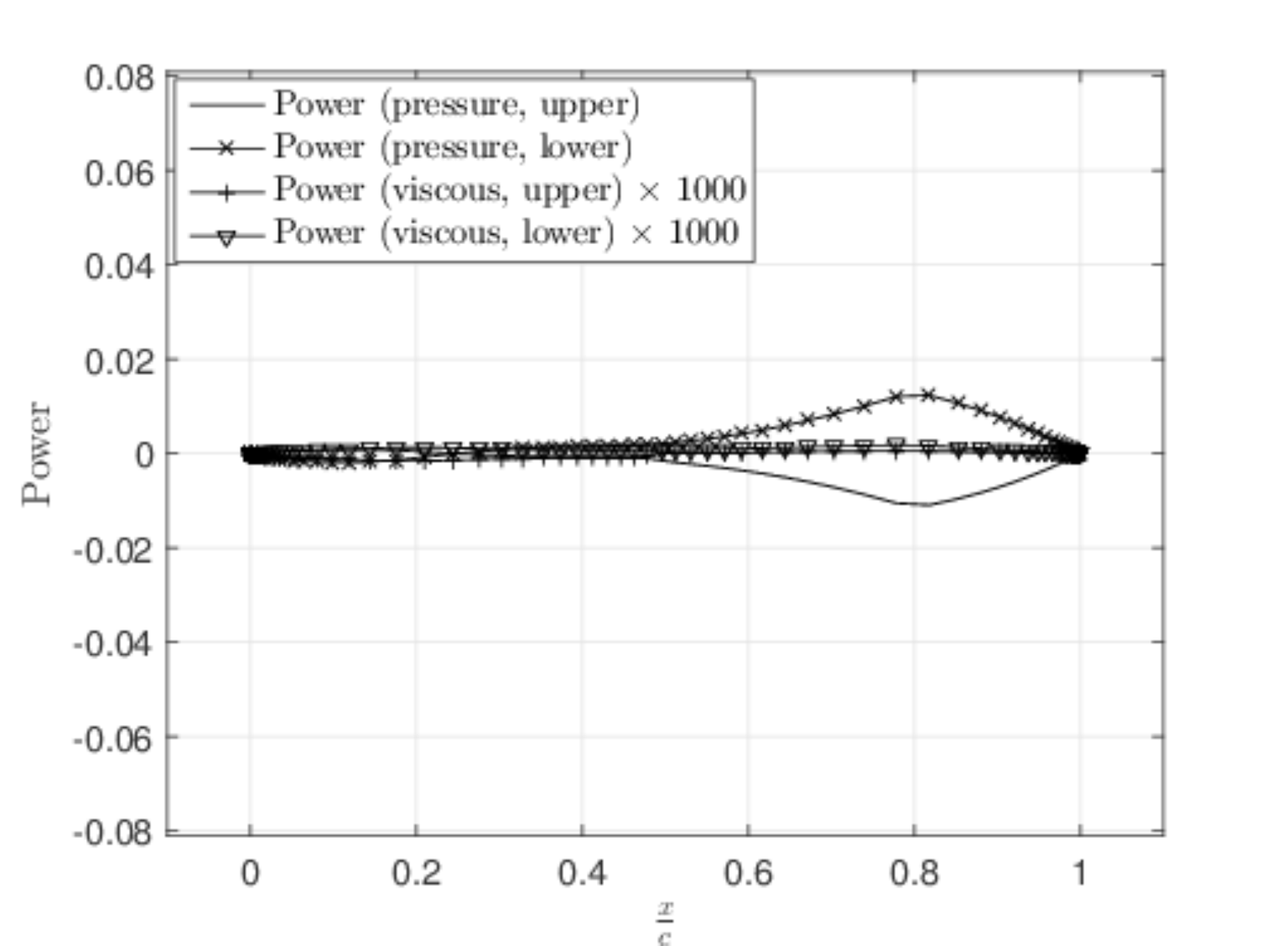}}
\subfigure[$\alpha = 4.58^\circ, \varphi = 144.84^\circ$]{\label{u3_52}\includegraphics[scale=0.40]{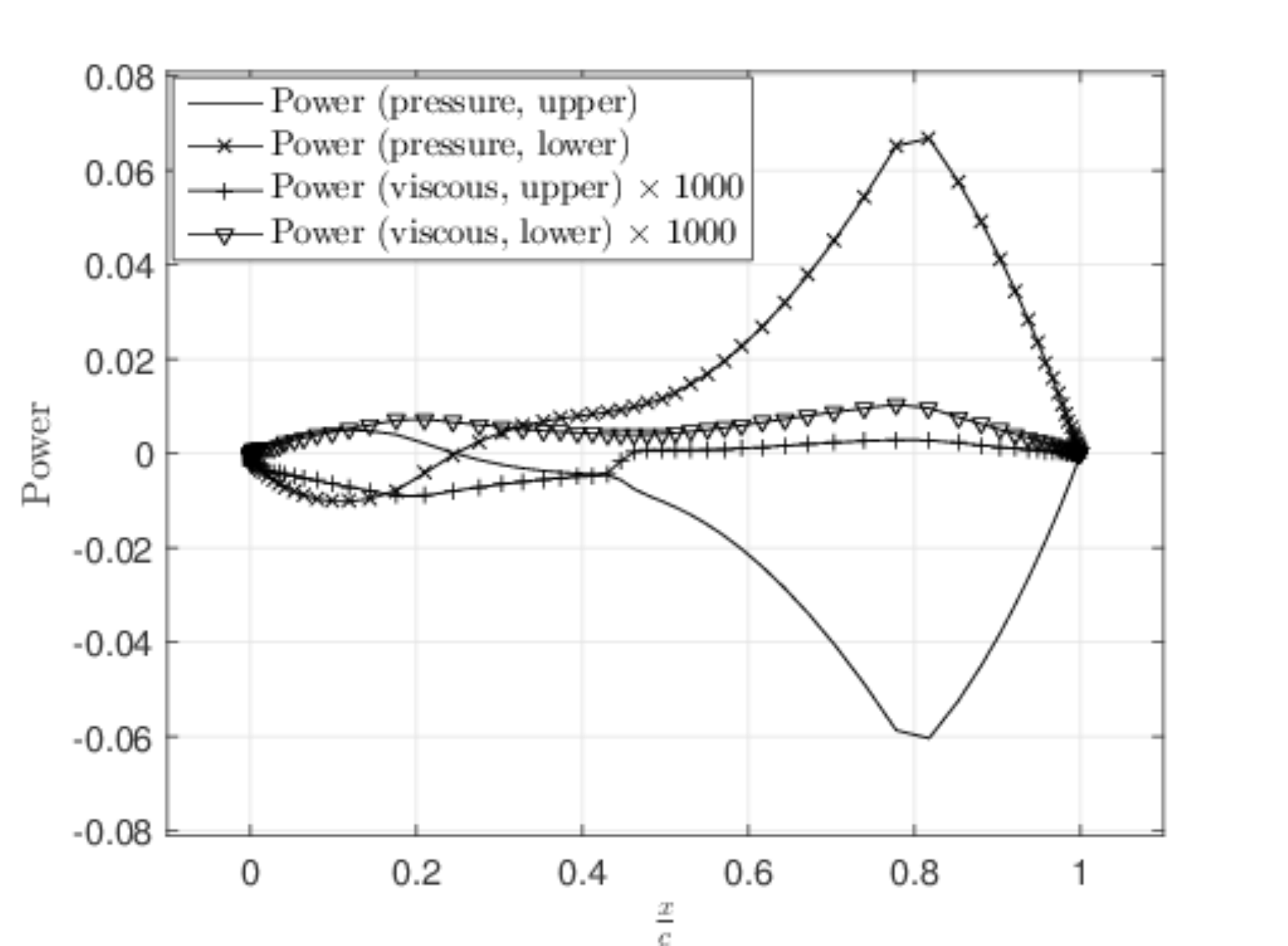}}
\subfigure[$\alpha = 4.08^\circ, \varphi = 175.10^\circ$]{\label{u3_53}\includegraphics[scale=0.40]{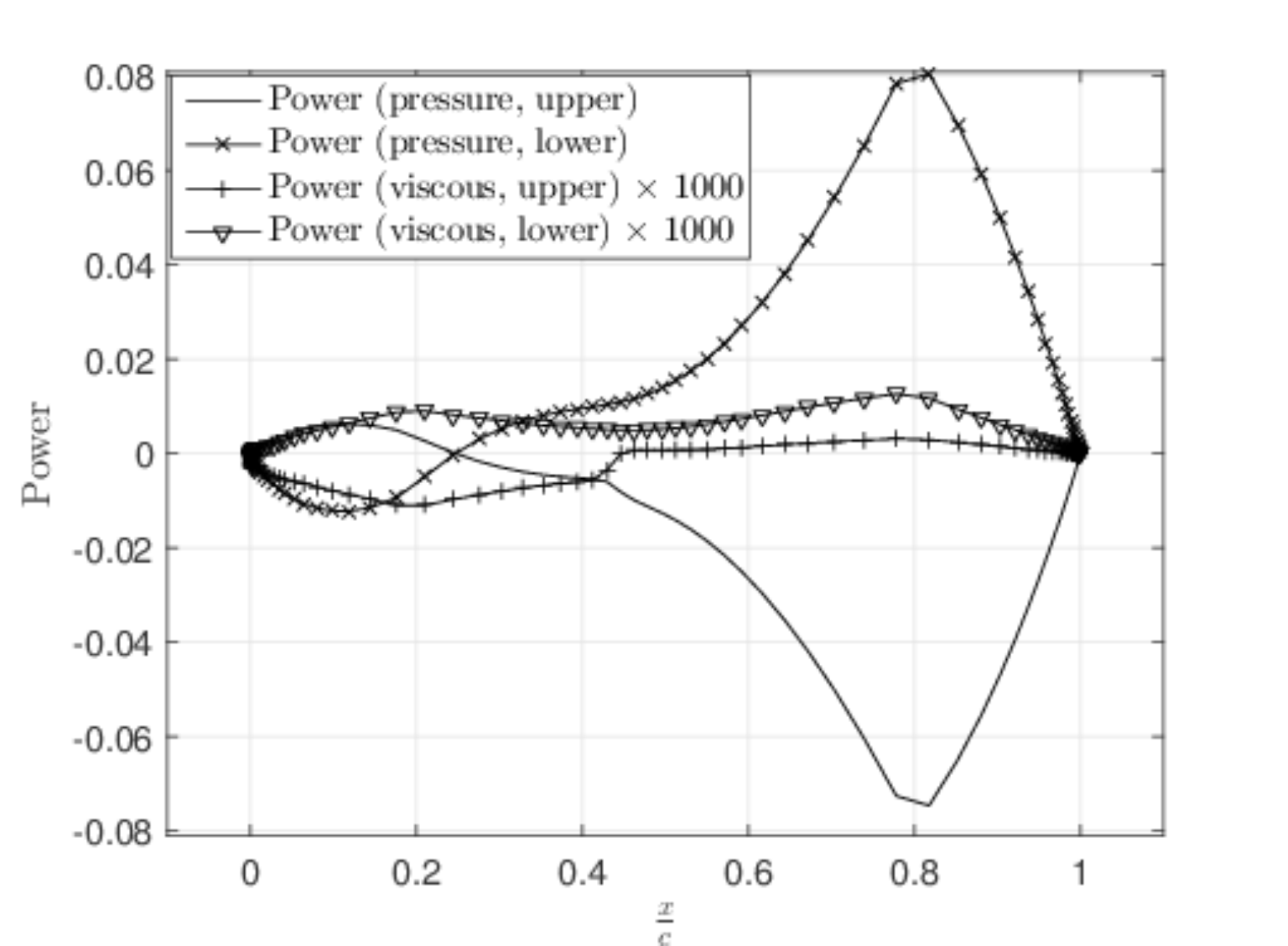}}
\subfigure[$\alpha = 3.58^\circ, \varphi = 204.61^\circ$]{\label{u3_54}\includegraphics[scale=0.40]{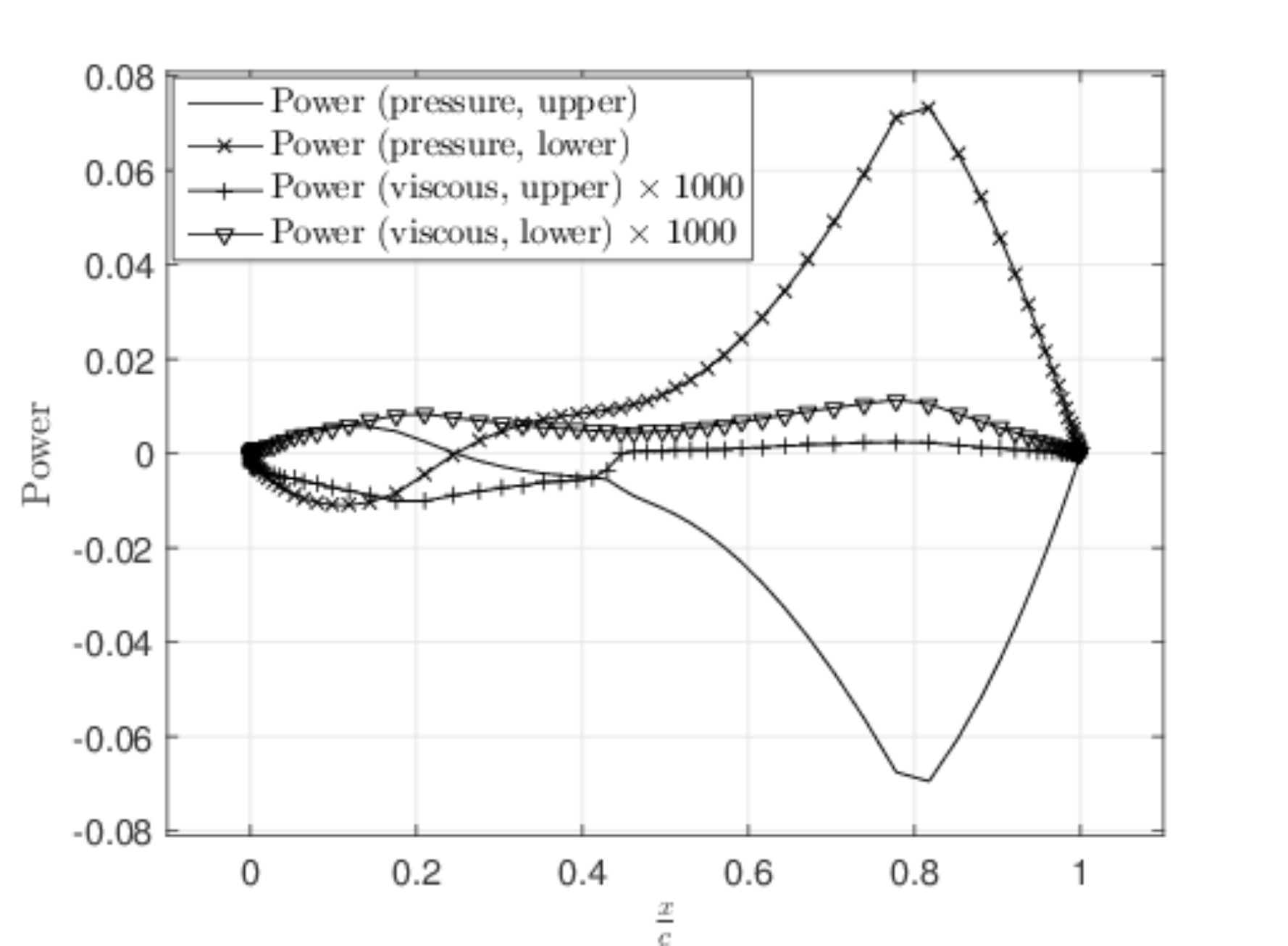}}
\subfigure[$\alpha = 2.99^\circ, \varphi = 269.98^\circ$]{\label{u3_55}\includegraphics[scale=0.40]{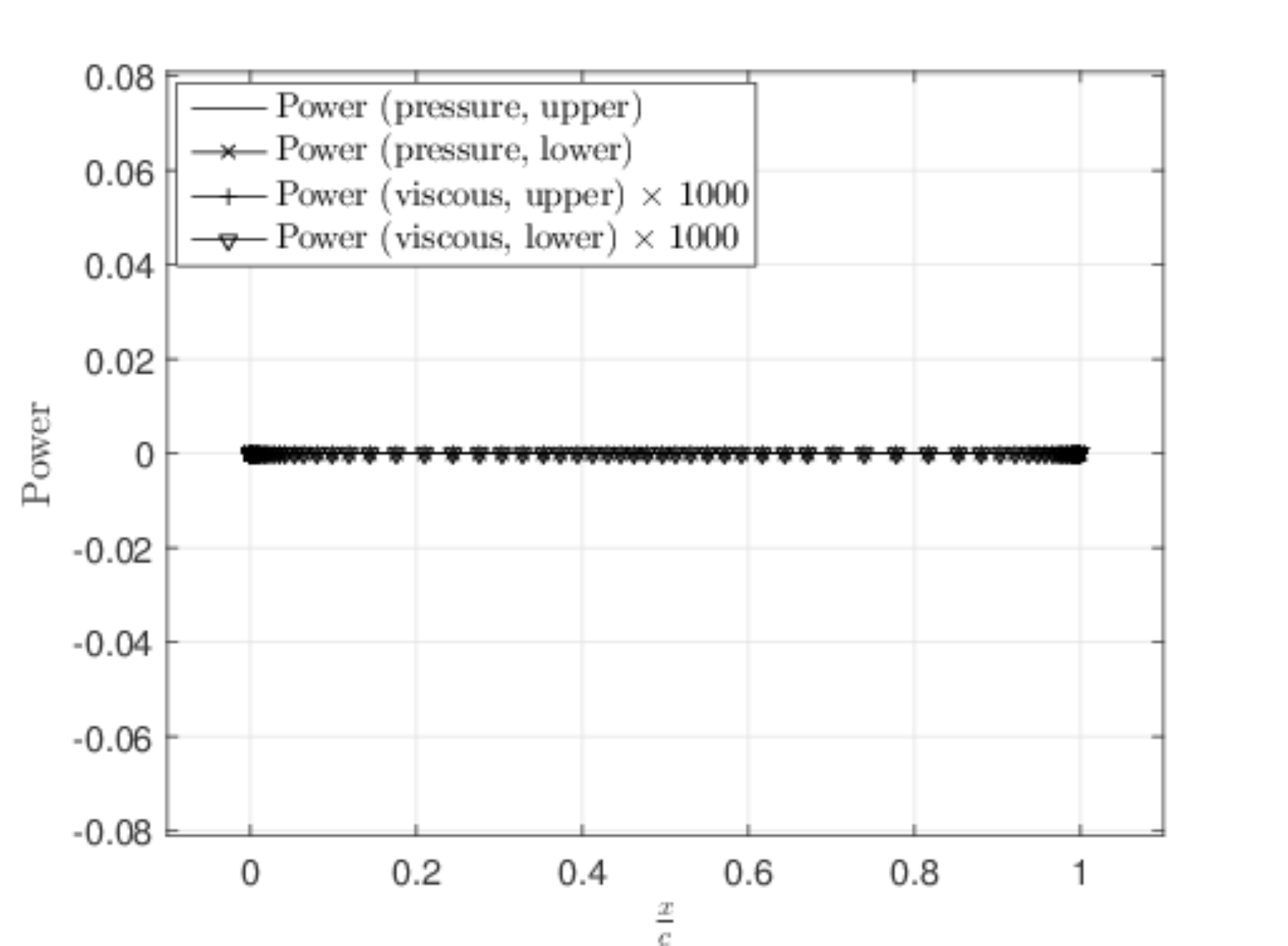}}
\subfigure[$\alpha = 3.51^\circ, \varphi = 331.69^\circ$]{\label{u3_56}\includegraphics[scale=0.40]{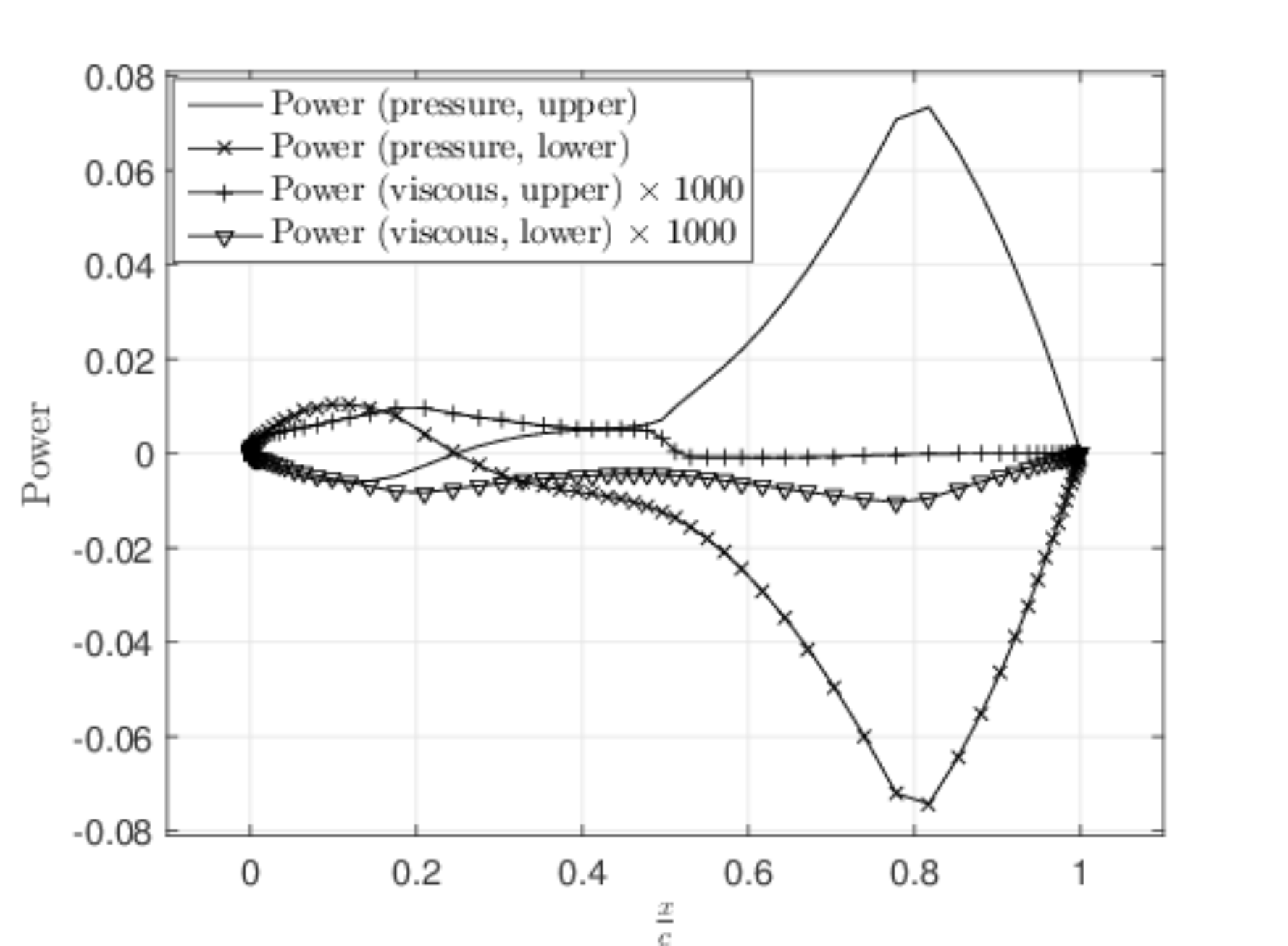}}
\caption{Instantaneous power distribution on airfoil upper and lower surface for shock induced boundary layer separation}
\label{unsteady3_inst_power}
\end{figure}

In order to understand the effects of pressure and viscous forces that influence the motion of the airfoil, power distribution on the airfoil surface at different instants are shown in Figure~\ref{unsteady3_inst_power}. The aerodynamic power calculated on the surface of the airfoil due to pressure and viscous forces are

\begin{equation}
\label{aerodynpower}
\begin{aligned}
\text{aerodynamic power due to pressure force}_{x,y} = (\Delta y x' - \Delta x y') p, \\
\text{aerodynamic power due to viscous force}_{x,y} = (\Delta x x' + \Delta y y') \frac{1}{2} \rho u_\infty^2 C_f.
\end{aligned}
\end{equation}

Here $\Delta x$ and $\Delta y$ are the increments in $x$ and $y$ directions on the airfoil surface over which the pressure $p$ and skin friction coefficient $C_f$ are computed numerically. $x'$ and $y'$ are the velocities in the $x$ and $y$ directions respectively on the airfoil surface at point $(x,y)$. The computed power at each point $(x,y)$ is non-dimensionalized by $\rho_\infty b^5 \omega^3$. 

\begin{figure}
\subfigure[$\alpha = 4.99^\circ, \varphi = 98.43^\circ$]{\label{u3_41}\includegraphics[scale=0.40]{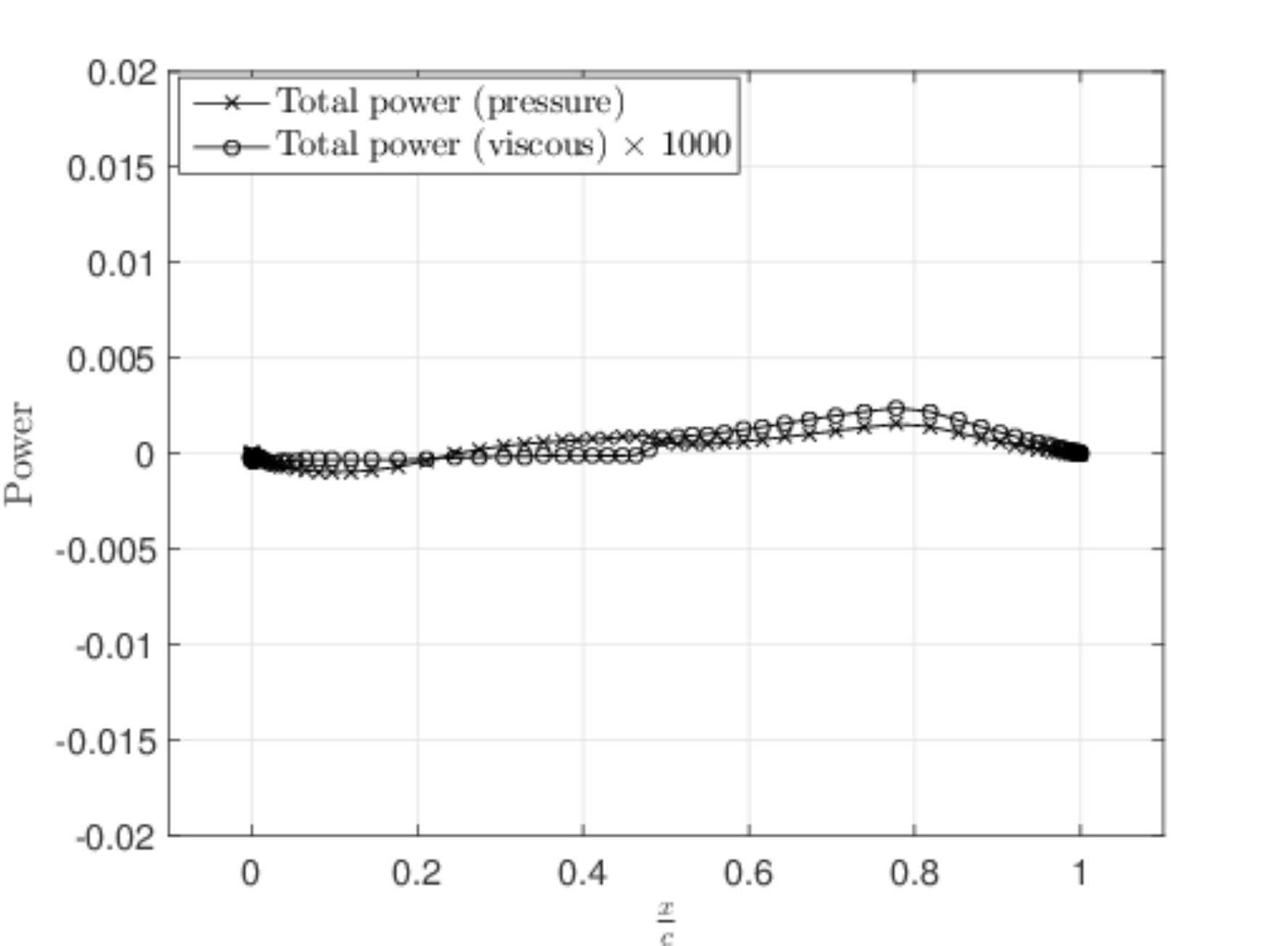}}
\subfigure[$\alpha = 4.58^\circ, \varphi = 144.84^\circ$]{\label{u3_42}\includegraphics[scale=0.40]{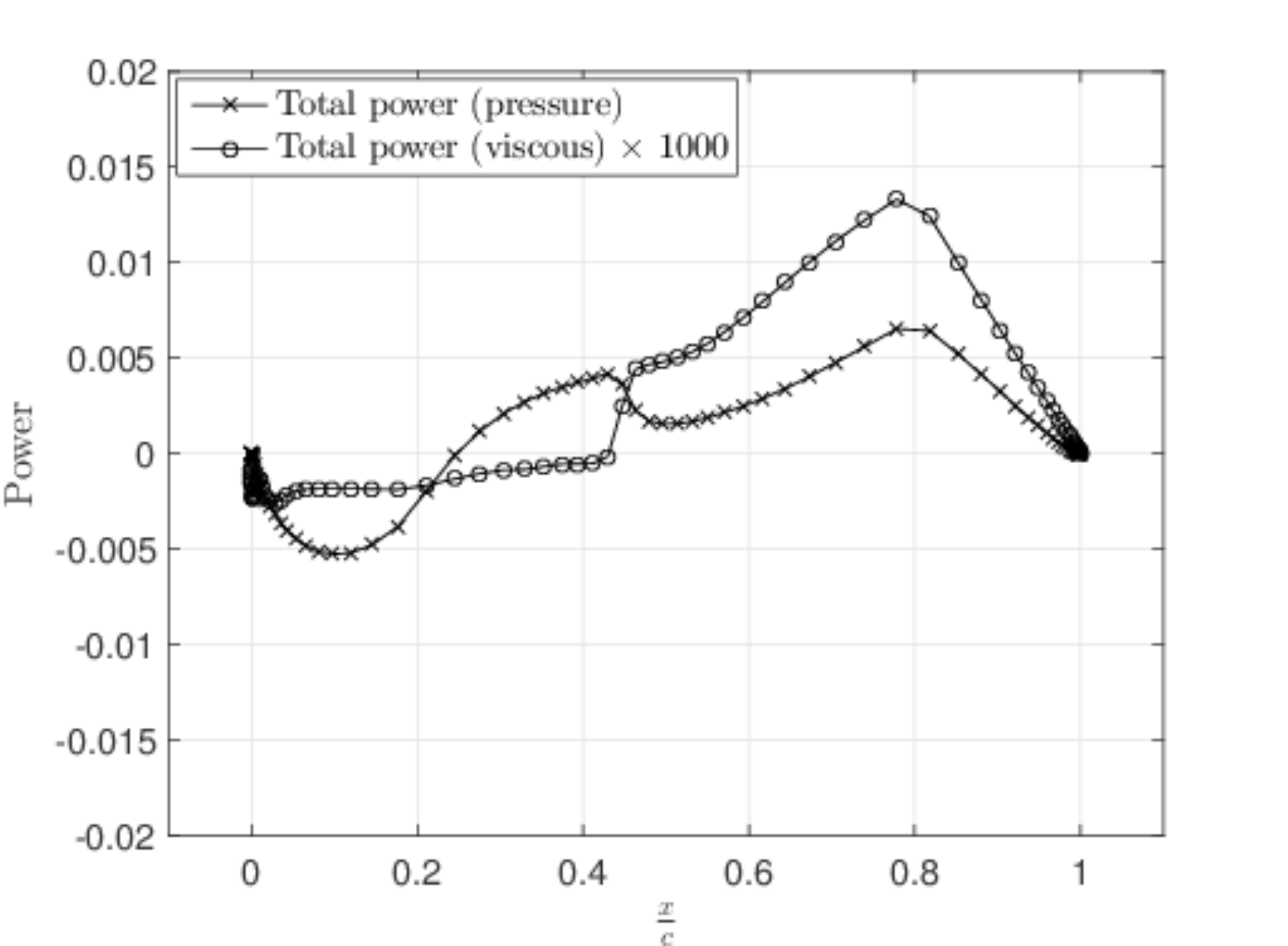}}
\subfigure[$\alpha = 4.08^\circ, \varphi = 175.10^\circ$]{\label{u3_43}\includegraphics[scale=0.40]{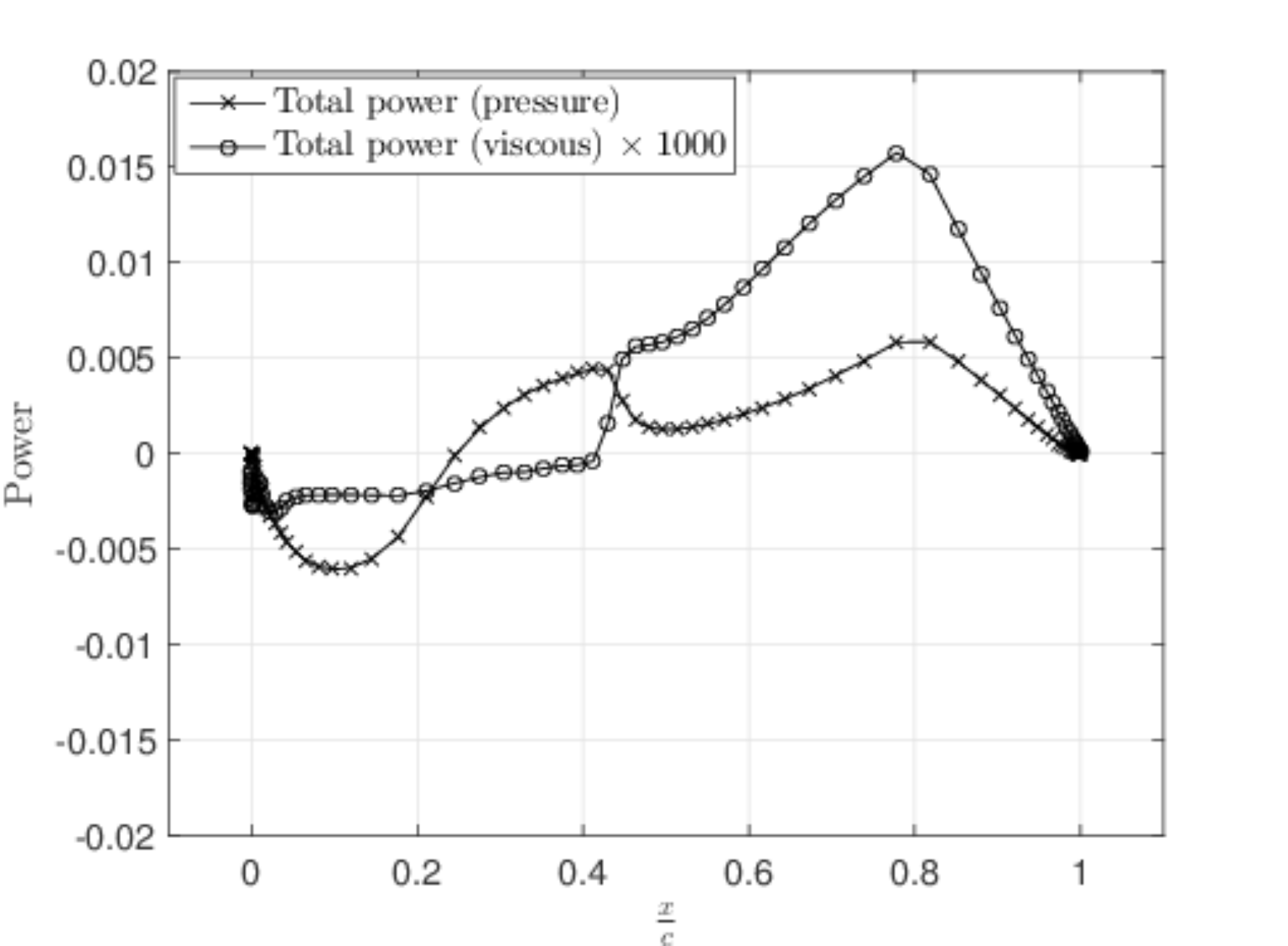}}
\subfigure[$\alpha = 3.58^\circ, \varphi = 204.61^\circ$]{\label{u3_44}\includegraphics[scale=0.40]{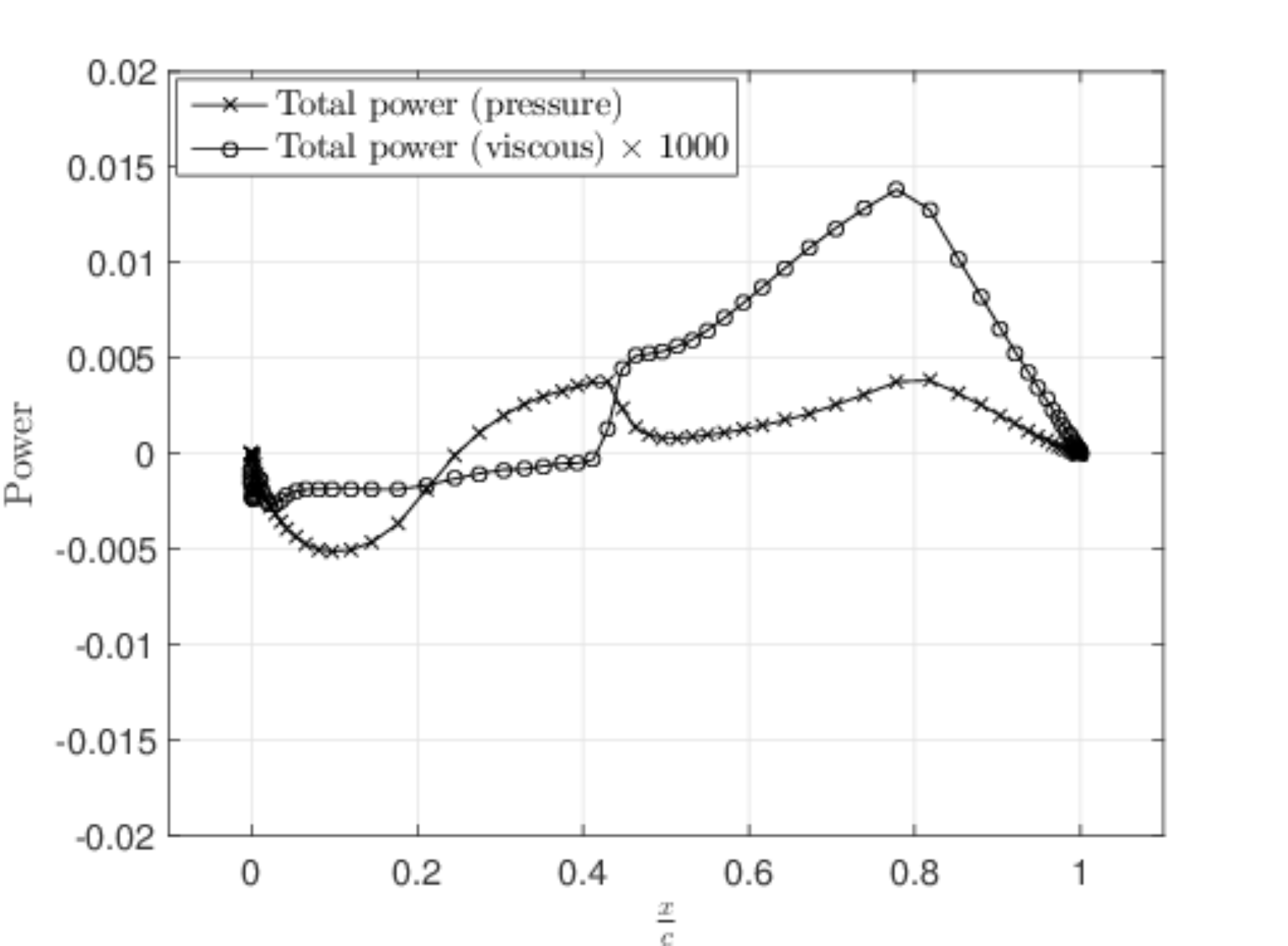}}
\subfigure[$\alpha = 2.99^\circ, \varphi = 269.98^\circ$]{\label{u3_45}\includegraphics[scale=0.40]{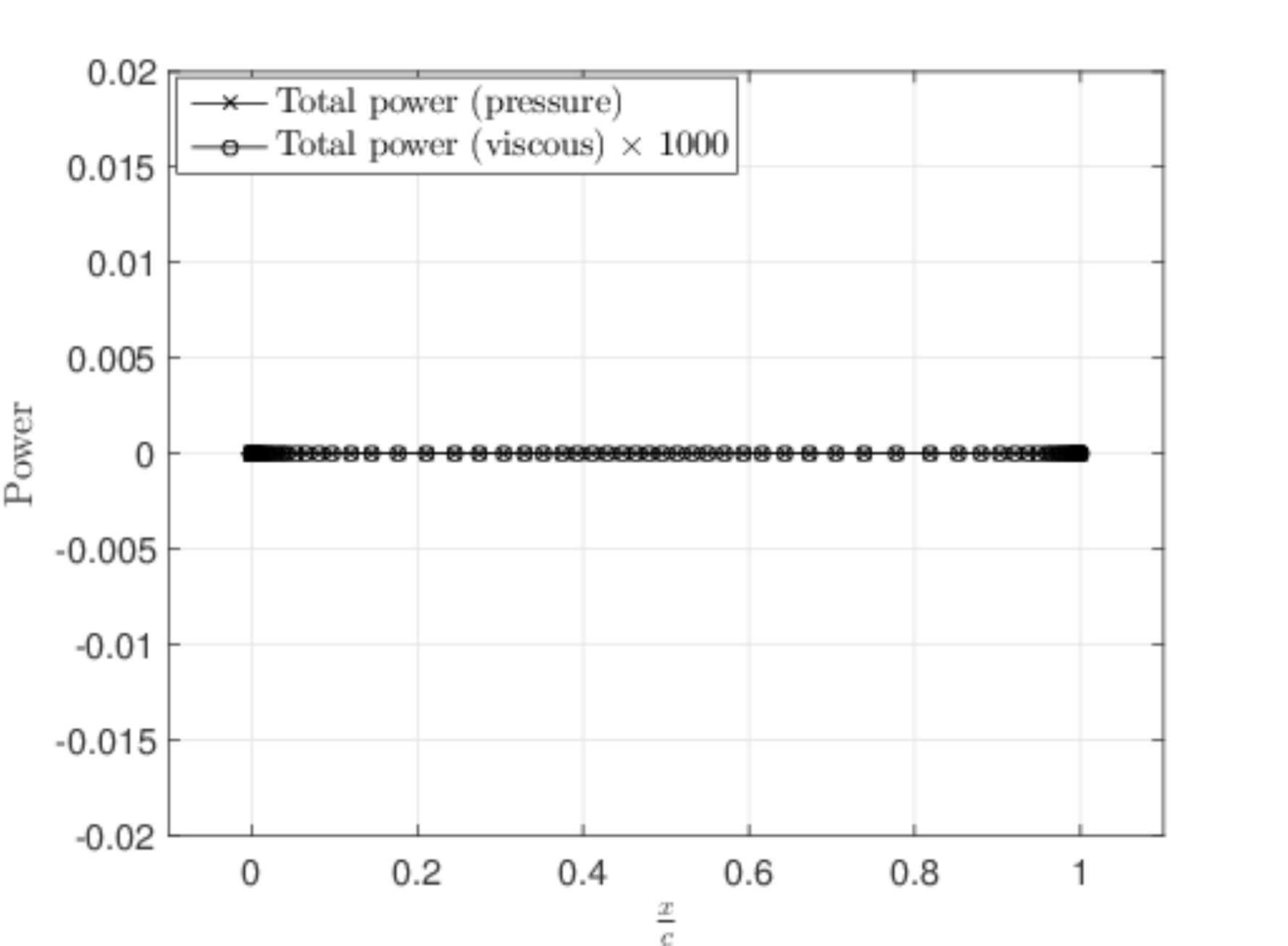}}
\subfigure[$\alpha = 3.51^\circ, \varphi = 331.69^\circ$]{\label{u3_46}\includegraphics[scale=0.40]{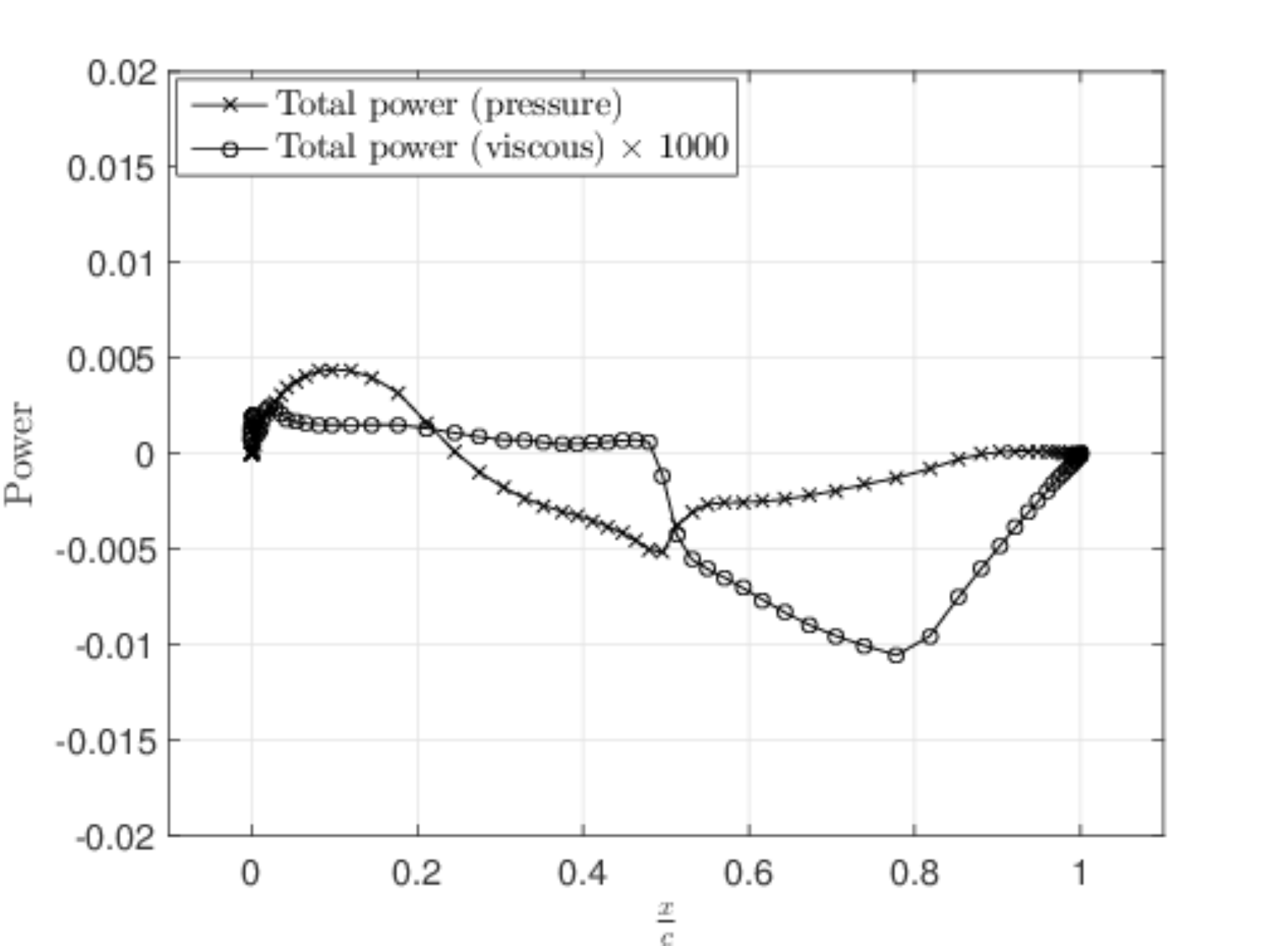}}
\caption{Instantaneous power distribution on airfoil for shock induced boundary layer separation}
\label{unsteady3_inst_tot_power}
\end{figure}

The total power is the rate of doing work by the pressure and viscous forces acting along the camber line of the airfoil. This power is equal to the sum of the rate of doing work by the respective forces on the upper and lower surfaces corresponding to the points on the camber line. The computed total power on the airfoil camber line is shown in Figure~\ref{unsteady3_inst_tot_power} at different instants. Observe that the power contribution by viscous forces is very less, typically three orders of magnitude less relative to the power contribution by the pressure forces. Positive aerodynamic power implies that the aerodynamic force is in the direction of motion of the airfoil. Figure~\ref{u3_51} to Figure~\ref{u3_55} and Figure~\ref{u3_41} to Figure~\ref{u3_45} shows the instants where the airfoil executes its motion from approximately its maximum nose up position, $\alpha_{max}$, to its minimum nose up position, $\alpha_{min}$. The power in Figure~\ref{u3_55} and Figure~\ref{u3_45} are zero all over the airfoil as the velocity of the airfoil at this instant is zero. The next instant shown in Figure~\ref{u3_56} and Figure~\ref{u3_46} is to indicate that the power distribution on the airfoil changes its sign relative to those shown in Figures~\ref{u3_54} and \ref{u3_44} as the velocity reverses direction. The power distribution due to pressure forces changes its sign at the quarter chord as the Cartesian velocity components change their direction at this point. A sudden change in the power distribution by the viscous forces can be observed at the point of boundary layer separation.

\begin{figure}
\subfigure[Pressure distribution on the airfoil]{\label{u3_11_e}\includegraphics[scale=0.40]{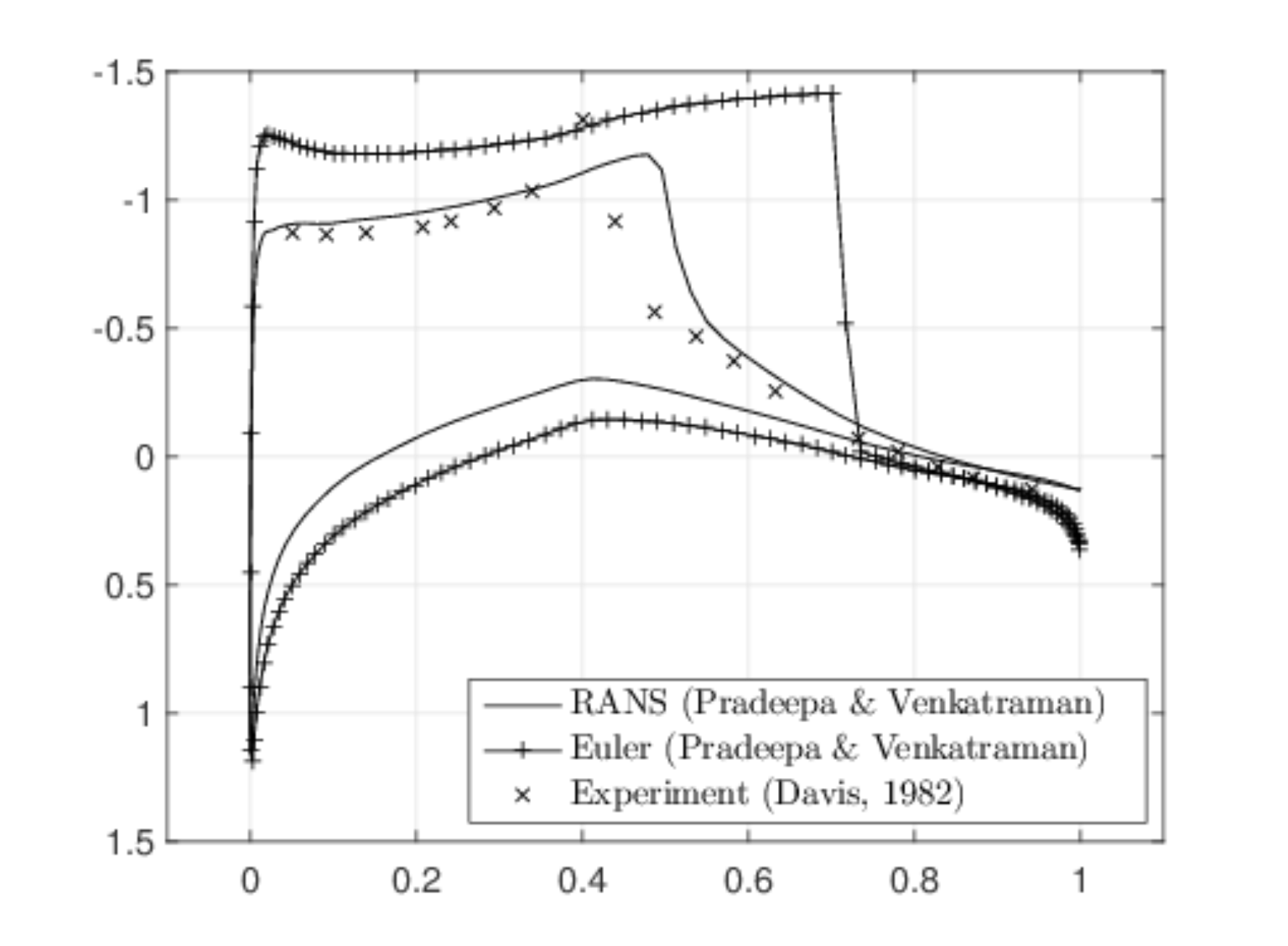}}
\subfigure[Mach contours computed with Euler solver]{\label{u3_31_e}\includegraphics[scale=0.40]{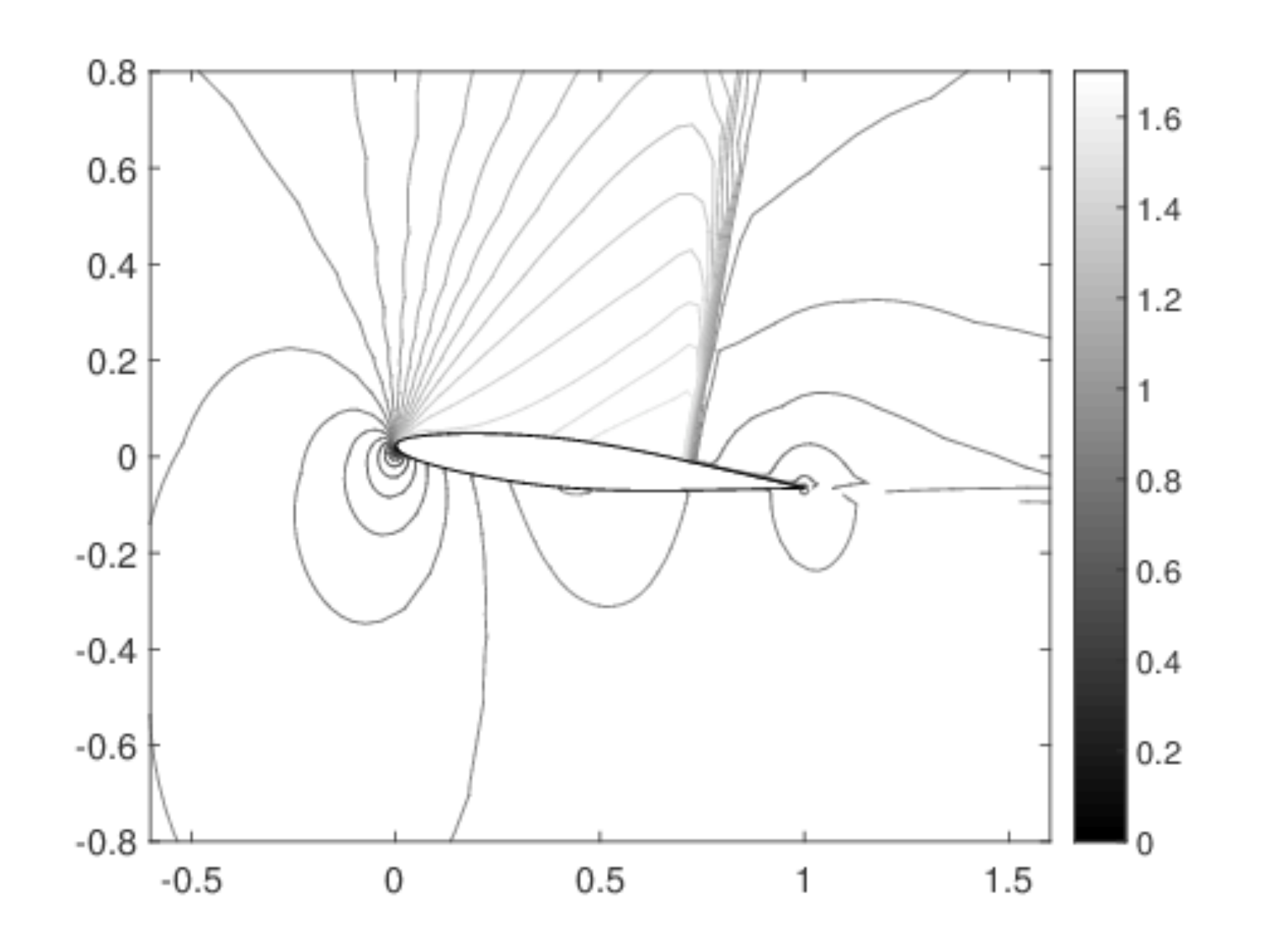}}
\caption{Comparison of Euler results with RANS results at $\alpha = 4.99^\circ, \varphi = 98.43^\circ$ for shock-stall test case }
\label{unsteady3_inst_compare_euler}
\end{figure}

\begin{figure}
\subfigure[Shock displacement variation]{\label{shock_disp}\includegraphics[scale=0.40]{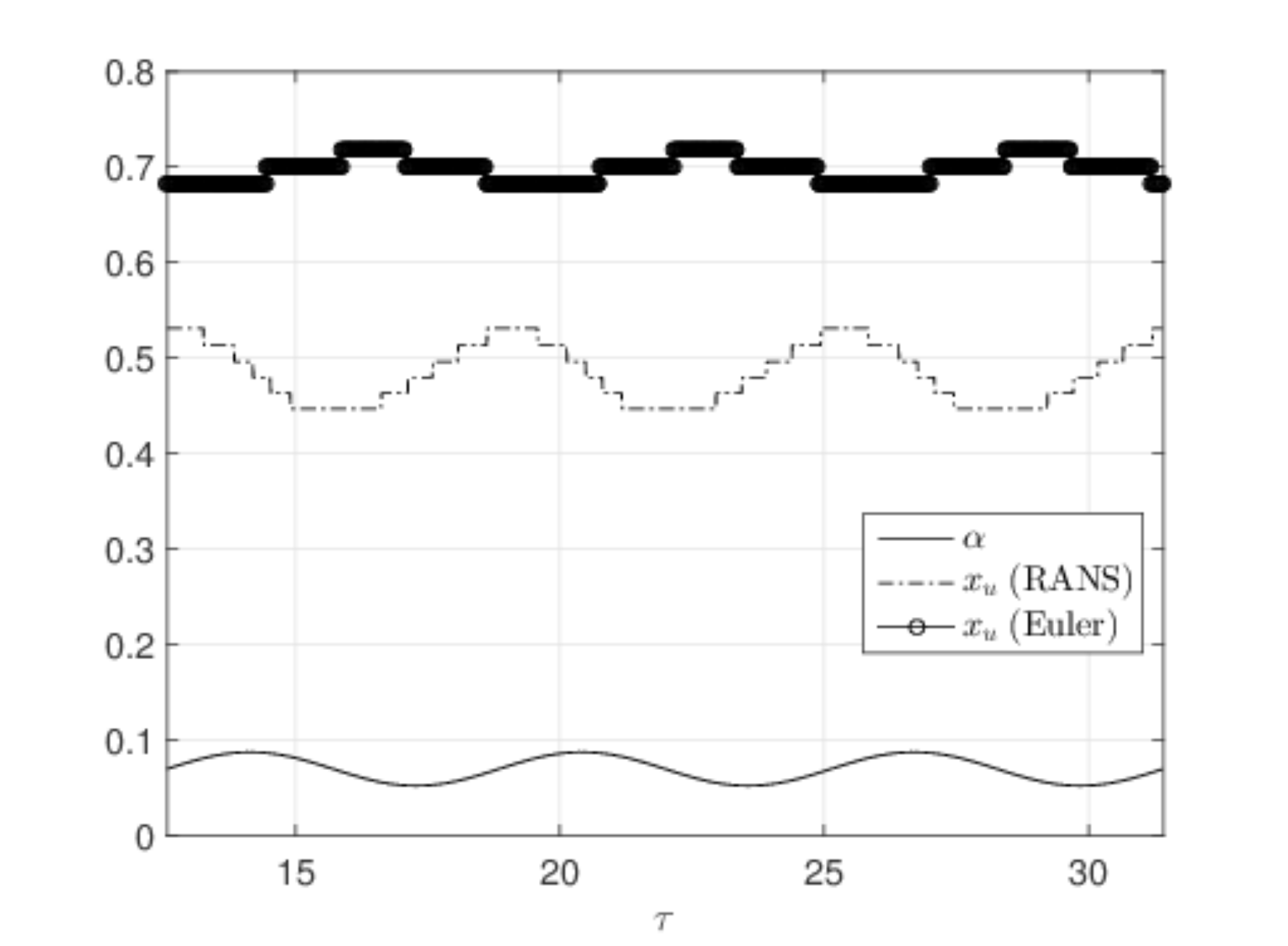}}
\subfigure[Shock strength variation]{\label{shock_strength}\includegraphics[scale=0.40]{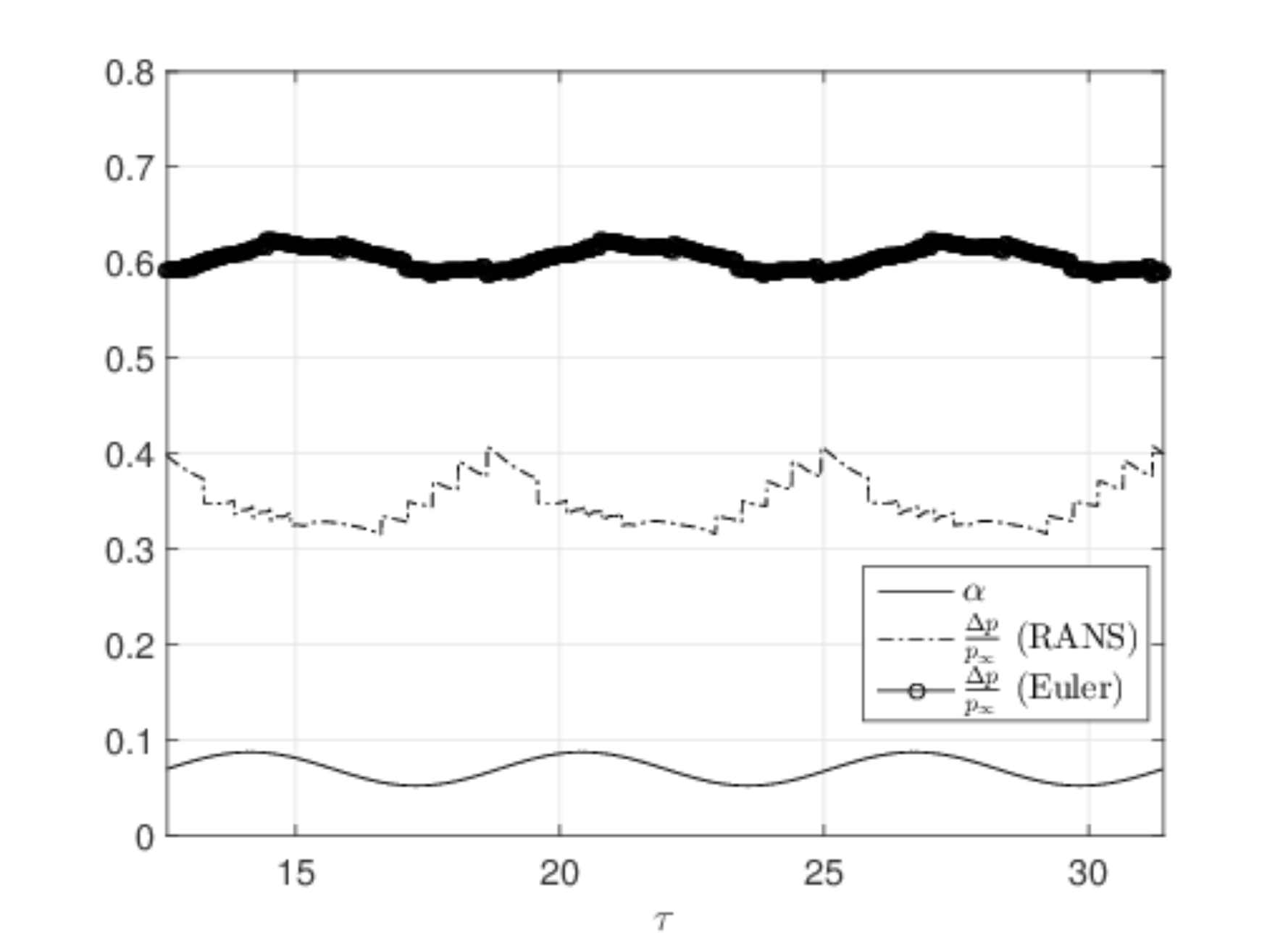}}
\caption{Comparison of shock dynamics of Euler with RANS results for shock-stall test case}
\label{unsteady3_shock_parameter}
\end{figure}

The influence of viscosity on unsteady transonic flow can be appreciated when the pressure distribution and Mach contours computed for an inviscid flow using Euler equations are compared with viscous simulation results and shown in Figure~\ref{unsteady3_inst_compare_euler}. The inviscid results do not capture the boundary layer and its separation. As a result the captured shock using Euler equations is stronger compared to the one captured using RANS simulations and that from experiment. The inviscid shock location is shifted towards the trailing edge compared to the actual shock location from experiment as well as the RANS result. Therefore, though the viscous forces are insignificant in terms of their contribution to the total aerodynamic power---that is almost three orders of magnitude smaller relative to the power contributed by pressure forces---viscosity changes the flow characteristics significantly by shifting the shock location as a consequence of shock-induced flow separation, thus changing the power distribution on the airfoil by the pressure forces.

The shock displacement on the upper surface of the airfoil and its strength are shown as a function of non-dimensional time $\tau = \omega t$ in Figure~\ref{unsteady3_shock_parameter}. The shock location in Figure~\ref{shock_disp} computed by an inviscid solver is always much aft of the shock location computed by our RANS solver. But the amplitude of shock displacement is less in the inviscid computations when compared with the RANS result. The shock strength variation on the airfoil upper surface with time, Figure~\ref{shock_strength}, shows that the inviscid solver captures a much stronger shock compared to that calculated by the RANS solver. These results once again reiterate the strong influence of shock-boundary layer interaction on the shock motion and shock strength which then drastically changes the aerodynamic loading on the oscillating airfoil.

The final observation in this study of the interaction of viscosity and shock is the phase-difference between the unsteady shock motion as predicted by the RANS and Euler solvers as shown in Figure~\ref{unsteady3_shock_parameter}. The phase difference between the two shock motions is almost $\pi$ radians implying that the shock motion predicted by the RANS solver is exactly opposite to that predicted by the Euler solver. Therefore, inclusion of viscosity reverses the direction of motion of the shock. As an illustration, in the inviscid case, consider the instant where the nose-up motion of the airfoil leads to the motion of the shock towards the trailing edge with an increase in shock strength as the rate of expansion has increased. But in the viscous simulation, the boundary layer will not allow the shock to move towards the trailing edge as it cannot sustain the adverse pressure gradient posed by the shock. Instead the shock moves upstream leading to an early separation. This is the reason for the shock-reversal observed here. Apart from these boundary layer interactions on the shock motion, shock does not move in phase with the airfoil motion as it also depends on the frequency of pitching of the airfoil.

\section{Summary}
\label{sec:concl}

In transonic flows, even at low angles of incidence of the airfoil, and low amplitudes of oscillation, viscosity influences the aerodynamic forces by changing the profile of the solid boundary as seen by the external fluid flow. Steady viscous flow computations predict an early and weak shock when compared to the one predicted by an inviscid solver. The boundary layer on the solid surface ensures a slower expansion of the flow leading to an early weak shock. A diffused shock-foot in viscous computation is caused by the boundary layer that allows information to pass upstream within it making the shock-foot diffused. Small amplitude oscillations about a small mean angle of incidence of the airfoil in a transonic flow does not result in a shock induced flow separation. The shock location and strength captured by the RANS solver are closer to those from experiment than those from inviscid computations. The inviscid flow solver captures a stronger shock aft of the experimentally predicted shock location. 

Shock boundary layer interactions in transonic flows, however, become significant and exhibit interesting behaviour at moderate mean angles of attack but with small amplitude motions superposed on it. The adverse pressure gradient induced because of the strong shock in the flow, as well as the moderately high angle of incidence, leads to flow separation. The shock strength predicted by RANS simulation as well as experiment are weaker when compared with those obtained from inviscid simulations. The inviscid shock location is shifted towards the trailing edge compared to the actual shock location from experiment as well as the RANS result. Power input to the airfoil by viscous forces was found to be very less, around three orders of magnitude lower than power due to pressure forces. Though viscosity does not contribute much to the total power distribution, it led to large variation in total power distribution on the airfoil surface by manipulating the shock location, in turn leading to a significant redistribution of power input by pressure forces. Shock motion, when there is boundary layer separation, as predicted by our RANS solver, was exactly opposite to that of the inviscid simulation. The boundary layer, being unable to sustain the adverse pressure gradient, was found to be responsible for shock-reversal.

The implications of these findings are critical in the understanding of transonic flutter of wings or lifting surfaces. The role of viscosity, through the boundary layer, dramatically changes and redistributes the energy pumped into the elastodynamics of the wing by the dominant pressure forces.

\bibliographystyle{aipauth4-1}
\bibliography{ptkkvarxivx2021}

\end{document}